\documentclass[aps,twocolumn,prc,superscriptaddress,showpacs,nofootinbib,floatfix,amssymb,amsfonts,amsmath]{revtex4-1}

\usepackage{graphicx}% Include figure files
\usepackage{dcolumn}% Align table columns on decimal point
\usepackage{bm}% bold math
\usepackage{xcolor}
\usepackage{amsmath}    % need for subequations
\usepackage{amsfonts}   %note how statements can be commented out
\usepackage{amssymb}
\usepackage{graphicx}   % for figuresManMan//
\usepackage{multirow} 

\begin{document}

\title{Towards accurate nuclear mass tables in covariant density functional 
        theory}

\author{A. Taninah}
\affiliation{Department of Physics and Astronomy, Mississippi
State University, MS 39762}
\affiliation{Department of Chemistry and Physics, Louisiana 
                State University, Shreveport, LA 71115.}

\author{B. Osei}
\affiliation{Department of Physics and Astronomy, Mississippi
State University, MS 39762}

\author{A. V. Afanasjev}
\affiliation{Department of Physics and Astronomy, Mississippi
State University, MS 39762}

\author{U. C. Perera}
\affiliation{Department of Physics and Astronomy, Mississippi
State University, MS 39762}

\author{S. Teeti}
\affiliation{Department of Physics and Astronomy, Mississippi
State University, MS 39762}

\date{\today}

\begin{abstract}

 The current investigation focuses on detailed analysis of the anchor
based optimization approach (ABOA), its comparison with alternative 
global fitting protocols and on the global analysis of the truncation of 
basis effects in the calculation of binding energies. It is shown that 
ABOA provides a solution which is close to that obtained in alternative
approaches but at small portion of their computational time. The
application of softer correction function after few initial iterations of
ABOA stabilizes and speeds up its convergence. For the 
first time,  the numerical errors in the calculation of binding energies 
related to the truncation of bosonic and fermionic bases have been 
globally investigated with respect of asymptotic values corresponding 
to the infinite basis in the framework of covariant density functional
theory (CDFT).  These errors typically grow up with the increase of the 
mass and deformation of the nuclei.  To reduce such errors  in bosonic sector 
below 10 keV for  almost all nuclei with proton number $Z<120$
one should truncate the bosonic basis at $N_B=28$ instead of 
presently used $N_B=20$. The reduction of the errors in binding 
energies due to the  truncation of  the fermionic basis in CDFT is 
significantly more numerically  costly. For the first  time it is shown 
that the pattern and the speed of the convergence of binding energies 
as a function of the size of fermionic basis given by $N_F$ depend 
on the type of covariant energy density functional. The use of 
explicit density dependence of the meson-nucleon coupling constants 
or point couplings slows down substantially the speed of convergence of binding 
energies as a function of $N_F$. The present paper clearly 
indicates the need for accounting of infinite basis corrections for
accurate calculation of binding energies and for fitting more
precise next generation covariant energy density functionals: note that 
such  corrections are neglected in the present generation of the
functionals. A new procedure for finding the asymptotic values of 
binding energies is suggested in the present paper: it allows
better control of numerical errors.
\end{abstract}

\maketitle

%%%%%%%%%%%%%%%%%%%%
\section{Introduction}
%%%%%%%%%%%%%%%%%%%%

   Nuclear binding energies (or masses) is one of the most fundamental properties
of atomic nuclei.  In experiment they are studied with extreme relative mass 
precision  of  $\delta m/m = 10^{-9}$ and in some nuclei even with precision of 
$10^{-10}$ \cite{MLO.03,DBBE.18}. In absolute terms this corresponds to the 
accuracy of the measurements of the binding energies better than 1 keV. 
Accurate knowledge of binding energies is important for nuclear structure,
neutrino physics and nuclear astrophysics.  The measurements of nuclear 
masses far from stability provide a fundamental test for our understanding 
of nuclear structure \cite{MLO.03} and the constraints on model predictions
for the boundaries of nuclear landscape (see Ref.\ \cite{Eet.12,AARR.13,AARR.14}).
The measurements of mass differences ($Q$ values) of specific isotopes play a 
key role in neutrino physics: in order to determine the neutrino mass on a sub-eV 
level one should measure the $Q$ values of certain $\beta$ transitions better 
than 1 eV \cite{DBBE.18}. Different nuclear astrophysical processes [such as r-process 
(see Refs.\ \cite{AGT.07,CSLAWLMT.21}) or the processes in the crust of neutron 
stars (see Refs.\ \cite{CH.08,LBGABDG.18})] also sensitively depend on nuclear masses 
since they affect nuclear reaction rates.

   There was a substantial and continuous effort in non-relativistic models to
improve the functionals and global reproduction of experimental masses
(see, for example, Refs.\ \cite{TGPO.00,SGHPT.02,GCP.13,BCPM,MSIS.16,UNEDF0,UNEDF1,UNEDF2}
and references quoted therein).  Existing global fits of the model 
parameters to the ground state masses reach an accuracy of  $\approx 0.5$ MeV 
in the global reproduction of experimental masses in the microscopic+macroscopic 
(mic+mac) approach \cite{MSIS.16} and in Skyrme density functional theory (DFT) 
\cite{GCP.13} and around 0.8 MeV in the Gogny  DFT \cite{D1M}. These fits 
take into account the correlations beyond mean field by either adding rotational and 
vibrational corrections in a phenomenological way or by employing five-dimensional 
collective Hamiltonian (5DCH). Note that the global fits at the mean field level in 
the Skyrme DFT lead to $1.4 - 1.9$ MeV accuracy in the global description of 
experimental binding energies (see Refs.\ \cite{UNEDF0,UNEDF1,UNEDF2}).

   In contrast, the attempts to create covariant energy density functionals (CEDF)
based on the global set of data on binding energies are very limited in covariant
density functional theory (CDFT). It is only in Refs.\ \cite{AGC.16,TA.23} that they
were undertaken (see Sec.\ \ref{global-fitting} for more detail).  This is due to 
the fact that the calculations within the CDFT are significantly more numerically 
challenging and more time-consuming as compared with those carried
out in nonrelativistic DFT due to relativistic nature of the CDFT.  First, the 
wavefunctions of the single-particle states  in the CDFT are represented by 
Dirac spinors the large and small components of which have opposite parities. 
As a consequence, the basis for a given maximum value of principal quantum 
number $N$ in the basis set expansion has approximately double size in the 
CDFT  as  compared with non-relativistic  models.
As a result, the numerical calculations of matrix elements and the diagonalization 
of the matrices requires substantially more time than in the case of nonrelativistic DFTs.
A rough estimate provided in Ref.\ \cite{NL1} for spherical symmetry indicates that 
the calculations in the Skyrme DFT and CDFT for a given maximum value of principal
quantum number $N$ in the basis set expansion differ by approximately an order of 
magnitude. Second, the most widely used classes of CEDFs contain the mesons
in addition to fermions \cite{VALR.05,AARR.14}. This requires the solution of 
Klein-Gordon equations which adds numerical cost to the calculations. Third, in the 
CDFT the nucleonic potential ($\approx -50$ MeV/nucleon) emerges as the sum of 
very large attractive scalar $S$ ($\approx -400$ MeV/nucleon)
 and repulsive vector 
$V$ ($V \approx 350$ MeV/nucleon) potentials (see Ref.\ \cite{VALR.05}). In the nucleus 
with mass $A$, these values are multiplied by $A$ and this leads to a cancellation of 
very large quantities. This may lead to a slower convergence of binding energies as a 
function of the size of the basis in the CDFT as compared with non-relativistic DFTs.
 
     Many of the fits of the relativistic and non-relativistic 
functionals (including global ones) have been carried out using computer codes employing
basis set expansion approach in which the wave functions are expanded into the 
basis of harmonic oscillator wave functions. This expansion is precise for the infinite 
size of the basis. However, due to numerical reasons the basis is truncated in the 
calculations.  Unfortunately, a rigorous analysis of numerical errors in calculated 
binding energies emerging from the truncation of basis is missing in global 
calculations. However, there are some indications that they are not negligible. For 
example, the relativistic Hartree-Bogoliubov (RHB) calculations of the ground state in 
$^{208}$Pb with DD-ME2 functional indicate that binding energies calculated with 
$N_F=20$ fermionic shells deviate from asymptotic value of binding energy $B_{\infty}$ 
by approximately 250 keV (see discussion of Fig. 2 in Supplemental Material to Ref.\ 
\cite{TA.23}). Note that this difference represents only 0.015\% error in the description 
of total binding energy.  In a similar way, total binding energies of the $^{120}$Sn and 
$^{102,110}$Zr nuclei obtained in the Skyrme DFT calculations with the SLy4 force in 
the basis of $N=20$ and $N=25$ harmonic oscillator (HO) shells differ by  110-150 
keV \cite{DSN.04,PSFNSX.08}. Ref.\ \cite{PSFNSX.08} also indicates that the HO 
basis with $N=25$ is needed in order to describe the binding energies of the nuclei 
in this mass region with an accuracy of the couple of tens keV.

   Thus, the present paper aims at the investigation of several issues related to
the calculations of binding energies with high precision in the CDFT framework. One 
of the goals is to understand the numerical errors in the calculations of binding 
energies. In particular, their global dependence on the truncation of the basis both 
in the fermionic and  bosonic sectors of the CDFT will be investigated for the first time.
We will also study how these numerical errors depend on the type of the 
symmetry (spherical versus axially deformed) of the nuclear shape and how the 
convergence of binding energies as a function of the size of the basis depends on 
the type of functional.  The second goal is to understand  to which extent the procedures 
for the definition of asymptotic binding energies employed earlier in non-relativistic 
functionals are applicable in the CDFT. The third goal is to further understand the role 
and the features of the anchor based optimization method suggested in Ref.\ \cite{TA.23} 
as an alternative to other methods of global fitting of the functionals.

   The paper is organized as follows. A short review of the approaches to global 
fitting  of the energy density functionals (EDF) is presented in Sec.\ \ref{global-fitting}. 
Sec.\  \ref{Theory} provides a brief outline of theoretical formalism and the discussion 
of major classes of CEDFs under study. A discussion of the anchor based approach 
to the optimization of energy density functionals and its comparison with alternative 
methods for global fitting of the functionals is given in Sec.\ \ref{anchor-further}. 
The impact of the truncation of basis in bosonic and fermionic sectors of the CDFT 
on binding energies as well as asymptotic behavior of the calculated binding energies 
are analysed in Sec.\ \ref{truncation-effects}. A new method of defining asymptotic 
values  of calculated binding energies is presented in Sec.\ \ref{fermion-sect-asymptot-rel}. 
Finally, Sec.\ \ref{Concl} summarizes the  results of our paper. 
 
%%%%%%%%%%%%%%%%%%%%%%%%%%%%%%%%%
\section{A review of approaches to global fitting of functionals} 
\label{global-fitting}
%%%%%%%%%%%%%%%%%%%%%%%%%%%%%%%%%
 
   The analysis of literature reveals that at present there are three 
different approaches to global fitting of energy density functionals.
These are
  
\begin{itemize}

\item
   {\bf Fully Global Approach (further FGA)}. In the FGA the optimization of 
   the functional is performed using computer codes which include either
   axial or triaxial deformation. It allows to use in 
   fitting protocol the quantities which depend on deformation such as fission 
   barriers which are not accessible in alternative fitting protocols. 

\item
    {\bf Reduced Global Approach (further RGA)} has been outlined in Sec. II
    of Ref.\ \cite{TGPO.00}.  
    The specific feature of this approach is the fact that the deformation effects
    are taken into account in deformed and transitional nuclei by calculating 
    the deformation energy $E_{def}$ 
\begin{eqnarray} 
E_{def}(Z,N) = E_{sph}(Z,N) - E_{eq}(Z,N)  
\label{def-energy} 
\end{eqnarray}   
where $E_{eq}$  is the energy at the equilibrium deformation and $E_{sph}$
is the energy of spherical configuration.   The  $E_{def}$ values are calculated 
using axially deformed code and then all experimental binding energies are 
renormalized to their "equivalent spherical configuration" values. Then, the rms errors of the
current iteration of the functional are calculated using a spherical  code by comparing 
the binding energies it gives with these {\it renormalized} experimental binding energies. Thus, each 
iteration of optimization consists of defining the $E_{def}$ in deformed code and 
fitting the functional in spherical code.  However, to reach an optimal fit one should 
repeat iterations until full convergence of $E_{def}(Z,N)$ is reached.

\item
    {\bf Anchor Based Optimization Approach (further ABOA).} In the ABOA  
   (see Ref.\ \cite{TA.23}), the optimization of the parameters of EDFs is carried out 
   for a selected set of spherical anchor nuclei, the physical observables of which are 
   modified by the correction function 
\begin{eqnarray} 
E_{corr}(Z,N) = \alpha_i(N-Z) + \beta_i(N+Z) + \gamma_i
\label{corr-func} 
\end{eqnarray} 
which takes into account the global performance of EDFs. Here $i$ is 
the counter of  the iteration in the anchor-based optimization. The parameters 
$\alpha_i$ and $\beta_i$ correct the deficiencies of EDF in isospin and in mass number
while $\gamma_i$ is responsible for a global shift in binding energies\footnote{Note that similar 
in spirit approach is used in Ref.\ \cite{D1M*} in which the quadrupole zero-point energy 
is replaced by a constant binding energy shift [similar to the $\gamma_i$ term in Eq.\ 
(\ref{corr-func})] which is fixed by minimizing the global rms deviation.}.
\end{itemize}  
  Each of these approaches has its own advantages and disadvantages 
which are discussed below.
  
      The FGA potentially allows to avoid all uncertainties related 
to the selection of nuclei used in the fitting protocol and allows to use the physical 
quantities which depend on deformation (such as the height of fission barriers)
in fitting protocol. However, the calculations in deformed codes are extremely
numerically expensive and computational time does not depend drastically on 
the selection of the functional (see Table \ref{table-compt-time-deform}). 
For example, within the CDFT framework the numerical calculations in deformed 
RHB code  are by more than two orders of magnitude more time consuming than 
those in spherical RHB one for the fermionic basis characterized by full $N_F=16$
fermionic shell (see columns 4 and 5 in Table \ref{table-compt-time-compar}). 
Further increase of fermionic basis leads to a drastic increase of the ratio of 
the calculational time in deformed and spherical RHB calculations: this ratio
is approaching three orders of magnitude at $N_F=18$ (see Table 
\ref{table-compt-time-compar}).The simplification of the code by the transition 
to the relativistic mean field (RMF)+BCS framework does not change much 
this ratio. Similar situation holds also in non-relativistic Skyrme and Gogny 
DFT.

%%%%%%%%%%%%%%%%%%%%%%%%%%%%%%%%%%%%%%%%%%%%%%%%%%
\begin{table}[htb]
\centering
\caption{Average computational time per nucleus (in CPU-hours)  in axially
              deformed RHB code as a function of the size of fermionic basis 
              (given by $N_F$) for indicated CEDFs. It is defined for the ground 
              states of the set of 100 nuclei more or less equally distributed over 
              the part of nuclear chart for which experimental data are available.
              Note that as a rule computational time increases for the deformations
              of the basis deviating from the equilibrium one.              
\label{table-compt-time-deform}
              }
\begin{tabular}{ |c |c |c |c |c |} 
\hline
\multirow{2}{*}{$\mathrm{N_F}$} & \multicolumn{4}{c|}{Computational time per nucleus [CPU-hours]} \\ \cline{2-5}
                                & DD-MEX  & DD-PC1  & NL5(E)  & PCPK1  \\ \hline
                            16  & 0.26    & 0.15    &  0.19   & 0.19   \\
                            18  & 0.51    & 0.35    &  0.40   & 0.39   \\
                            20  & 1.05    & 0.68    &  0.68   & 0.68   \\
                            22  & 2.02    & 0.98    &  1.43   & 1.22   \\
                            24  & 3.36    & 2.02    &  2.57   & 2.66   \\
                            26  & 5.17    & 4.08    &  4.48   & 4.20   \\
                            28  & 8.21    & 6.39    &  7.77   & 7.49   \\
                            30  & 15.94   & 11.11   &  13.34  & 11.66  \\                             
\hline
\end{tabular}
\end{table}
%%%%%%%%%%%%%%%%%%%%%%%%%%%%%%%%%%%%%%%%%%%%%%%%%%

%%%%%%%%%%%%%%%%%%%%%%%%%%%%%%%%%%%%%%%%%%%%%%
\begin{table}[h!]
\label{table-compt-time-compar}
\centering
\caption{Columns (2) and (3): the same as in Table \ref{table-compt-time-deform} but 
              for the calculations in spherical RHB code for the same set of 100 nuclei. 
              Columns (4) and (5) show the ratio of computational time in deformed and 
              spherical RHB calculations.  
}
\begin{tabular}{ |c |c |c |c |c |} 
\hline
\hline
\multirow{3}{*}{$\mathrm{N_F}$} & \multicolumn{2}{c|}{Computational time} 
& \multicolumn{2}{c|}{Ratio} \\
\multirow{3}{*}{} & \multicolumn{2}{c|}{per nucleus [CPU-hours]} 
& \multicolumn{2}{c|}{} 
\\ \cline{2-5}
                                 &  2 & 3& 4& 5 \\
                                  & DD-MEX  & DD-PC1 & DD-MEX  & DD-PC1  \\ \hline
                            16  & 0.000440  & 0.000238  & 590.91  & 630.25  \\
                            18  & 0.000596  & 0.000365  & 855.70  & 958.90  \\
                            20  & 0.000646  & 0.000590  & 1625.39 & 1152.54 \\
                            22  & 0.001233  & 0.000700  & 1638.28 & 1400.00 \\
                            24  & 0.001654  & 0.000953  & 2031.44 & 2119.62 \\
                            26  & 0.002163  & 0.001264  & 2390.20 & 3227.85 \\
                            28  & 0.002401  & 0.001634  & 3419.41 & 3910.65 \\
                            30  & 0.002758  & 0.001983  & 5779.55 & 5602.62 \\                          
  \hline
\end{tabular}
\end{table}
%%%%%%%%%%%%%%%%%%%%%%%%%%%%%%%%%%%%%%%%%%%%%%

    The FGA has been used in the number of studies but there 
are two possible ways of its application. In the first method, a large basis is used to 
reduce numerical errors in the calculations of binding energies but this restricts the 
number of the nuclei which can be included in the fitting protocol.  However, the lack 
of strict mathematical procedure in the selection of nuclei leads to some subjectivity 
and possibly to some biased errors. This problem can be reduced by the inclusion of 
a larger set of nuclei into the fitting protocol.  However, the computational cost raises 
drastically so that  the increase  of the number of nuclei leads to a substantial reduction 
of  the basis size used in the calculations because of the limitations imposed on 
availability of computational resources. This, in turn requires the use of approximate 
relations to find the asymptotic value of the binding energy corresponding to infinite size 
of basis (see Sec.\ \ref{truncation-effects} for details). Thus,  a very large number of the 
nuclei is used in the fitting protocol of the second method but this requires the use of 
moderate basis in numerical calculations and approximate relations to find the asymptotic 
value of the binding energy corresponding to infinite size of basis.
 
   For example, the first method was used in Refs.\ \cite{UNEDF0,UNEDF1,UNEDF2}  in 
Skyrme DFT for the development  of the UNEDF class of the functionals. Note that fitting protocol of the 
UNEDF1 and UNEDF2 functionals has been supplemented by  information on fission 
barriers and single-particle energies, respectively. In all optimizations, the large spherical 
basis with 20 harmonic oscillator shells ($N=20$) was used which led to quite 
substantial computational time in the calculations with deformed HFBTHO code.  As a 
consequence, a limited number of nuclei (approximately 28 spherical and 46 deformed) 
were used in these optimizations.

  The second method has been used in the fitting protocol of  the BCPM  (Barcelona-Catania-Paris-Madrid) 
functional (see Ref.\ \cite{BCPM}). It included 579 measured  masses of even-even nuclei and  
the calculations have been carried out in a reduced basis which takes into account eleven harmonic 
oscillator shells for the $Z<50$ nuclei, 13 shells for the $50 < Z < 82$ nuclei, and 
15 shells for $Z>82$ nuclei. Another example is the D1M and D1M* functionals 
which were fitted to 2149 and 620 nuclei, respectively, in axially deformed calculations 
corrected by quadrupole corrections to total binding energy and infinite-basis 
corrections $\Delta B_{\infty}$ [see Eq.\ (\ref{B-infty}) below] in Refs.\ \cite{D1M,D1M*}. 
Note that only even-even nuclei were considered in the fit of the D1M* functional.  The 
axially deformed calculations for these two functionals were  performed in the basis 
with $N\leq 14$ major oscillator shells.

   There is also a hybrid method which was used in the fitting of the 
DD-MEB1 and DD-MEB2 CEDFs in Refs.\ \cite{AGC.16}: this fitting protocol
contains 2353 experimental masses. The minimization of 
these functionals was performed in axially deformed RHB code with $N_F=16$ 
fermionic  shells but with calculated  binding energies corrected before and after 
minimization by the difference between the results obtained with $N_F=16$ and 
$N_F=20$ (see Ref.\ \cite{Goriely.23-private}).

   The RGA  has been used in the fitting protocol of the BSk1-32 series of 
the Skyrme functionals using very large bases (see, for example, Refs.\ 
\cite{SGHPT.02,GCP.13,GCP.16,Goriely-private}).  For instance, it consisted of the 
$N=25$ HO shells in spherical calculations and $N=21$ HO 
shells in deformed calculations  in Ref.\ \cite{SGHPT.02}.  This allows to keep numerical 
error in calculated masses relatively low without involving the correction for infinite basis 
size. Note that fitting protocols of these functionals typically include all nuclei for which 
experimental data on binding energies is available.
 
   The ABOA has been introduced in Ref.\ \cite{TA.23}. It was shown 
that the use of this approach leads to a substantial improvement in the global description of 
binding energies for several classes of CEDFs (see also Sec.\ \ref{conv-improv} 
below). This approach leads to the results which are comparable to those obtained in 
RGA (see Ref.\ \cite{TA.23}  and the discussion in Sec.\ \ref{ABOA-vs-RGA}). The computational 
cost of defining a new functional within this approach is drastically lower as compared with FGA 
and RGA and this aspect is discussed in detail in Sec.\ \ref{comp-appr} below. It turns out that the 
use of  twelve anchor spherical nuclei distributed over the nuclear chart in the optimization  
is sufficient for this  approach (see Ref.\ \cite{TA.23}). Note that existing optimizations 
of CEDFs within the ABOA have been obtained in the RHB calculations with $N_F=20$ 
fermionic  and $N_B=20$ bosonic shells (see Ref.\ \cite{TA.23}). Such a truncation of 
the basis is used also in other recent CEDF optimizations\footnote{Unfortunately, the majority of the papers dedicated to the optimization of CEDFs do not explicitly 
provide information on the size of the basis used in the fitting protocols.} (see, for example, Refs.\ \cite{PC-PK1,TAAR.20}). 
However, there are also the cases when smaller basis was used: for example, the 
DD-PC1 functional has been defined in deformed calculations with $N_F=16$ 
fermionic shells \cite{DD-PC1}.

It is necessary to mention that in these investigations the numerical errors in the binding 
energies associated with the truncation of basis or the definition of their asymptotic values  
have been either discussed only for a few selected nuclei or not 
mentioned at all.  Thus, at present there is no published information on how these errors
evolve across the nuclear landscape and how they depend on the type of the functional
or theory employed. Thus, one of the goals of the present study is to close this gap in
our knowledge in the CDFT framework.

%%%%%%%%%%%%%%%%%%%%%%%%%%%%%%%%%%%%
\section{Theoretical framework} 
\label{Theory}
%%%%%%%%%%%%%%%%%%%%%%%%%%%%%%

   The calculations were performed in the framework of relativistic 
Hartree-Bogoliubov (RHB) approach the technical details on the solution of which 
can be found in Refs.\ \cite{VALR.05,AARR.14,DIRHB-code.14}. The calculations 
were carried out with two versions of the RHB code suitable for spherical and axially 
deformed nuclei, respectively.  Both codes allow the calculations with a wide range 
of covariant energy density functionals. We employ a parallel version of the axial 
RHB computer code developed in Ref.\ \cite{AARR.14}: it allows simultaneous 
calculations for a significant number of nuclei and deformation points in each 
nucleus. The unconstrained calculations are performed in axial RHB code using 
four initial deformations of the basis $\beta_2=-0.2, 0.0, 0.2$ and 0.4. As follows 
from the comparison with the calculations constrained on $\beta_2$ this procedure 
guarantees finding the lowest in energy minimum but it is substantially less 
numerically costly as compared with constrained calculations.

   Both spherical and axial RHB computer codes are based on the expansion  
of the Dirac spinors and the meson fields in terms of harmonic oscillator wave 
functions with respective symmetry [so-called basis set expansion method] (see 
Ref.\ \cite{GRT.90,DIRHB-code.14}). We  illustrate here this expansion for 
the case of axially deformed nuclei (see Refs.\ \cite{GRT.90,DIRHB-code.14} for 
more details and for the case of spherical symmetry).

The single-particle wave function $\psi_i (\vec{r},s,t)$ of the nucleon in the state $i$, which follows
from the solution of the Dirac equation, is given by
\begin{eqnarray}
\psi_i (\vec{r},s,t) = \binom{f_i(\vec{r},s,t)}{ig_i(\vec{r},s,t)}
%\psi_i (\vec{r},s,t) = \left( \begin{array}.  f_i(\vec{r},s,t) \\ ig_i(\vec{r},s,t) \end{array} \right)
\end{eqnarray}
where the large ($f_i(\vec{r},s,t)$) and small ($g_i(\vec{r},s,t)$) components of a Dirac 
spinor are expanded independently in terms of the HO eigenfunctions 
$\Phi_{\alpha}(\vec{r},s)$
\begin{eqnarray}
f_i(\vec{r},s,t) = \sum_{\alpha}^{\alpha_{max}} f_{\alpha}^{(i)} \Phi_{\alpha}(\vec{r},s) \chi_{t_i}(t) 
\label{Eq-large}
\\
g_i(\vec{r},s,t) = \sum_{\tilde{\alpha}}^{\tilde{\alpha}_{max}} f_{\tilde{\alpha}}^{(i)} \Phi_{\tilde{\alpha}}(\vec{r},s) \chi_{t_i}(t). 
\label{Eq-small}
\end{eqnarray}
The HO eigenfunctions have the form (see Ref.\ \cite{DIRHB-code.14}) 
\begin{eqnarray} 
\Phi_{\alpha}(\vec{r},s) = \varphi_{n_z} (z_,b_z) \varphi_{n_r}^{\Lambda} (r_{\perp}, b_{\perp})
\frac{e^{i\Lambda\phi}}{\sqrt{2\pi}} \chi(s)
\end{eqnarray} 
where $n_z$ and $n_r$ are the number of nodes in the $z$- and $r_{\perp}$- directions, 
respectively, and $\Lambda$ and $m_s$ are projections of the orbital angular momentum 
and spin on the intrinsic $z$-axis, respectively. $b_z$ and $b_{\perp}$ are oscillator lengths
in respective directions.

The quantum numbers 
$\left| \alpha \right> = \left| n_z n_r \Lambda s \right>$ and  
$\left| \tilde{\alpha} \right> = \left| \tilde{n_z} \tilde{n_r} \tilde{\Lambda} \tilde{s} \right>$ in the 
sums of Eqs. \ (\ref{Eq-large}) and (\ref{Eq-small}) run over all permitted combinations of 
$n_z$ and $n_r$, $\Lambda$ and $m_s$ (and their tilted counterparts) defined by the 
truncation of the basis. 
In absolute majority of the CDFT calculations, full fermionic HO shells are used in the 
basis set expansion and we follow this prescription in the present paper. Thus, $\alpha_{max}$ 
is selected in such
a way that all basis states corresponding to full $N_F=2n_r +|\Lambda| +n_z$ fermionic 
shells are included into the basis of large components. The basis of small components
includes $N_F+1$ full fermionic shells to avoid spurios contributions to the RHB equation
(see Ref.\ \cite{GRT.90} for details). Note that large and small components of the Dirac 
spinor have opposite parities: this leads to an approximate doubling of the basis in the 
CDFT as compared with non-relativistic theories.  

   The solution of the Klein-Gordon equation  is obtained by expanding 
mesonic fields in a complete set of basis states [HO eigenfunctions] 
along [$ \varphi_{n_z} (z_,b_z)$] and perpendicular [$\varphi_{n_r}^0 (r_{\perp}, b_{\perp})]$
the axis of symmetry of the nucleus
\begin{eqnarray}
\phi (z, r_{\perp})  = \sum_{n_z n_r}^{N_B} \phi_{n_z n_r} 
                                 \varphi_{n_z} (z_,b_z) \varphi_{n_r}^0 (r_{\perp}, b_{\perp})
\end{eqnarray}
This expansion is truncated at full $N_B$ bosonic shells. 

The spherical RHB code is not parallelized but it allows the calculations to
a very large value of $N_F=38$ (see Sec.\ \ref{fermion-sect}  below).
Thus, the convergence of the results of the calculations in spherical nuclei as a 
function of $N_F$ has been tested  using this code. In contrast, the axial RHB 
code works only up to $N_F=30$ due to memory allocation problems 
at larger values of $N_F$. Note that similar to Fig.\ 2 of Supplemental Material 
to Ref.\ \cite{TA.23} we verified for selected set of spherical and deformed nuclei 
that the codes developed in our group and those existing in the DIRHB package 
of the RHB codes (see Ref.\ \cite{DIRHB-code.14}) provide almost the same 
(within a few keVs) results for binding energies as a function of 
$N_F$ and $N_B$. Note that only DD-ME* and DD-PC* types of the functionals
are included into the DIRHB codes of Ref.\ \cite{DIRHB-code.14}.

     In order to understand the dependence of the convergence properties  
on the functional we consider three major classes of CEDFs used at 
present. These are (i) those based on meson exchange with non-linear meson 
couplings (NLME) (see Ref.\ \cite{BB.77}), (ii) those based on meson exchange 
with density dependent meson-nucleon couplings (DDME) (see Ref.\ \cite{TW.99}), 
and finally (iii) those based on point coupling (PC) models containing various 
zero-range interactions in the Lagrangian (see Ref.\ \cite{PC-F1}).

  The Lagrangians of these three major classes of CEDFs can be written as:
$\mathcal{L} = \mathcal{L}_{common} + \mathcal{L}_{model-specific}$
where the $\mathcal{L}_{common}$ consist of the Lagrangian of the
free nucleons and the electromagnetic interaction.
It is identical for all three classes of functionals and is written as
\begin{eqnarray}
\mathcal{L}_{common} = \mathcal{L}^{free} + \mathcal{L}^{em}
\end{eqnarray}
with
\begin{eqnarray}
\mathcal{L}^{free} = \bar{\psi}(i\gamma_{{\mu }}\partial^\mu - m ) \psi
\label{lagrfree}
\end{eqnarray}
and
\begin{eqnarray}
\mathcal{L}^{em} = -\frac{1}{4} F^{\mu\nu} F_{\mu\nu} - e\frac{1-\tau_3}{2}\bar{\psi} \gamma^\mu\psi A_\mu.
\label{lagrem}
\end{eqnarray}

  For each model there is a specific term in the Lagrangian: for the DDME models
we have
\begin{eqnarray}
\mathcal{L}_{DDME} &=&  \frac{1}{2}(\partial{\sigma})^2 - \frac{1}{2} m_{\sigma}^2 \sigma^2 - \frac{1}{4}\Omega_{\mu\nu}\Omega^{\mu\nu} + \frac{1}{2} m_\omega^2\omega^2
\notag\\
&-&\frac{1}{4}\vec{R}_{\mu\nu}\vec{R}^{\mu\nu} + \frac{1}{2}m_\rho^2 \vec{\rho}^{\,2}-g_{\sigma}(\bar{\psi}\psi)\sigma
\notag\\
&-&g_\omega(\bar{\psi}\gamma_\mu\psi)\omega^\mu - g_\rho (\bar{\psi}\vec{\tau}\gamma_\mu\psi)\vec{\rho}^{\,\mu}
\label{lagrddme}
\end{eqnarray}
with the density dependence of the coupling constants given by
\begin{eqnarray}
g_i(\rho) &=& g_i(\rho_0) f_i(x)~~~{\rm for}~i=\sigma,\omega \\
g_\rho(\rho) &=& g_\rho(\rho_0)\exp[-a_\rho(x-1)]
\label{Eq:DD}
\end{eqnarray}
where $\rho_0$ denotes the saturation density of symmetric nuclear matter and $x=\rho/\rho_0$.
The functions $f_i(x)$ are given by the Typel-Wolter ansatz \cite{TW.99}
\begin{eqnarray}
f_i(x) = a_i \frac{1+b_i(x+d_i)}{1+c_i(x+d_i)}.
\label{Eq:Typel-Wolter}
\end{eqnarray}
Because of the five conditions $f_i(1) = 1$, $f''_i(1)=0$, and $f''_\sigma(1)=f''_\omega(1)$,
only three of the eight parameters $a_i$, $b_i$, $c_i$, and $d_i$ are independent and we finally
have the four parameters $b_\sigma$, $c_\sigma$, $c_\omega$, and $a_\rho$ characterizing
the density dependence. In addition we have the four parameters of the Lagrangian $\mathcal{L}_{DDME}$
$m_\sigma$, $g_\sigma$, $g_\omega$, and $g_\rho$. As usual  the masses of the
$\omega$- and the $\rho$-meson are kept fixed at the values $m_\omega=783$ MeV and $m_\rho=763$ MeV
\cite{DD-ME2,DD-MEdelta}. We therefore  have $N_{par}=8$ parameters in the DDME class of the models. The DD-MEX \cite{TAAR.20}, DD-ME2 \cite{DD-ME2} and DD-MEY \cite{TA.23} functionals have been used in the present study. 

   The NLME class of the functionals has the same model specific Lagrangian
as the DDME  class except that the coupling constants $g_\sigma$, $g_\omega$, and $g_\rho$ are constants
and there are extra terms for a non-linear $\sigma$ meson coupling. These couplings are important for the description
of surface properties of finite nuclei, especially the incompressibility~\cite{BB.77} and for nuclear
deformations~\cite{GRT.90}.
\begin{eqnarray}
\mathcal{L}_{NLME}  = \mathcal{L}_{DDME}   - \frac{1}{3} g_2 \sigma^3 - \frac{1}{4}g_3 \sigma^4
\label{lagrnl5}
\end{eqnarray}
For the NLME class we have $N_{par}=6$ parameters $m_\sigma$, $g_\sigma$, $g_\omega$, $g_\rho$, $g_2$, and $g_3$. In the present study, we use the
NL5(E) CEDF from Ref.\ \cite{AAT.19}.

   In general,  the Lagrangian of the PC models contains three parts:
\\
(i) the four-fermion point coupling terms:
\begin{eqnarray}
\begin{aligned}
\mathcal{L}^{4f} = & - \frac{1}{2}\alpha_S (\bar{\psi}\psi)(\bar{\psi}\psi) - \frac{1}{2}\alpha_V (\bar{\psi}\gamma_{\mu}\psi)(\bar{\psi}\gamma^{\mu}\psi) & \\
                  & - \frac{1}{2}\alpha_{TS} (\bar{\psi}\vec{\tau}\psi)(\bar{\psi}\vec{\tau}\psi)
                  - \frac{1}{2}\alpha_{TV} (\bar{\psi}\vec{\tau}\gamma_{\mu}\psi)(\bar{\psi}\vec{\tau}\gamma^{\mu}\psi),&
\end{aligned}
\label{lagr4f}
\end{eqnarray}

(ii) the gradient terms which are important to simulate the effects of finite range:
\begin{eqnarray}
\begin{aligned}
\mathcal{L}^{der} = &- \frac{1}{2}\delta_S \partial_{\nu}{(\bar{\psi}\psi)}\partial^{\nu}{(\bar{\psi}\psi)} & \\
& - \frac{1}{2}\delta_V \partial_{\nu}{(\bar{\psi}\gamma_{\mu}\psi)}\partial^{\nu}{(\bar{\psi}\gamma^{\mu}\psi)}& \\
&- \frac{1}{2}\delta_{TS} \partial_{\nu}{(\bar{\psi}\vec{\tau}\psi)}\partial^{\nu}{(\bar{\psi}\vec{\tau}\psi)}&\\
& - \frac{1}{2}\delta_{TV} \partial_{\nu}{(\bar{\psi}\vec{\tau}\gamma_{\mu}\psi)}\partial^{\nu}{(\bar{\psi}\vec{\tau}\gamma^{\mu}\psi)},&
\end{aligned}
\label{lagr4f-a}
\end{eqnarray}
(iii) The higher order terms which are responsible for 
the effects of medium dependence
\begin{eqnarray}
\begin{aligned}
\mathcal{L}^{hot}  = & - \frac{1}{3}\beta_S (\bar{\psi}\psi)^3 - \frac{1}{4}\gamma_S (\bar{\psi}\psi)^4&   \\
                    & - \frac{1}{4}\gamma_V[(\bar{\psi}\gamma_{\mu}\psi)(\bar{\psi}\gamma^{\mu}\psi)]^2. &
\end{aligned}
\label{lagrhot}
\end{eqnarray}
  
 However, the structure of two PC functionals, namely, PC-PK1 and DD-PC1, employed in the present
paper depends on particular realization of this scheme. The PC-PK1 CEDF
closely follows this scheme but it neglects the scalar-isovector channel, i.e. 
$\alpha_{TS}=\delta_{TS}=0$. This is because the information on masses 
and charge radii of finite nuclei does not allow one to distinguish the effects of the 
two isovector mesons $\delta$ and $\rho$ (see Ref.\ \cite{DD-MEdelta}).
Thus, for this PC CEDF  we have $N_{par}=9$ parameters $\alpha_S$, $\alpha_V$, 
$\alpha_{TV}$, $\delta_S$, $\delta_V$, $\delta_{TV}$, $\beta_S$, $\gamma_S$, $\gamma_V$.  The DD-PC1 functional \cite{DD-PC1} introduces the density 
dependence of the $\alpha_S$, $\alpha_V$ and $\alpha_{TV}$ constants via
\begin{eqnarray} 
\alpha_i(\rho) = a_i+(b_i+c_i x) e^{-d_i x} \qquad ({\rm for} \,\,\, i=S,V, TV), \nonumber \\
\end{eqnarray}
and eliminates the terms proportional to $\alpha_{TS}$, $\delta_{TS}$ [this
is similar to the PC-PK1 functional] and $\delta_V$, $\delta_{TV}$, $\beta_S$,
$\gamma_S$ and $\gamma_V$ in Eqs.\ (\ref{lagr4f}-\ref{lagrhot}). The introduced
density dependence  of the $\alpha_S$, $\alpha_V$ and $\alpha_{TV}$ constants
partially takes care of the impact of omitted terms. Note that in the isovector channel 
a pure exponential dependence is used, i.e. $a_{TV}=c_{TV}=0$. Thus, the DD-PC1
functionals contains the $N_{par}=10$ parameters, i.e. $a_S$, $b_S$, $c_S$,
$d_S$, $a_V$, $b_V$, $c_V$, $d_V$, $b_{TV}$ and $d_{TV}$.

   Note that in the deformed RHB calculations we use $ngh=30$ and $ngl=30$
integration points in  the Gauss-Hermite and Gauss-Laguerre  quadratures in the 
$z>0$ and $r_{\perp}>0$ directions, respectively,  in order to properly take
into account the details of the single-particle wave functions. The same $ngh=30$ 
is used in the Gauss-Hermite quadratures for $r>0$ in spherical RHB calculations.

  The experimental binding energies of twelve anchor spherical nuclei 
($^{16}$O, $^{40,48}$Ca, $^{72}$Ni, $^{90}$Zr, $^{116,124,132}$Sn, 
$^{204,208,214}$Pb and $^{210}$Po) and charge radii of nine of them (namely, 
$^{16}$O, $^{40,48}$Ca, $^{90}$Zr, $^{116,124}$Sn and $^{204,208,214}$Pb) 
are used in the anchor based optimization approach (see Ref.\ \cite{TA.23} for more 
details).  The correction function of Eq.\ (\ref{corr-func}) is defined in ABOA using 
855 even-even nuclei for which experimental binding energies are available in 
Ref.\ \cite{AME2016-third}.  
  
%%%%%%%%%%%%%%%%%%%%%%%%%%%%%%%%%%%%
\section{Further discussion of anchor-based optimization approach} 
\label{anchor-further}
%%%%%%%%%%%%%%%%%%%%%%%%%%%%%%%%%%%%

   The anchor-based optimization of energy density functionals is defined 
in Ref.\ \cite{TA.23} and the basic results for this approach have been discussed 
in this reference and in Supplemental Material to it. It is based on the 
introduction of the correction function $E_{corr}(Z,N)$ given in 
Eq.\ (\ref{corr-func}) 
by which the binding energies of spherical anchor nuclei are redefined for 
global performance of the EDFs (see Ref.\ \cite{TA.23}) for full details). 
The 
behavior of these parameters during the iterative procedure has been 
illustrated for the DD-MEX1, DD-MEY and NL5(Y) functionals in Tables I, 
II and III of the Supplemental Material to Ref.\ \cite{TA.23}.  In addition,  
the evolution of the rms deviations  $\Delta E_{rms}$ between calculated 
and experimental binding energies $E$ for these functionals during the 
iterative procedure  was also illustrated in Fig.\ 1 of Supplemental Material 
to Ref.\ \cite{TA.23}.

     In this section we extend the discussion of the ABOA to other 
functionals, introduce new improved strategy for the fitting of the functionals
within ABOA, provide an estimate of the computation time for the ABOA and 
RGA approaches and compare the results obtained for the DD-MEY type of
the functionals in the RGA and ABOA.

%%%%%%%%%%%%%%%%%%%%%%%%%%%%%%%%%%%%
\subsection{Further improvement of the convergence of the ABOA}
\label{conv-improv} 
%%%%%%%%%%%%%%%%%%%%%%%%%%%%%%%%%%%%

\begin{figure}[htb]
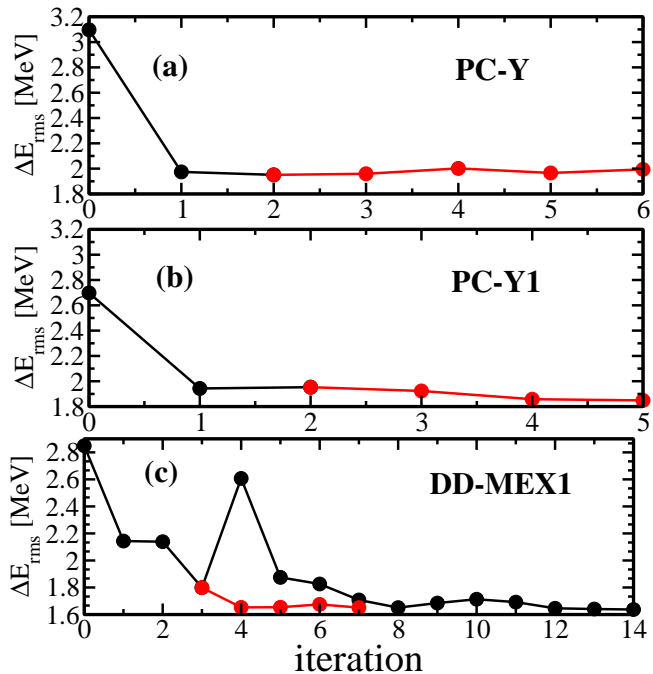

\centering
\includegraphics*[width=8.5cm]{fig-1-a.eps}
\includegraphics*[width=8.5cm]{fig-1-b.eps}
\vspace{1em}
    \caption{The evolution of the rms deviations $\Delta E_{rms}$ 
between calculated and experimental binding energies as a function of 
iteration counter $i$ in the anchor based optimization method for the indicated 
functionals. Note that $\Delta E_{rms}$ is defined for 855 even-even
nuclei  for which experimental binding energies are known (see Ref.\ \cite{TA.23} 
for more details).  The steps of iterative procedure for which the correction functions 
of Eqs.\ (\ref{corr-func}) and (\ref{eq-abg=mod}) are used are shown by black
and red colors, respectively. 
\label{conv-anchor}
}
\end{figure}

   Figure 1 of Supplemental Material to Ref.\ \cite{TA.23} 
[see also black curve in Fig.\ \ref{conv-anchor}(c)]  shows a significant 
increase in global $\Delta E_{rms}$ at the 4$^{th}$ iteration of the convergence 
process.  This indicates that the push provided by the correction function of 
Eq.\ (\ref{corr-func}) is too strong.  It turns out that this problem can be eliminated 
by switching to a softer correction function
\begin{eqnarray}  
E_{corr}^{mod}(Z,N) = \frac{\alpha_i}{2}(N-Z) + \frac{\beta_i}{2}(N+Z) + \frac{\gamma_i}{2}
\label{eq-abg=mod} 
\end{eqnarray}
after few initial iterations in the convergence process. Note that the convergence 
of the iterative procedure is reached in the limit $\alpha_i \rightarrow 0$, 
$\beta_i \rightarrow 0$ , and $\gamma_i \rightarrow 0$.

  The advantage of this modified procedure is illustrated in Fig.\ 
\ref{conv-anchor}(c) on the example of the DD-MEX1 functional. The use of 
correction function $E_{corr}(Z,N)$ leads to the convergence curve for 
$\Delta E_{rms}$ shown in black color which is characterized by a large peak 
at the $4^{\rm th}$ iteration and which requires 14 iterations for a convergence.
However, the switch to $E_{corr}^{mod}(Z,N)$ at the 3$^{\rm d}$ iteration eliminates 
this peak and reduces the number of required iterations to seven (see red 
curve in Fig.\ \ref{conv-anchor}(c) and Table \ref{table-DDMEX1}).   Thus,  the 
combination of these two correction functions leads to a rapid and smooth 
convergence process: the $E_{corr}(Z,N)$ function provides a significant
push so that only one to three iterations are needed to drive the functional
towards its minimum in the parameter hyperspace
 in the beginning of iterative process  and $E_{corr}^{mod}(Z,N)$ 
guaranties the smoothness of convergence process at further iterations.

   This combined procedure has also been used in the fit of the PC-Y and
PC-Y1 functionals defined in Ref.\ \cite{TA.23}. Figs.\ \ref{conv-anchor}(a) 
and (b) and Tables \ref{table-PCY} and \ref{table-PCY1} illustrate that 
drastic improvement of the global performance of these functionals was
reached after first iteration and then additional few iterations provide
only minor improvement of the functionals.

%%%%%%%%%%%%%%%%%%%%%%%%%%%%%%%%%%%%%%%%%%
\begin{table}[htb]
\begin{center}
\caption{The values of the parameters $\alpha_i$, $\beta_i$ and 
$\gamma_i$ of Eqs.\ (\ref{corr-func}) and (\ref{eq-abg=mod}) defined 
at the $i$-th iteration during  the iterative procedure of the anchor based optimization 
method. The  results are presented for iterative procedure of the DD-MEX1 functional.
$\Delta E_{rms}$ is global rms deviation of binding energies at $i$-th 
iteration.  The steps of iterative procedure for which the correction functions 
of Eqs.\ (\ref{corr-func}) and (\ref{eq-abg=mod}) are used are shown in bold
and plain text, respectively.
}
\begin{tabular}{ccccc}
\hline \hline
    iteration            & $\Delta E_{rms}$ &    $\alpha_i$  &  $\beta_i$   &  $\gamma_i$     \\  
       $i$                 &    [MeV]                &                       &                   &                     \\ \hline                       
           0                & {\bf 2.849} &  {\bf -0.075}    &  {\bf 0.0235}      &    {\bf -3.29}     \\  
           1                & {\bf 2.144} &  {\bf 0.044}     &   {\bf -0.028}       &   {\bf 2.55}  \\ 
           2                & {\bf 2.139} &  {\bf -0.044}    &   {\bf 0.026}       &    {\bf -2.12}   \\ 
           3                & {\bf 1.799} &   0.045    &   -0.020       &     1.70  \\
           4                & 1.653 &   0.005    &    0.000       &     0.05     \\ 
           5                & 1.654 &   0.005    &    0.000       &     0.05     \\
           6                & 1.675 &  -0.010    &    0.000       &     0.00     \\
           7                & 1.651 &   0.000    &    0.000       &     0.00     \\ \hline
\end{tabular}
\label{table-DDMEX1}
\end{center}
\end{table}
%%%%%%%%%%%%%%%%%%%%%%%%%%%%%%%%%%%%

%%%%%%%%%%%%%%%%%%%%%%%%%%%%%%%%%%%%
\begin{table}[htb]
\begin{center}
\caption{The same as Table I but for the PC-Y functional.
}
\begin{tabular}{ccccc}
\hline \hline
    iteration            & $\Delta E_{rms}$ &    $\alpha_i$  &  $\beta_i$   &  $\gamma_i$     \\  
       $i$                 &    [MeV]                &                       &                   &                     \\ \hline                       
           0                & {\bf 3.097} &   {\bf +0.086}  & {\bf -0.036}   &  {\bf 1.00}      \\  
           1                & {\bf 1.974} &   {\bf -0.003}  & {\bf +0.002}   & {\bf -0.25}      \\ 
           2                & {\bf 1.951}  &   -0.002  & +0.001   & -0.15     \\ 
           3                & 1.959 &    0.000  &  0.000   & -0.19      \\  
           4                & 2.002 &   -0.007  &  0.001   & -0.13      \\  
           5                & 1.950 &    0.009  &  0.004   & -0.14      \\  
           6                & 1.951 &   -0.0099 &  0.004   & 0.02       \\  \hline
\end{tabular}
\label{table-PCY}
\end{center}
\end{table}
%%%%%%%%%%%%%%%%%%%%%%%%%%%%%%%%%%%%

%%%%%%%%%%%%%%%%%%%%%%%%%%%%%%%%%%%%%%%
\begin{table}[htb]
\begin{center}
\caption{The same as Table I but for the PC-Y1 functional.
}
\begin{tabular}{ccccc}
\hline \hline
    iteration            & $\Delta E_{rms}$ &    $\alpha_i$  &  $\beta_i$   &  $\gamma_i$     \\  
       $i$                 &    [MeV]                &                       &                   &                     \\ \hline                       
         0                  & {\bf 2.698} & {\bf +0.065}  & {\bf -0.030}   & {\bf 1.10}   \\  
         1                  & {\bf 1.943} & {\bf -0.010}  &  {\bf 0.000}   &  {\bf 0.20}   \\  
         2                  & {\bf 1.953} &   -0.010  & -0.001   &  0.15    \\ 
         3                  & 1.923 &   +0.024  & -0.001   & -0.03    \\ 
         4                  & 1.858 &   +0.017  & -0.002   &  0.01    \\ 
         5                  & 1.849 &   +0.001  & -0.0001  & -0.001      \\  \hline
\end{tabular}
\label{table-PCY1}
\end{center}
\end{table}
%%%%%%%%%%%%%%%%%%%%%%%%%%%%%%%%%%%%

%%%%%%%%%%%%%%%%%%%%%%%%%%%%%%%%%%
%%%%%%%%%%%%%%%%%%%%%%%%%%%%%%%%%%
 
%%%%%%%%%%%%%%%%%%%%%%%%%%%%%%%%%% 
\subsection{The dependence of computational time  on approach}
\label{comp-appr}
%%%%%%%%%%%%%%%%%%%%%%%%%%%%%%%%%%
    
   It is interesting to compare the computer resources required for definition
of the functional in different approaches. The numbers quoted below are based on the 
RHB calculations with $N_F=20$ and $N_B=20$. Numerical calculations are performed
at Orion of High Performance Computing Collaboratory (HPC$^2$) of Mississippi State 
University (see https://www.hpc.msstate.edu/computing/hpc.php) and they reflect the 
computation time allocation of this facility. The largest standard allocation per user is 
400 CPU with 48 hours execution time limit; it is utilized in most of our calculations.

Let us start with anchor based optimization approach. The first step in each iteration
of ABOA is the optimization of the parameters for 12 anchor spherical nuclei. This is carried out within 
the simplex  method\footnote{The 
simplex method of the minimization is one of the simplest and fastest
ones (see  Ref.\ \cite{NumRec}). The minimizations by the simplex method are 
prone to end in local minima and that is a reason why, in general,  it is not recommended 
for the search of the global minimum. However, it is our experience that by starting from
a reasonable number of randomly generated initial points in parameter hyperspace
we always find a global minimum as defined by alternative approaches such as
simulated annealing method (see also Refs.\ \cite{TAAR.20,TA.23}). Note that this number 
depends on the size of parameter hyperspace. In contrast, the simulated annealing 
method is extremely costly and numerically unstable for large parameter hyperspaces:
in our experience it works reasonably well only in a narrow parameter hyperspace around
the global minimum and even then it is by an order of magnitude more time consuming
than the simplex method. Note that the simplex method also provides the access to 
parametric correlations in the functionals (see Ref.\ \cite{TAAR.20}).
} starting from 200 randomly generated initial points in the large
parameter hyperspace at the 0$^{th}$ iteration of ABOA. At subsequent iterations 
of ABOA the number of randomly generated initial points is decreased to 20 since 
the parameter  hyperspace, which is centered around minimum found at previous 
iteration and in which the solution is sought, is made smaller.

  On average, each minimization of the parameters for 12 spherical anchor nuclei
requires approximately 3200 iterations in the optimization process and 8 hours of 
CPU time are allocated for 
it\footnote{The optimization is performed in parallel computer code. 
Thus, the computation time for each optimization iteration is defined by the nucleus
the calculation of which takes the longest time.  This is because the penalty
function is defined  at the end of each iteration the calculation of which
requires the information on each nucleus included in the fitting protocol.}.
Thus, the total allocated computation time for this step is 12 nuclei times 20 
runs times 8 CPU hours per run which results into 1920 CPU hours for 
iterations 1, 2, ... of ABOA. Note that computation time allocation is ten 
times larger for the $0^{th}$ iteration of ABOA.

  The calculations of each of 855 deformed nuclei requires the allocation 
of 4  CPU hours per run and they are carried out for 4 deformations of basis. 
This results in the allocation of $855\times 4 \times 4 = 13680$ CPU hours.

   Thus, the 0$^{th}$ iteration and each subsequent iteration of ABOA require 
the allocation of  19200+13680=32880 and 1920+13680=15600 CPU 
hours, respectively. The experience with several types of the functionals 
tells us that approximately six iterations of ABOA are required to define 
the functional (see Sec.\ \ref{conv-improv} and Supplemental Material to 
Ref.\ \cite{TA.23}). Thus, in total approximately 126400 CPU hours
are required  for the definition of the functional in the anchor-based 
optimization.  This is only approximately six times more time-consuming 
than in the typical fitting protocol restricted to 12 spherical nuclei used in 
Ref.\ \cite{TAAR.20} and which corresponds to the optimization
of anchor spherical nuclei in the 0$^{th}$ iteration of ABOA.
 
   The next approach is the RGA one. Because of the process of job 
allocation at  HPC$^2$, it is possible to calculate  maximum 400 nuclei in 
parallel. The  optimization of 400 "spherical" nuclei in the RGA approach is 
significantly more time consuming than twelve ones in ABOA because of three
reasons discussed below. First, the number of "spherical" nuclei is approximately 33 times 
larger. Second, the CPU time per nucleus in optimization process   is significantly 
larger in the RGA approach than in the ABOA one. This time is defined in parallel 
calculations by the nucleus the calculation of which takes the longest time since 
the penalty function is defined at the end of iteration when the information 
on all nuclei are collected. In that respect, the 12 spherical nuclei used in
ABOA are extremely well behaved with fast convergence in numerical 
calculations in part because the pairing collapses either in one or in both
subsystems.  In contrast, the pairing is present in most of the nuclei used
in the RGA approach and numerical calculations converge slowly in some 
of these nuclei. Third, because of larger set of nuclei in the RGA approach
as compared with ABOA, the optimization of "spherical" nuclei requires 
larger number of iterations in the minimization process (around 8000) and 
the  allocation of 24 hours per run.  Note that similar to ABOA, we use simplex 
method for minimization with 20 randomly generated initial points in the 
parameter hyperspace. As a result, one round of optimization of spherical 
nuclei in the RGA  requires $400\times 20 \times 24 = 192000$ CPU hours.

  The calculations of deformed nuclei in the RGA are less numerically 
costly  but they have to be done twice: the deformed RHB calculations without 
constraint on deformation for each nucleus with four initial deformations of basis for the 
definition of the energy of global minimum and deformed RHB calculations 
with constraint on spherical shape and with only spherical deformation of 
basis for the definition of the energy of spherical solutions. Thus the 
calculational time for deformed nuclei is 400 nuclei times 4 CPU hours per 
run times 4 runs for different deformations of basis for finding  global minima 
plus 400 nuclei times 4 CPU hours  per run for finding spherical solutions; 
altogether that is 8000 CPU hours.

   Thus, one iteration of optimization in the RGA  for 400 nuclei requires the allocation 
 of approximately 200000 CPU hours. This is somewhat larger than the time 
 required for defining the functional in the ABOA. If to scale this number to 855 
 nuclei (as in the ABOA), this would lead to approximately 427500 CPU hours.

%%%%%%%%%%%%%%%%%%%%%%%%%%%%%%%%%%%%%%%%%%%
\begin{figure}[htb]
  \begin{center}
    \centering
    \includegraphics*[width=8.0cm]{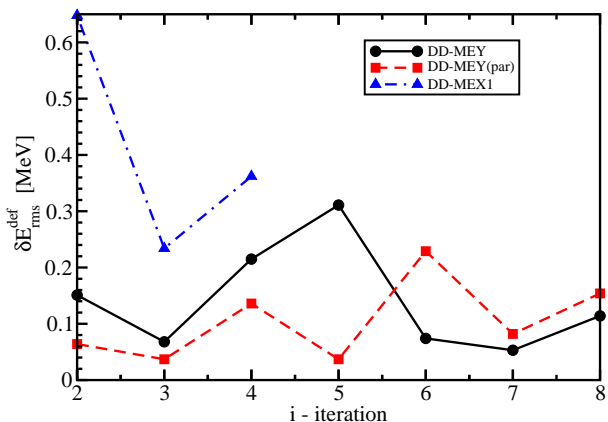}
\caption{The convergence of the error in deformation energy in the RGA calculations
with indicated functionals. At $i$-th iteration of the RGA, the deformation energy  
$E_{def}(Z,N)$ is calculated for the  $n = 855$ even-even nuclei using Eq.\ 
(\ref{def-energy}). The error (rms deviation) in deformation energy between the 
$i$-th and $(i-1)$th iterations of the RGA are then defined as 
$\delta E^{def}_{rms} = \sqrt{\frac{1}{n}\Sigma_{k=1}^{n}{[(E_{def})_{i}(Z,N) - (E_{def})_{i-1}(Z,N)]^2}}$.
\label{conv-def} 
}
\end{center}
\end{figure}
%%%%%%%%%%%%%%%%%%%%%%%%%%%%%%%%%%%%%%%%%%%
 
\begin{figure}[htb]
\centering
\includegraphics[width=6.1cm,angle=-90]{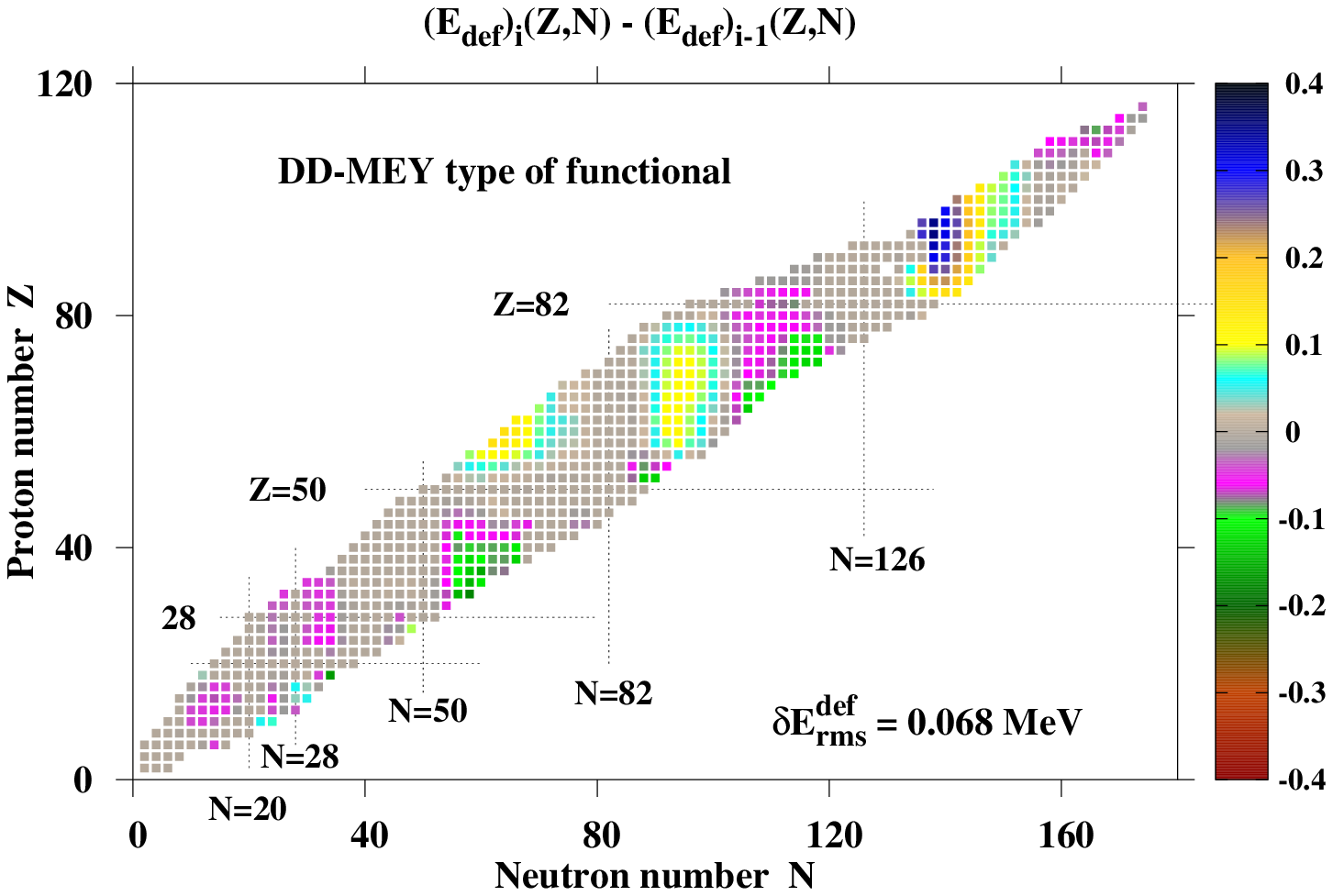}
\caption{The distribution of the $(E_{def})_{i}(Z,N) - (E_{def})_{i-1}(Z,N)$ values
across the nuclear chart for the $i=3$ iteration of the RGA calculations with the 
DD-MEY type of the functional characterized by $\delta E^{def}_{rms}(Z,N)=0.068$
MeV (see Fig.\  \ref{conv-def} and its caption for additional information).
\label{E-def-RGA}
}
\end{figure}

  However, to achieve an optimal solution one should perform a number of iterations 
of RGA  to ensure that the deformation energies $E_{def}$ converge. In order 
to find how many iterations of RGA are needed for that we carried out the 
calculations for the CEDFs the fitting protocols of which are the same
as for  the DD-MEX1 and DD-MEY functionals defined in Ref.\ \cite{TA.23}.  The
convergences of the error $\delta E^{def}_{rms}$ in deformation energies for these 
functionals are shown  in Fig.\ \ref{conv-def}. It was very slow for the DD-MEX1 
type  of the functional. The best $\delta E^{def}_{rms}=0.258$ MeV value has been 
achieved at the $3^{\rm d}$ iteration of the RGA (see Fig.\ \ref{conv-def}). However,
this value is large and that is a reason why the process has been interrupted after 
four iterations for this functional. The convergence
of the $\delta E^{def}_{rms}$ errors is better in the DD-MEY type of the functional 
(see Fig.\ \ref{conv-def}) but even for this functional no full convergence of 
$\delta E^{def}_{rms}$ is reached after eight iterations of RGA. Note that the 
absence of full convergence of deformation energies affects mostly the nuclei 
in which the deformation is well pronounced (see Fig.\ \ref{E-def-RGA}).
Fig.\ \ref{conv-def} clearly illustrates that substantial number of additional 
iterations of RGA would be required for a full convergence of deformation 
energies. Note that iterative processes shown in this figure were
interrupted because of  extreme numerical cost: one iteration of RGA  
is as expensive as a full fit of the functional in ABOA. 
 
   It was shown in Refs.\ \cite{AAT.19,TAAR.20} that there are correlations between
the parameters of state-of-the-art CEDFs: one can speak of parametric correlations 
between the parameters $p_k$ and $p_j$ when the parameter $p_k$, with a reasonable 
degree of accuracy, can be expressed as a function of the parameter $p_j$. The 
simplest type of such correlations which exists in CEDFs is a linear one given by
\begin{eqnarray}
f (p_k) =af(p_j) +b
\end{eqnarray}
where
\begin{eqnarray}
f(p_i) = \frac{p_i}{p_i^{opt}}.
\end{eqnarray}
Here $p_i$ is the value of the parameter $i$ in the functional variation and
$p_i^{opt}$ is the value of the parameter in the optimal functional
(see Ref.\ \cite{TAAR.20} for detail).

   The removing of such correlations allows to reduce the number of free 
parameters in CEDFs (see Refs.\ \cite{AAT.19,TAAR.20}).  Here by
using the DD-MEY type of the functionals as an example we will evaluate 
the impact of the removing of  parametric correlations on computational time 
and convergence of deformation energies in RGA. Using the procedure of 
Ref.\  \cite{TAAR.20} one can establish the following correlations between 
the $b_\sigma$, $c_\omega$ and $c_\sigma$ parameters
\begin{eqnarray}
       f(b_\sigma)    =    0.97303*f(c_\sigma) + 0.028846,  \label{1-par} \\
        f(c_\omega)    =    1.14180*f(c_\sigma) - 0.131090. \label{2-par} 
\end{eqnarray}
Thus,  the removing of parametric correlations leads to a reduction of 
the number of free parameters in the DD-MEY type of functionals from 
8 to 6. The functional with built-in correlations of Eqs.\ (\ref{1-par}) and 
(\ref{2-par}) is called as DD-MEY(par).

  As a consequence of this reduction of the number of adjustable parameters, 
the time required for one round of optimization of  spherical nuclei is reduced 
by approximately 30\%. However, the results labelled as "DD-MEY(par)" in 
Fig.\  \ref{conv-def} show that the removing of parametric correlations does 
not improve substantially the convergence of deformation energies.

%%%%%%%%%%%%%%%%%%%%%%%%%%%%%%%%%%%%%
%%%%%%%%%%%%%%%%%%%%%%%%%%%%%%%%%%%%%%

     Taking all these facts into account one can conclude that RGA is by
more than one order of magnitude numerically more time consuming than ABOA
and its performance strongly depend on how many iterations of RGA are
required for the convergence of deformation energies. However, the latter is strongly
dependent on the functional.
 
     FGA approach requires the use of deformed code for the calculations
 of all nuclei included in the fitting protocol.  At present, we did not make 
 any calculations in the FGA but a crude estimate can be obtained by 
 comparing the calculational time for the same set of nuclei carried out in spherical 
 and deformed RHB codes. As discussed earlier, the most of numerical cost 
 in the ABOA and RGA calculations are related to the calculations of spherical 
 nuclei.  In contrast, the numerical cost in FGA is defined by the cost of the
 calculations of deformed nuclei.  Let us assume that (i) the same set of the
 nuclei is used in the RGA and FGA and (ii) the number of iterations
 in the optimization is comparable for the calculations with spherical and
 deformed RHB codes. Then,  the results presented in the columns 4 and 
 5 of Table \ref{table-compt-time-compar} will clearly indicate that the numerical 
 cost of FGA will be larger  than that in a single iteration of RGA\footnote{Here
 we compare full FGA with a single iteration of RGA since FGA does not
 require the calculation of deformation energies.}
 by more than 
 three orders of magnitude at fermionic basis with  $N_F=20$ and this cost 
 difference will rapidly increase with increasing $N_F$.

    Note that utilized calculational time is substantially lower than 
allocated time since many processes finish before the end of allocated time.
Utilization rate defined as a ratio of actual calculational time versus allocated 
one can be improved by a further optimization of the structure of computer 
codes to the structure of run/time allocation at the HPC$^{2}$. However, in 
no way it will significantly change the ratio of the computational times of 
the ABOA, RGA and FGA.

  There is an additional issue related to the stability of  numerical 
calculations for the nuclei included into fitting protocol which have 
not been discussed in the literature on CDFT.  The importance of
this factor can be illustrated by comparing the 
parameters of the PC-PK1 and PC-X functionals in Table 3 of 
Supplemental Material in Ref.\  \cite{TAAR.20}. These two functionals 
have the same fitting protocols.  However, both $\delta_S$ and $\delta_{TV}$ 
parameters have opposite signs in these functionals and the penalty function obtained in 
PC-X is almost by a factor of two better than the one obtained in PC-PK1. 
A careful analysis reveals that numerical solutions for the pair of very 
light nuclei included into fitting protocol are unstable at some steps of 
minimization which in the end leads to non-convergence of the minimization 
process in a specific volume of parameter hyperspace. Note that this problem 
has been taken care in the minimization of 
the PC-X functional by using simplex method starting from large set of randomly
generated initial points in the parameter hyperspace. This example clearly
indicates that it is beneficial to use in fitting protocols the nuclei in which
the numerical solutions behave well. This not only ensures the stability
of the minimization process but also enlarges the parameter hyperspace
in which the optimum solution can be sought and reduces computational
time of minimization. In this respect,  a very small set of twelve nuclei included 
in the fitting protocol of ABOA are extremely well behaved in numerical 
calculations.

%%%%%%%%%%%%%%%%%%%%%%%%%%%%%%%%%
\subsection{DD-MEY functional: ABOA versus RGA}
\label{ABOA-vs-RGA} 
%%%%%%%%%%%%%%%%%%%%%%%%%%%%%%%%%

   It is important to understand to which extent the ABOA could be 
a reasonable substitute to a significantly more numerically 
time-consuming RGA. To do that we carried out RGA calculations using the
same experimental information (restricted to binding energies and charge
radii) and the isospin-dependent pairing as in the fitting protocol of the 
DD-MEY CEDF in ABOA (see Ref.\ \cite{TA.23} for details). The
computational procedure discussed in Sec.\ \ref{comp-appr} was
used in these calculations.

%%%%%%%%%%%%%%%%%%%%%%%%%%%%%%%%%%%%%%%%%%
\begin{table}[htb]
\begin{center}
\caption{The properties and parameters of indicated functionals. The DD-MEY
functional, obtained in ABOA, is taken from Supplemental Material to Ref.\ \cite{TA.23}.
DD-MEY(3) and DD-MEY(6) are the functionals obtained at the 3$^d$ and $6^{th}$
iterations of the RGA.   The global rms deviations 
$\Delta E_{rms}$($\Delta (r_{ch})_{rms}$) between experimental and calculated 
binding energies (experimental and calculated charge radii)  are given by 
$\Delta E_{rms}$($\Delta (r_{ch})_{rms}$). They are defined using
all (855 for binding energies and 305 for charge radii) even-even nuclei for 
which experimental data are available in Refs.\ \cite{AM.13,AME2016-third}.  
The results for the energy per particle  $E/A$, the density $\rho_0$, the 
incompressibility  $K_0$, the symmetry energy $J$ and its slope $L_0$ are 
displayed in the middle of the table.  The parameters of the functionals are 
presented at the bottom of the table. See text for additional details. 
}
\begin{tabular}{cccc}
\hline \hline
                             &  DD-MEY    &   DD-MEY(3)      &   DD-MEY(6)       \\  
                             &   ABOA       &    RGA                &    RGA                  \\  \hline 
         1                  &        2          &           3               &        4                  \\   \\                  
$\Delta E$   [MeV]         &    1.734        &     1.672(0.068)  &  1.725(0.074)      \\
$\Delta (r_{ch})_{rms}$ [fm]    &      0.0264             &         0.0294                 &       0.0303                        \\ 
                                                                                                                                                       \\                                                                                                                                                       
$E/A$  [MeV]                    & -16.09           & -16.044             &       -16.017    \\
$\rho_0$ [fm$^{-3}$]                           & 0.1529           & 0.1522              &       0.1545     \\
$K_{0}$ [MeV]               &  265.8           &    261.6              &       248.5       \\ 
 $J$       [MeV]               &  32.8             &    31.6                &        30.1        \\
$L_{0}$  [MeV]              &   51.8             &    42.6                &       29.5         \\
                                                                                                                                                        \\
 $m_{\sigma}$ [MeV]       &  551.321796  &  552.291521     &   551.989284    \\
$g_{\sigma}$        &  10.411867    &   10.398001       &  10.315743         \\
$g_{\omega}$       &  12.803298   &   12.758552       & 12.648663         \\
$g_{\rho}$             &   3.692170    &    3.548208        &  3.325439          \\
$b_{s}$                 &    2.059712   &    2.249514         &    2.148238       \\
$c_{s}$                 &    3.210289   &    3.551463         &    3.464883       \\
$c_{o}$                 &    3.025356   &    3.437551         & 3.284163            \\
$a_{r}$                  &    0.532267   &   0.652200          &    0.859341         \\ \hline \hline
\end{tabular}
\label{table-ABOA-vs-RGA}
\end{center}
\end{table}
%%%%%%%%%%%%%%%%%%%%%%%%%%%%%%%%%%%%%%%%

    When comparing the results of the RGA and ABOA calculations
one should keep in mind that due to limited computational power the results
of the RGA are always a subject of an error $\delta E^{def}_{rms}$ in 
deformation energies. However, these errors are acceptable (below 150 keV 
for $\delta E_{rms}^{def}$ in considered cases) so that the comparison of these two 
approaches is feasible. Table 
\ref{table-ABOA-vs-RGA} presents such a comparison between ABOA and
RGA results for two iterations of RGA [with respective functionals
labelled as DD-MEY(2) and DD-MEY(5)] 
characterized by $\delta E^{def}_{rms}$ values of 0.068 and 0.074 MeV, 
respectively. 

   One can see that the parameters $m_{\sigma}$, $g_{\sigma}$ and $g_{\omega}$
presented in Table \ref{table-ABOA-vs-RGA} are very similar for the series 
of the DD-MEY* functionals obtained  in different calculational 
schemes. Such a localization of these parameters in parameter hyperspace is 
typical for  all functionals which include meson exchange and it is a consequence 
of the fact that nucleonic potential is defined as a sum of very large attractive
scalar $S$ and repulsive vector $V$ potentials (see Refs.\ \cite{AAT.19,TAAR.20}).
In contrast, other parameters such as $g_{\rho}$, $b_s$, $c_s$, $c_o$ and 
$a_r$ in Table \ref{table-ABOA-vs-RGA} show somewhat larger relative spreads as
compared with $m_{\sigma}$, $g_{\sigma}$ and $g_{\omega}$ since for acceptable 
functionals they are less  localized in the parameter hyperspace  (see Refs.\ 
\cite{AAT.19,TAAR.20}). 

   However, 
considering different sets of the nuclei included in the fitting protocols 
of RGA (400 nuclei) and ABOA (12 nuclei) and  the remaining errors in the deformation 
energies in the RGA calculations, one can conclude that both ABOA and RGA 
lead to comparable functionals which provide  similar global accuracy of 
description of  binding energies ($\Delta E_{rms} \approx 1.7$ MeV) and 
charge radii ($\Delta (r_{ch})_{rms} \approx 0.03$ fm) (see Table 
\ref{table-ABOA-vs-RGA}).  Note
also that the predicted nuclear matter properties (NMP) of these functionals
are within the ranges  $\rho_0 \approx 0.15$ fm$^{-3}$, $E/A \approx -16$ MeV, 
$K_0= 190-270$, $J=25-35$ MeV ($J=30-35$ MeV) and $L_0=25-115$ 
($L_0=30-80$) for the SET2a (SET2b) sets of the constraints on the 
experimental/empirical ranges for the quantities of interest recommended for 
relativistic functionals in Ref.\ \cite{RMF-nm}. This is the first time that a
good reproduction of strict SET2b constraints has been achieved in CDFT
by fitting only binding energies and charge radii of finite nuclei: the fitting 
protocol of DD-MEY* does not contain any information on NMP (see 
Supplemental Material to Ref.\ \cite{TA.23}).

%%%%%%%%%%%%%%%%%%%%%%%%%%%%%%%%
\subsection{Charge radii in ABOA}
\label{charge-radii}
%%%%%%%%%%%%%%%%%%%%%%%%%%%%%%%%  

   A general expression for a charge radius $r_{ch}$ in the CDFT is  given by 
\cite{HP.12,Kurasawa.19}\footnote{The analysis of the contributions of the last two terms 
of this expression in non-relativistic DFTs is presented in Refs.\ \cite{FN.75,RN.21,NCLX.21}.
There are some differences  between relativistic and 
non-relativistic treatments of these terms (see detailed discussion in Ref.\ 
\cite{Kurasawa.19}), however, in general  a comparable  modification of charge radii 
is generated by such terms in relativistic and non-relativistic DFTs.}
\begin{eqnarray}
r^2_{ch} = \left< r^2_p \right>  + r_p^2 
                  + \frac{N}{Z}  r^2_n +                   
                   \left< r^2_p \right>_{SO} +
                    \frac{N}{Z} \left< r^2_n \right>_{SO}
\label{r_charge_general}
\end{eqnarray} 
where $<r_p^2>$ stands for point-proton mean square radius as emerging
from the CDFT calculations, $r_p$ and $r_n$ are mean-square charge radii
of single proton and neutron, respectively, and $\left< r^2_p \right>_{SO}$ 
and $\left< r^2_n \right>_{SO}$ are proton and neutron spin-orbit contributions 
to the charge radius.

   The situation with neutron mean-square charge radius is currently 
settled: the weighed average of few experiments provides $r_n^2 = -0.1161 \pm 0.0022$ fm$^2$ 
\cite{RevPartPhys.20} and the value $r_n^2 = -0.1161$ fm$^{2}$ is used in recent studies of 
charge radii within Skyrme DFT and CDFT frameworks (see Refs.\ \cite{HP.12,Kurasawa.19,RN.21}).
In contrast, there are some uncertainties in the definition of the proton  mean-square 
charge radius which are generally discussed in the literature as {\it proton radius puzzle} (see Refs.\ 
\cite{Sick.18,GV.22}).  This puzzle is known as a discrepancy between proton radius obtained from 
muonic hydrogen spectroscopy which provides $r_p = 0.8409\pm 0.004$ fm (see Ref.\ 
\cite{RevPartPhys.20,GV.22}) and that derived from elastic electron-proton scattering which give a 
larger value of $r_p = 0.887 \pm 0.012$ fm \cite{Sick.18}.
It is  still not fully resolved and 
future experiments are required to remove this discrepancy in $r_p$ (see Ref.\ \cite{GV.22}). 

%%%%%%%%%%%%%%%%%%%%%%%%%%%%%%%%%%%%%%%%%%%%%%%%
\begin{table}[htb]
\caption{The rms deviations $\Delta(r_{ch})_{rms}$ between calculated
and experimental charge radii $r_{ch}$ for different calculational 
schemes. See text for more details.
\label{table-radii}
}
\begin{tabular}{|c|c|c|c|c|}
\hline
&\multicolumn{2}{c|}{DDMEY}& \multicolumn{2}{c|}{NL5(Y)} \\ \hline
Calculational & Anchor &Global& Anchor &Global\\  
    scheme &  & &  & \\ \hline
\multicolumn{5}{|c|}{$r_{p} = 0.8$ fm} \\ \hline
Scheme-A &  0.0143   &  0.0218  &  0.0179   &  0.0294  \\ 
Scheme-B &  0.0162   &  0.0253  &  0.0190   &  0.0308  \\ 
Scheme-C &  0.0186   &  0.0253  &  0.0217   &  0.0308  \\ \hline
\multicolumn{5}{|c|}{$r_{p} = 0.8409$ fm} \\ \hline
Scheme-A &  0.0201  & 0.0241  &  0.0228  &  0.0295  \\ 
Scheme-B &  0.0129  & 0.0223  &  0.0163  &  0.0297  \\ 
Scheme-C &  0.0144  & 0.0223  &  0.0203  &  0.0297  \\ \hline
\multicolumn{5}{|c|}{$r_{p} = 0.887$ fm} \\ \hline
Scheme-A &  0.0284      &  0.0291     &   0.0304      &  0.0334      \\ 
Scheme-B &  0.0151      &  0.0217     &    0.0181     &  0.0272      \\ 
Scheme-C &  0.0147      &  0.0217     &   0.0185      &  0.0272       \\ \hline
\end{tabular}
\end{table}
%%%%%%%%%%%%%%%%%%%%%%%%%%%%%%%%%%%%%%%%%%%%%%%%

  To our knowledge all existing CEDFs  were fitted with $r_p=0.8$ fm
and only with first two terms in Eq.\ (\ref{r_charge_general}) (see discussion in Ref.\ 
\cite{PAR.21}): this corresponds to Scheme-A discussed below.
Thus, it is important to evaluate  the significance of these limitations 
as well as those connected with proton radius puzzle on the results for the nuclei included 
in the fitting protocol and on global results. This is done in Table \ref{table-radii} which 
compares the results obtained with three calculational schemes, namely,
\begin{itemize} 
\item 
      {\bf Scheme-A}: only first two terms of Eq.\ (\ref{r_charge_general}) 
                               are included,
  
\item 
      {\bf Scheme-B}: only first three terms of Eq.\ (\ref{r_charge_general}) 
                          are included,
    
\item 
    {\bf Scheme-C}: all terms of Eq.\ (\ref{r_charge_general}) are taken into account 
    in the calculations of charge radii of twelve spherical anchor nuclei included in anchor-based 
    optimization. Such results are shown in the columns labelled as "ABOA" in Table 
    \ref{table-radii}. In contrast, only three first terms of Eq.\ (\ref{r_charge_general})  
    are taken into account in the global calculations which also include transitional and 
    deformed nuclei. This is based on the assessment of Ref.\ \cite{RN.21} that such 
    approximate relation is sufficient in the applications which aim at global description 
    of charge radii\footnote{The spin-orbit contribution to charge radii decreases with 
    increasing  the mass of nuclei and it almost does not depend on the CEDF. These 
    features are  illustrated in Table II of Ref.\ \cite{HP.12}. Since the calculations of this 
    reference are restricted to spherical  shape and neglect the pairing correlations,  the 
    values quoted for spin-orbit  contribution to charge radii  in this table for non-doubly 
    magic nuclei  have to be considered as an upper limit. This is because the deformation 
    and pairing give rise to the mixing of different spherical subshells 
   in the structure of the single-particle states
    which results in a smoothing of the spin-orbit correction to charge radii as a function of particle 
    number \cite{RN.21}. As a consequence, the charge radii in only restricted
    number of light nuclei are moderately affected by spin-orbit contributions (see 
    Refs.\ \cite{HP.12,Kurasawa.19}). Thus, Ref.\ \cite{RN.21} suggested to 
    ignore last two terms of Eq.\ (\ref{r_charge_general}) in global analysis of charge radii.}. 
    As a result, the global results obtained in Scheme-B and Scheme-C 
    calculations are identical.
        
\end{itemize}
and with three values of $r_p = 0.8$, 0.8409 and 0.887 fm. In addition, the calculations 
are performed with two CEDFs, namely, DD-MEY and NL5(Y) in order to see the 
dependence of the results on the functional.

   The column "Anchor" of Table \ref{table-radii} shows the results for 
the nuclei included in typical fitting protocol of anchor-based optimization approach 
(see Ref.\ \cite{TA.23}) which contains 12 nuclei and only 10 of them, namely, 
$^{16}$O, $^{40,48}$Ca, $^{90}$Zr, $^{116,124,132}$Sn and $^{204,208,214}$Pb
provide the information on charge radii. In the column "Global" we compare the results 
of the calculations for the set of 261 even-even nuclei in which experimental data
on charge radii exist. Note that our set is reduced as compared with the 351 
even-even nuclei listed in the compilation of Ref.\ \cite{AM.13} because of the 
reasons discussed below. For example, we exclude the nuclei with proton number 
$Z>83$ since with the exception of uranium nuclei there are no direct experimental
measurements of  the absolute charge radii of such nuclei: the data for these nuclei 
provided in the compilation of Ref.\ \cite{AM.13} is based on extrapolations. Moreover,
we exclude from consideration the nuclei in which beyond mean field and shape 
coexistence effects have substantial impact on charge radii. These are, 
for example, the $Z<10$ nuclei, the Pb and Hg isotopes with $N=100-106$, the Sr, 
Kr, and Mo isotopes with $N<50$ (see Ref.\ \cite{PAR.21}). Thus, we focus on the
nuclei for which the absolute values of charge radii are experimentally 
measured\footnote{It is sufficient  to measure the absolute value of 
charge radius of a single nucleus in isotopic  chain and then to establish the 
charge radii of other nuclei in the chain by means
of laser spectroscopy (see Ref.\ \cite{CMP.16}).} and for which
mean field approximation is expected to be a reasonable well
justified.

    The analysis of the results  presented in Table \ref{table-radii} leads 
to the following conclusions. First, the comparisons of the results obtained with
$r_p = 0.8409$ fm and 0.887 fm in different calculational schemes  clearly indicate 
that the uncertainties in the value of $r_p$ lead to the uncertainties in rms charge 
radii. The latter are the largest for calculational Scheme-A in which they are located
between 0.0039 fm and 0.0083 fm  dependent on the functional and  on whether 
Anchor or Global results are compared. This provides a
potential limit on the accuracy with which the charge radii can be meaningfully 
calculated at present. Further improvement of the description of charge radii below this 
limit requires the resolution  of proton radius puzzle.

   Second, Table  \ref{table-radii} reveals clear correlations between rms 
deviations $\Delta(r_{ch})_{rms}$ obtained in the Anchor and Global calculations: 
for a given functional and $r_p$ the calculational scheme with the 
lowest  $\Delta(r_{ch})_{rms}$ value in Anchor calculations provides 
either the lowest or comparable with the lowest $\Delta(r_{ch})_{rms}$ 
value in Global calculations.  This clearly indicates that the use of twelve
anchor spherical nuclei in ABOA ten of which contain experimental 
information on charge radii is sufficient  for constraining  the properties 
of  charge radii globally. 

  Note that the extension of the dataset of the nuclei with charge radii used 
in the fitting protocol beyond these ten nuclei will not necessary improve the 
agreement with experiment on a global scale because of increased theoretical 
uncertainties in the description of charge radii of the ground states. Indeed, 
charge radii sensitively depend on the underlying single-particle structure (see 
Refs.\ \cite{PAR.21,NOSW.23,PA.23}) the accuracy of the description of which 
typically reduces on going on  from spherical to deformed nuclei (see, for 
example, Refs.\  \cite{BQM.07,LA.11,DABRS.15}). In addition,  in the 
case of the nuclei  with shape coexistence or soft potential energy surfaces 
the beyond  mean effects have to be taken into account and their inclusion 
in model description is extremely numerically costly. Thus, the restriction of 
the dataset of the nuclei with charge radii used in the fitting protocol to  
spherical predominantly single and doubly magic ones reduces theoretical 
uncertainties related to the shape of the nucleus and underlying single-particle 
structure.  However, for these nuclei the inclusion of the last three terms of
Eq.\ (\ref{r_charge_general}) is important for further improvement of the 
functionals.
      
%%%%%%%%%%%%%%%%%%%%%%%%%%%%%%%%%%%%%%%%
\section{Asymptotic behavior of the binding energies: basis truncation effects}
\label{truncation-effects}
%%%%%%%%%%%%%%%%%%%%%%%%%%%%%%%%%%%%%%%%  

%%%%%%%%%%%%%%%%%%%%%%%%%%%%%%%%%%%%%%%%%%%%
\begin{figure*}[htb]
\centering
\includegraphics*[width=\textwidth,height=18cm,keepaspectratio]{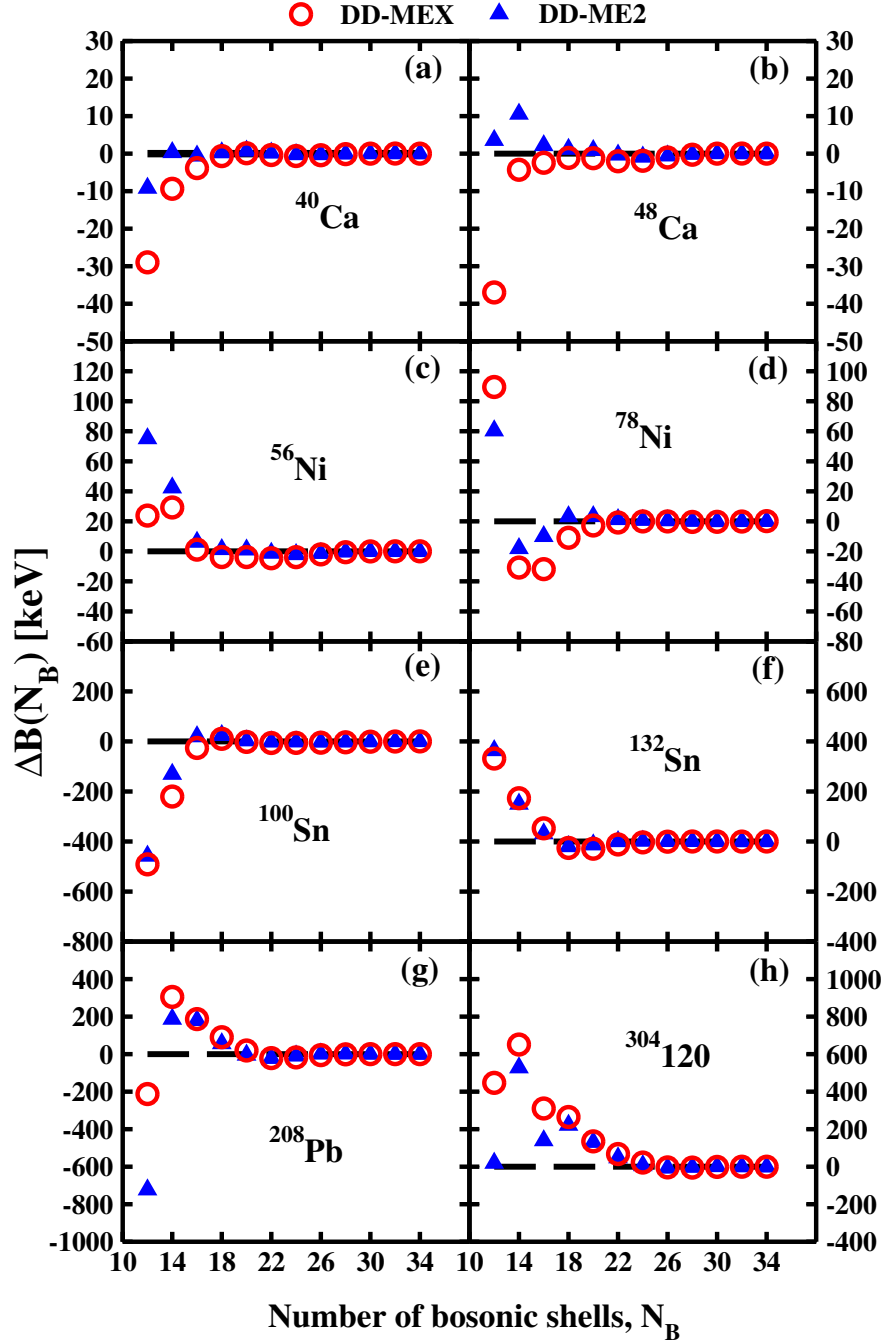}
\caption{The dependence of the difference $\Delta B(N_B) = B(N_B) - B_{\infty}$ 
on the  number of bosonic shells $N_B$. The asymptotic limit is shown by black 
dashed line.  The $\Delta B(N_B)$ values are shown by blue solid triangles 
and open red circles for the DD-ME2 and DD-MEX CEDFs,  respectively. 
\label{bos-sect-conv}
}
\end{figure*}
%%%%%%%%%%%%%%%%%%%%%%%%%%%%%%%%%%%%%%%%%%%%

   The values of $N_F$ and $N_B$ employed in the RMF+BCS or 
RHB calculations depend on available computational power [which increases with 
time]  and on the approach of theory group to handling the numerical errors in physical 
observables. In absolute majority of such calculations, $N_B$ is
fixed at 20 since the calculations in bosonic sector of the models are relatively 
cheap.  In contrast, the size of the fermionic basis shows substantial variations 
and dependence on time of publication, the sophistication of theoretical 
approach and theory group since the cost of numerical calculations in the
RMF+BCS and RHB approaches is mostly defined by this sector.

   Let us briefly review truncation schemes of fermionic basis of existing 
global mass calculations in the CDFT framework. Note that we specify $N_B$ below 
only if it differs from 20. The $N_F=12$ value has been used in global RMF+BCS mass 
calculations with NL3 CEDF in Ref.\ \cite{LRR.99}  
with a larger number of fermionic  shells used in very heavy nuclei.
The RMF+BCS calculations of Ref.\ \cite{GTM.05} have been carried with 
$N_F = N_B=14$ and TMA CEDF. Global RMF+BCS calculations with CEDFs NL3, 
TM1, FSUGold and BSR4 were performed in Ref. \cite{RA.11} with $N_F=18$.  
$N_F=20$ was used in global RHB calculations with the DD-ME2, DD-PC1, NL3* 
and DD-ME$\delta$ functionals in Ref.\  \cite{AARR.14}. Ref.\ \cite{ZNLYM.14} 
presents the results of the global RMF+BCS calculations with phenomenological  
treatment of the dynamical correlation energy and CEDF PC-PK1: it uses $N_F=14$ 
for $Z<60$ nuclei  and $N_F=18$ for heavier ones.  The transition to microscopic
beyond mean field approaches leads to a significant increase of computational
time. As a consequence, the fermionic basis is reduced in such studies. For example,
the first global investigation within five-dimensional collective Hamiltonian (5DCH)
based on triaxial RMF+BCS calculations with PC-PK1 CEDF employs $N_F=12$,
14 and 16 for the nuclei with $Z<20$, $20\leq Z \leq 82$, and $Z\geq 82$, 
respectively (see Ref.\ \cite{LLLYM.15}). The same fermionic basis is used in the 
5DCH studies based on triaxial RHB calculations with PC-PK1 CEDF presented in 
Ref.\ \cite{YWZL.21}.

While the majority of these schemes of the truncation of the fermionic basis provide a sufficient 
accuracy for absolute majority of physical observables (such as radii, deformations,
etc)  it is still an open question on how accurate they are 
for the calculations of the binding energies.  This is because the assessment of the 
accuracy of the truncation of the basis in these publications has been either 
not carried or performed by comparing the solutions obtained with $N_F$ and
$N_F+2$ (or $N_F+6$ as in Ref.\ \cite{AARR.14}) for very restricted set of nuclei. 
 As a consequence,  a number of the issues has not even been
considered in earlier publications. For example, how the truncation errors in 
binding energies depend on the type of the functional?  Or what is the evolution of 
truncation errors with proton and neutron numbers? Are those errors gradually
increase with particle numbers or there are some local fluctuations caused by
the shell effects?

  To address these questions one should consider an asymptotic limit 
for the binding energies $B(Z,N)$\footnote{$B(Z,N)$ is the (negative) binding 
energy of the nucleus with $Z$ protons and $N$ neutrons.}
which corresponds to $N_F \rightarrow \infty$ and
$N_B \rightarrow \infty$. The asymptotic 
(negative) binding energy  corresponding to these limits is defined as 
$B_{\infty}(Z,N) = B(N_F \rightarrow \infty, N_B \rightarrow \infty)$ and 
it serves as a benchmark with respect  of which the truncation errors 
in binding energies  in the basis with ($N_F, N_B$) have to be defined. 
The consideration of this limit and related truncation errors are presented 
below for bosonic and fermionic sectors separately.

%%%%%%%%%%%%%%%%%%%%%%%%%%%%%%%%%%%%%
\subsection{Bosonic sector}
\label{bos-sect}
%%%%%%%%%%%%%%%%%%%%%%%%%%%%%%%%%%%%%

Since the bosonic basis size grows modestly with $N_B$ 
the absolute majority of the RMF+BCS and RHB
calculations have been performed with $N_B=20$ starting from 90ites of
the last century. For most of physical observables, this basis provides 
a required accuracy. However, we are not aware of any publication in which 
the convergence of binding energies as a function of $N_B$ has been 
presented.  Fig.\ \ref{bos-sect-conv} and Table \ref{table-bos-conv} fills this gap 
in our knowledge and shows the dependence of the difference  
$\Delta B(N_B) = B(N_B) - B_{\infty}$   between the  calculated binding energies
$B(N_B)$ and $B_{\infty}$ on the  number of bosonic shells $N_B$ for selected 
set of spherical nuclei ranging from light to superheavy ones. The binding energies 
obtained in the calculations with $N_B=32$ and $N_B=34$ do not differ by more
than 1 keV both in spherical and deformed calculations. Thus, their average is 
treated as $B_{\infty}$. Note that fermionic basis is fixed at $N_F=20$ in these 
calculations. To illustrate the dependence of the convergence of binding energies 
with increasing $N_B$ on the functional, the calculations are carried out with for 
CEDFs DD-ME2 \cite{DD-ME2} and DDMEX \cite{TAAR.20}, which are similar in 
structure but somewhat different in numerical values of the parameters.

   The convergence pattern of binding energies as a function of $N_B$ 
depends  on the nucleus. For example, the $^{40}$Ca and $^{100}$Sn nuclei 
are more bound at low $N_B$ ($\Delta B(N_B)<0)$ as compared with asymptotic 
limit which they gradually  reach with increasing $N_B$ [see Fig.\ \ref{bos-sect-conv}(a)
and (e)]. In contrast, the $^{132}$Sn nucleus is less bound ($\Delta B(N_B)>0)$
at $N_B\leq 16$ than asymptotic limit, then it becomes slightly more bound for $N_B=18-24$, and finally approaches asymptotic limit [see Fig.\ \ref{bos-sect-conv}(f)]. Similar but inverted behavior is seen in $^{78}$Ni [see Fig.\ \ref{bos-sect-conv}(d)].
If to exclude the results at $N_B=12$, the $^{208}$Pb nucleus is
less bound than in asymptotic limit for $N_B=14-20$ but calculated binding energies 
are close to asymptotic limit at higher $N_B$  [see Fig.\ \ref{bos-sect-conv}(g) and (h)].

 The convergence of the results to asymptotic limit only weakly depends on the 
functional [see Table \ref{table-bos-conv}]. Few keV accuracy
is achieved globally with $N_B=28$, in sub-lead ($Z<82$) region with $N_B=24$
and in tin and sub-tin ($Z\leq 50$) region with $N_B=20$. 
With increasing mass number, the accuracy of a given truncation of bosonic basis 
decreases.  For example, the truncation scheme with $N_B=20$ is accurate up to 
1 keV in $^{40}$Ca in both functionals. However, its accuracy drops to $\approx 130$ 
keV in superheavy $^{304}$120 nucleus. Note that the sign of $\Delta B(N_B)$ 
fluctuates across the nuclear chart for a given $N_B$: this means that in some nuclei 
a given truncation scheme provides more bound and in others less bound solutions 
as  compared with asymptotic ones (see also Fig.\ \ref{bos-sect-conv}).
 
   The numerical errors in the description of binding energies increase on 
transition from spherical to deformed nuclei. This is illustrated in Table 
\ref{table-bos-conv-def} on the example of selected  deformed rare-earth, actinide 
and superheavy nuclei. For example, the use of $N_B=20$, 24 and 28 in rare-earth 
$^{162}$Sm, $^{164}$Dy and $^{170}$Yb nuclei leads to $\approx 50$ keV, 
$\approx 10$ keV and  a few keV errors in the description of binding energies, 
respectively. These errors somewhat increase on transition to actinides ($^{240,250}$Pu) 
and superheavy ($^{272}$Ds, $^{278}$Cn) nuclei so that it is required to use $N_B=28$ 
basis to achieve $\approx 10$ keV errors in these nuclei. Note that the errors in 
the light and medium mass ($A<130$) nuclei which are not shown in Table 
\ref{table-bos-conv-def} are substantially lower than those in rare-earth region.

   Considering the patterns of the behavior of $\Delta B(N_B)$ with increasing
$N_B$ shown in Fig.\ \ref{bos-sect-conv} (such as those in $^{132}$Sn), the
extrapolation relations (similar to those discussed in Sec.\ \ref{fermion-sect}) 
based on small size of bosonic basis can suffer from numerical errors. Thus,
taking into account the modest increase of basis size with $N_B$, it is better to use
large basises (such as $N_B=28$) for precise mass calculations.  This will lead 
only to modest increase of computational time and memory requirements as
compared with $N_B=20$ case and can be easily handled on modern high-performance 
computers.

%%%%%%%%%%%%%%%%%%%%%%%%%%%%%%%%%%%%%%%%%%%
\begin{table}[h!]
\centering
\caption{The $\Delta B(N_B)$ values (in keV) for $N_B=20$, 24 and 28 for the CEDF DD-ME2 and 
               DD-MEX.
               \label{table-bos-conv} 
               }
\begin{tabular}{|  c| c| c| c| c| c| c|}
\hline
\hline
 \multirow{2}{3em}{Nuclei} & \multicolumn{3}{c|}{DD-ME2} & \multicolumn{3}{c|}{DD-MEX}\\  \cline{2-7}
  & $\Delta$B(20) & $\Delta$B(24) & $\Delta$B(28)& $\Delta$B(20) & $\Delta$B(24) & $\Delta$B(28) \\ \hline
  $\mathrm{^{40}}$Ca & 0.72 & -0.24 & 0.02         &  0.10 & -0.63 & -0.14    \\ 
  $\mathrm{^{48}}$Ca & 0.90 & -0.87 & -0.15          & -1.23 & -1.90 & -0.30    \\ 
  $\mathrm{^{56}}$Ni & 1.10 & -1.99 & -0.34           & -3.83 & -3.90 & -0.56    \\ 
  $\mathrm{^{78}}$Ni & 3.23 & 0.72 & 0.09              & -2.79 & -0.07 & -0.37   \\
  $\mathrm{^{100}}$Sn & 3.42 & -0.95 & -1.60        & -1.69 & -5.65  & -3.22   \\ 
  $\mathrm{^{132}}$Sn & -12.74 & 1.98 & 0.61       & -28.73 & -3.42 & 0.21   \\ 
  $\mathrm{^{208}}$Pb & -4.25 & -9.00 & 2.03        & 19.58 & -17.34 & -0.86 \\ 
  $\mathrm{^{304}120}$ & 130.96 & 6.59 & -2.09    & 135.33 & 21.95 & -4.98 \\  \hline
  \hline
\end{tabular}
\end{table}
%%%%%%%%%%%%%%%%%%%%%%%%%%%%%%%%%%%%%%%%%

%%%%%%%%%%%%%%%%%%%%%%%%%%%%%%%%%%%%%%%%%
\begin{table}[htb]
\centering
\caption{The same as in Table \ref{table-bos-conv} but for selected set of axially 
deformed nuclei.
\label{table-bos-conv-def}
}
\begin{tabular}{|  c| c| c| c| c| c| c|}
\hline
\hline
  \multirow{2}{3em}{Nuclei} & \multicolumn{3}{c|}{DD-ME2} & \multicolumn{3}{c|}{DD-MEX}\\  \cline{2-7}
  & $\Delta$B(20) & $\Delta$B(24) & $\Delta$B(28)& $\Delta$B(20) & $\Delta$B(24) & $\Delta$B(28) \\ \hline
  $\mathrm{^{162}}$Sm  &  -36.38   &   2.93   &   0.85  &  -48.83 &  -4.76  &  0.24       \\ 
  $\mathrm{^{164}}$Dy   &  -40.53   &   0.49   &   0.98  & -56.03  &  -9.89  &  -0.12    \\ 
  $\mathrm{^{170}}$Yb   &  -50.78   &  -1.59   &   1.59  & -53.59  & -13.79  &   0.73   \\ 
  $\mathrm{^{240}}$Pu   &  -33.94   & -25.39  &   0.00  &  -0.49  & -34.30  &  -4.40    \\
  $\mathrm{^{250}}$Pu   &   11.48   & -10.50   &  -5.01  &  35.89  & -10.74  &  -8.79  \\ 
  $\mathrm{^{272}}$Ds   &  -33.33   & -24.17   & -11.84  & -10.13  & -17.70  & -16.60 \\ 
  $\mathrm{^{278}}$Cn   &  -57.37   & -11.48   & -13.06  & -38.33  &  -6.47  & -15.63 \\ \hline
  \hline
\end{tabular}
\end{table}
%%%%%%%%%%%%%%%%%%%%%%%%%%%%%%%%%%%%%%%%% 

%%%%%%%%%%%%%%%%%%%%%%%%%%%%%%%%%%
\subsection{Fermionic sector}
\label{fermion-sect}
%%%%%%%%%%%%%%%%%%%%%%%%%%%%%%%%%%
  
%%%%%%%%%%%%%%%%%%%%%%%%%%%%%%%%%%
\subsubsection{Binding energies}
\label{fermion-sect-binding}
%%%%%%%%%%%%%%%%%%%%%%%%%%%%%%%%%%

%%%%%%%%%%%%%%%%%%%%%%%%%%%%%%%%%%%
\begin{figure*}[htb]
\centering
\includegraphics*[width=14.5cm,angle=0]{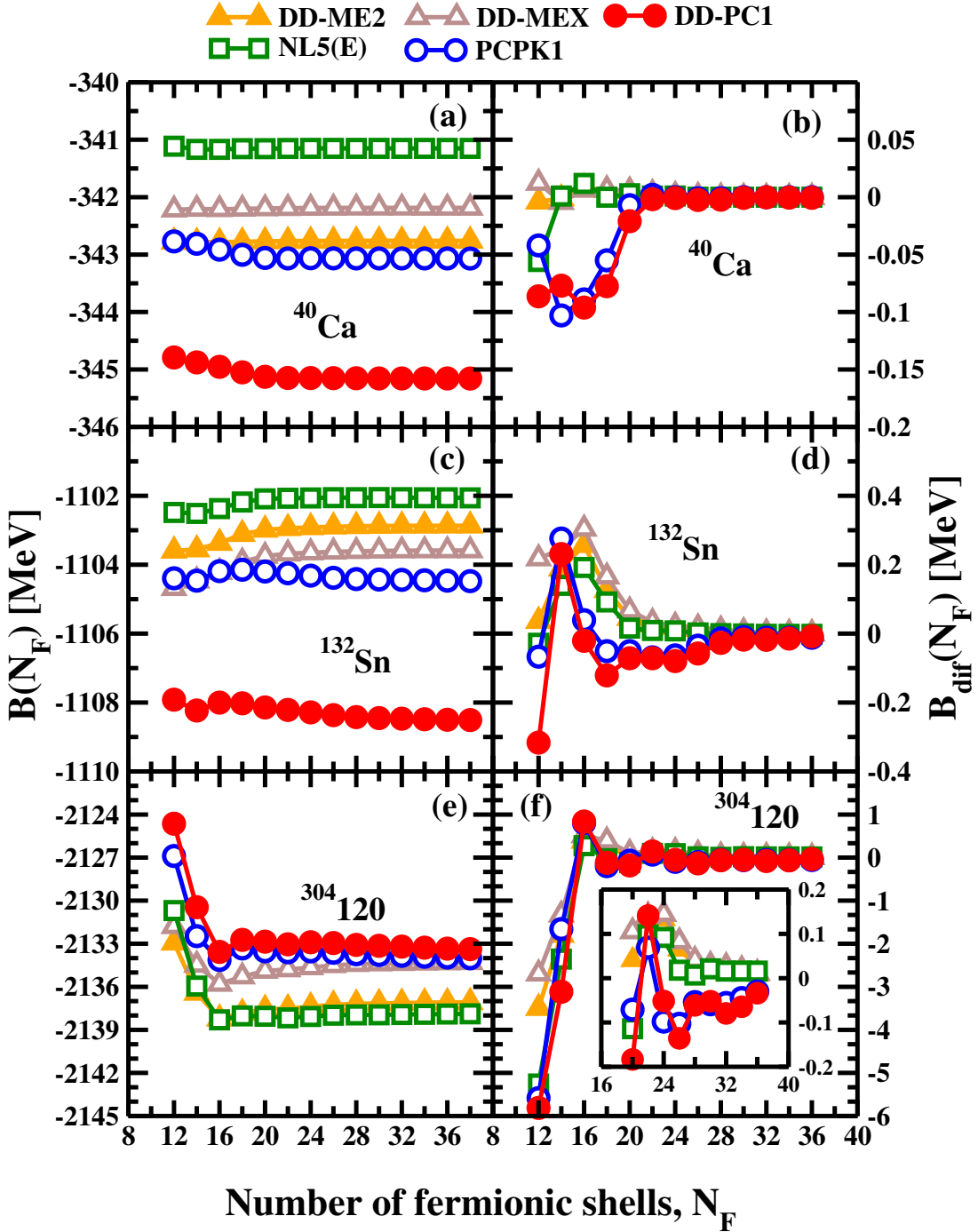}
\caption{The evolution of binding energies (left panels) and the $B_{dif}(N_F)$ quantities 
(right panels) in the $^{40}$Ca, $^{132}$Sn and $^{304}$120 nuclei under the constraint
to spherical shape for  indicated functionals. 
\label{E-ferm-conv-sph}
}
\end{figure*}
%%%%%%%%%%%%%%%%%%%%%%%%%%%%%%%%%%%%%%%

  The asymptotic values of binding energy are extremely  numerically expensive 
to calculate and we are not aware about any attempts in the CDFT framework apart 
of those presented in Refs.\ \cite{AATG.19,TA.23} for a few nuclei to  obtain or to 
approach these values by a drastic increase of the basis size beyond the one (typically 
$N_F=16-20$)  used in global calculations of binding energies.  To address this issue 
we carried out the calculations for selected sets of spherical and deformed nuclei, the 
representative cases of which are discussed below.

   The evolution of both the calculated binding energy $B(N_F)$ and the quantity 
$B_{dif} (N_F)$ [defined below] as a function of number of fermionic shells 
$N_F$ is considered here. The quantity
\begin{eqnarray}
B_{dif} (N_F) = B(N_F+2) - B(N_F) 
\end{eqnarray}
provides a more accurate measure of the convergence process since it compares the 
binding energies in the calculations with $N_F$ and $N_F+2$.  The convergence is 
reached when $B_{dif} (N_F) \rightarrow 0$ with increasing $N_F$.

\begin{figure*}[htb]
\centering
\includegraphics[width=14.5cm,angle=0]{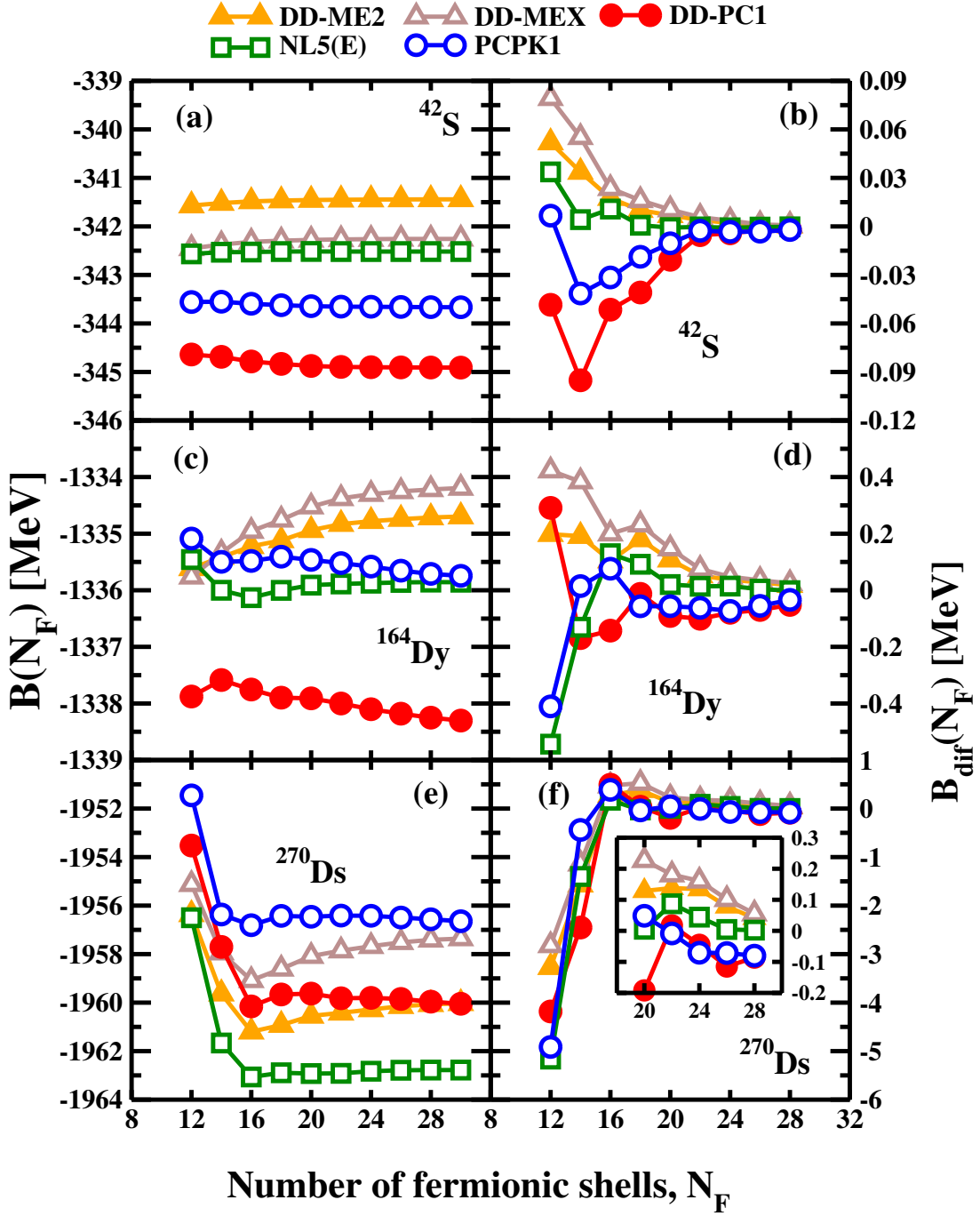}
\caption{The same as in Fig.\ \ref{E-ferm-conv-sph} but for deformed $^{42}$S, 
$^{164}$Dy and $^{270}$Ds nuclei.
\label{E-ferm-conv-def}
}
\end{figure*}

   The results of the calculations for the evolution of these quantities with the increase of 
$N_F$ for  a few selected spherical and deformed nuclei are presented in Figs.\ \ref{E-ferm-conv-sph} 
and \ref{E-ferm-conv-def}, respectively. A few general conclusions emerge from the analysis of the 
results of these calculations.

   First, the binding energy can either decrease or  increase on approaching its asymptotic 
value and this process depends on CEDFs.  For $N_F\geq 18$, the absolute value of binding energy increases 
for the functionals which contain point coupling (PC-PK1 and DD-PC1) while it decreases for CEDFs 
which contain meson exchange (NL5(E), DD-ME2 and DD-MEX).  This trend can be disturbed and reversed
at  lower $N_F$ in medium and heavy mass nuclei [see, for example, Fig.\ \ref{E-ferm-conv-sph}(e)]    
leading to the fluctuations of binding energies at $N_F\approx 18$.

  Second, the convergence speed depends on the functional (see right panels of 
Figs.\ \ref{E-ferm-conv-sph} and \ref{E-ferm-conv-def}). In a given nucleus, 
the fastest convergence (at lower $N_F$) is reached by the NL5(E) functional. This follows by 
the DD-ME2 and DD-MEX functionals which typically require a few extra fermionic shells as 
compared with NL5(E) for a full convergence. The point coupling functionals are characterized
by the slowest convergence since they require additional fermionic shells for a full convergence
as compared with the DD-ME* ones. Note that in superheavy $^{304}$120 nucleus no full 
convergence is reached  with PC functionals even with $N_F=36$ in spherical RHB 
calculations [see Figs.\ \ref{E-ferm-conv-sph}(e)].

%%%%%%%%%%%%%%%%%%%%%%%%%%%%%%%%%%%%%%%%%%%%%%%%%%%%%
%%%%%%%%%%%%%%%%%%%%%%%%%%%%%%%%%%%%%%%%%%%%%%%%%%%%%

  Third, the convergence speed depends on the mass of the nucleus (see right panels of 
Figs.\ \ref{E-ferm-conv-sph} and \ref{E-ferm-conv-def}).  For example, spherical solution
converges at $N_F \approx 18$, $\approx 20$ and $\approx 22$ in the calculations
with NLME, DDME and PC functionals in $^{40}$Ca [see Figs.\ \ref{E-ferm-conv-sph}(b)].  However, 
to reach a comparable level of convergence one should have $N_F \approx 24$,  $\approx 28$ 
and $\approx 36$ in $^{132}$Sn [see Fig.\ \ref{E-ferm-conv-sph}(d)]. The required number 
of shells increases further for superheavy $^{304}$120 nucleus: the convergent solution 
is approached at $N_F\approx 26$ in the calculations with NL5(E) functional and at $N_F=36$ 
in the calculations with the DDME functionals [see Fig.\  \ref{E-ferm-conv-sph}(f)]. Note that
the PC functionals do not fully converge even at $N_F=36$.

        Fourth, some additional increase of the size of fermionic basis  is required for 
a full convergence of the RHB solutions as a function of $N_F$ in deformed nuclei. 
This is illustrated in Fig.\ \ref{E-ferm-conv-def}. Note that because of the memory 
allocation limits one can carry out axially deformed RHB calculations only for  
$N_F\leq 30$.  The results presented  in Fig.\ \ref{E-ferm-conv-def}
suggest that the calculations with NL5(E) functional 
fully converge at $N_F\approx 20$ for masses $A\leq 160$;  at $N_F\approx 24$ for 
masses $160\leq A\leq 260$ but will require the $N_F > 26$ basis for a full 
convergence in superheavy nuclei.  The DDME functionals require $N\approx 20$, 
$\approx 26$ and $\approx 28$ bases for a full convergence in deformed nuclei with 
masses $A\leq 50$, $50 \leq A \leq 100$ and $100 \leq  A \leq  130$, respectively.  Higher $N_F$
is required for a full convergence of binding energies in heavier nuclei (see also 
Sec.\ \ref{fermion-sect-asymptot-rel}). 
The analysis of deformed calculations
with the PC functionals confirms that similar to spherical nuclei they require larger 
fermionic basis for a full convergence as compared with the NLME and DDME ones.   Because of 
the limit of $N_F=30$ in deformed RHB calculations full convergence of the PC
functionals is reached
only in light 
$A\leq 50$ nuclei [see Fig.\ \ref{E-ferm-conv-def}(b)] and its approach is seen in 
the $50\leq A \leq 160$ nuclei [see Fig.\ \ref{E-ferm-conv-def}(d)].  However, this 
size of the basis  is not sufficient to judge at which value of $N_F$ the asymptotic 
value of binding energy will be reached in heavier nuclei [see Fig.\ 
\ref{E-ferm-conv-def}(f)].  Note that these general conclusions following  from 
the results presented in Figs.\ \ref{E-ferm-conv-sph} and \ref{E-ferm-conv-def}  
are in line with the ones obtained in the global studies presented in Sec.\ 
\ref{fermion-sect-asymptot-rel}.

  It is interesting to compare the convergence of the binding energies as a function 
of $N_F$  for different nuclei obtained in CDFT with that in non-relativistic DFTs. 
Unfortunately, to our knowledge such information in latter type of models has been 
published for only  three nuclei, namely, for $^{120}$Sn and $^{102,110}$Zr
in Refs.\ \cite{DSN.04,PSFNSX.08}. The total binding energies of the $^{120}$Sn and 
$^{102,110}$Zr nuclei obtained in the Skyrme DFT calculations with the SLy4 force in 
the basis of $N=20$ and $N=25$ HO shells differ by  110-150 keV \cite{DSN.04,PSFNSX.08}. 
Ref.\ \cite{PSFNSX.08} also indicates that the HO basis with $N=25$ is needed in order 
to describe the binding energies of the nuclei in this mass region with an accuracy of 
the couple of tens keV.  Tables \ref{120Sn} and \ref{Zr-defor} show the differences 
$\Delta B = B(N_F)-B(N_F=20)$ between binding  energies of these three nuclei 
calculated at $N_F$ and  $N_F=20$ for a few selected values of $N_F$. One can 
see that as compared with Skyrme DFT results the convergence of binding energies 
as a function of $N_F$ is significantly better for the NL5(E) functional, slightly
better for the DD-MEX CEDF, and worse for the DD-PC1 one.

   Thus, the convergence of binding energies as a function of the basis size is 
in general comparable for these nuclei in the covariant and Skyrme DFTs.
Note that in  CDFT the nucleonic potential  ($\approx -50$ MeV/nucleon) emerges 
as the sum of very large attractive scalar  $S$  ($S\approx -400$ MeV/nucleon)  and
repulsive vector $V$ ($V\approx 350$ MeV/nucleon) potentials \cite{VALR.05}.  In 
the nucleus with mass $A$, these values are multiplied by $A$ and this leads to a 
cancellation of very large quantities. This is quite different from the structure of the 
Skyrme DFT \cite{Rei.89}. However, this difference in the structure on the models 
does  not have a significant impact on the convergence of binding energies.

%%%%%%%%%%%%%%%%%%%%%%%%%%%%%%%%%%%%%%  
\begin{table}[h!]
\label{table1}
\centering
\caption{The difference $\Delta B = B(N_F)-B(N_F'=20)$ between binding 
energies of spherical $^{120}$Sn nucleus calculated at $N_F$ and 
$N_F'=20$ for indicated CEDFs.
\label{120Sn} 
}
\begin{tabular}{ |c |c |c |c |} 
\hline
\multirow{2}{*}{$\mathrm{N_F}$} & \multicolumn{3}{c|}{$\Delta B$ [MeV]} \\ \cline{2-4}
                                & DD-MEX  & DD-PC1 & NL5(E) \\ \hline
24 & 0.065 & -0.137 & 0.006  \\
26 & 0.076 & -0.201 & 0.006  \\
30 & 0.091 & -0.284 & 0.005  \\
38 & 0.097 & -0.318 & 0.005  \\                 
  \hline
\end{tabular}
\end{table}
%%%%%%%%%%%%%%%%%%%%%%%%%%%%%%%%%%%%

%%%%%%%%%%%%%%%%%%%%%%%%%%%%%%%%%%%%
\begin{table}[h!]
\label{table2}
\centering
\caption{The same as Table \ref{120Sn} but for deformed $^{102,110}$Zr 
               nuclei. 
\label{Zr-defor}
}
\begin{tabular}{ |c |c |c |c |c |c |c |} 
\hline
\multirow{2}{*}{$\mathrm{N_F}$} & \multicolumn{3}{c|}{$\mathrm{^{102}Zr}$: $\Delta B$ [MeV]} & \multicolumn{3}{c|}{$\mathrm{^{110}Zr}$: $\Delta B$ [MeV]} \\ \cline{2-7}
   & DD-MEX    & DD-PC1    & NL5(E)    & DD-MEX    & DD-PC1    & NL5(E) \\ \hline
24 &  0.077   & -0.129   & -0.015   &  0.035   & -0.118   & -0.033 \\
26 &  0.092   & -0.176   & -0.017   &  0.031   & -0.165   & -0.037 \\
30 &  0.101   & -0.229   & -0.022   &  0.028   & -0.215   & -0.040 \\                 
  \hline
\end{tabular}
\end{table}
%%%%%%%%%%%%%%%%%%%%%%%%%%%%%%%%%%%%%  

%%%%%%%%%%%%%%%%%%%%%%%%%%%%%%%%%%%%%%%%%%%%
\begin{figure*}[htb]
\centering
\includegraphics*[width=\textwidth,height=15cm,keepaspectratio]{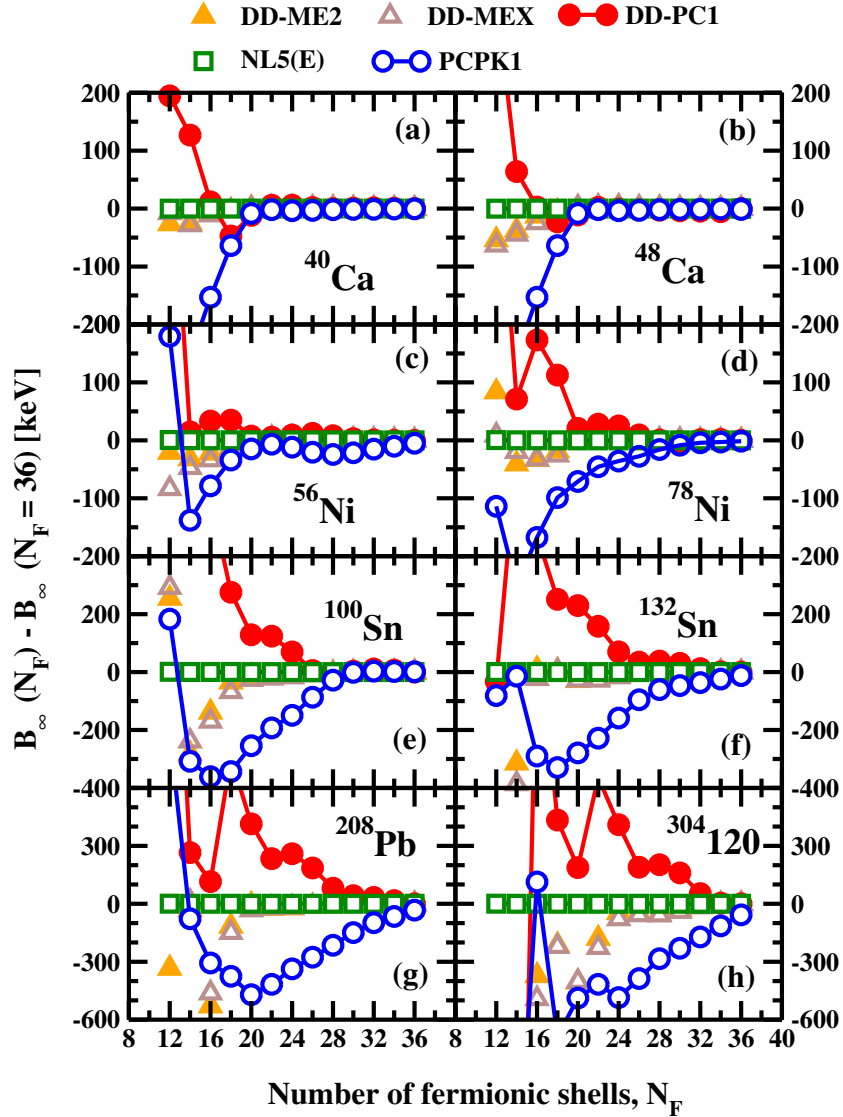}
\caption{The dependence of the  $B_{\infty}(N_F) - B_{\infty}(N_F=36)$
quantity on the  number of fermionic shells $N_F$. The asymptotic limit is shown by 
black  dashed line.  
\label{ferm-sect-conv-sph}
}
\end{figure*}
%%%%%%%%%%%%%%%%%%%%%%%%%%%%%%%%%%%%%%%%%%%%

%%%%%%%%%%%%%%%%%%%%%%%%%%%%%%%%%%%%%%%%%%%%
\begin{table*}[htb]
\centering
\caption{Asymptotic values of binding energy $B_{\infty}(N_F)$ for a set of 
spherical nuclei defined at the values of $N_F=16$, 20 and 36 using prescription defined
in Sec.\ \ref{fermion-sect-asymptot}.   Column 5 shows the binding energies $B(N_F=38)$ 
obtained in the calculations with $N_F=38$. Note that these values are rounded at 0.1 keV. 
The calculations are performed with CEDF DD-MEX. The differences $B_{\infty}(N_F)-B(N_F=38)$
are shown in the parantheses in columns 2-4.
\label{table-asympt-DD-MEX}
}
\begin{tabular}{|  c| c| c| c| c|}
\hline
Nuclei & $B_{\infty}(N_F=16)$ [MeV] & $B_{\infty}(N_F=20)$ [MeV] &  $B_{\infty}(N_F=36)$ [MeV] & $B(N_F=38)$ [MeV] \\ 
 1 & 2 & 3 & 4 & 5 \\ 
$\mathrm{^{40}}$Ca   &  -342.1970 (-0.0111) &  -342.1856 ( 0.0030) &  -342.1860 (-0.0001)  &  -342.1859 \\
$\mathrm{^{48}}$Ca   &  -415.2169 (-0.0246) &  -415.1901 ( 0.0022) &  -415.1924 (-0.0001)  &  -415.1923 \\
$\mathrm{^{56}}$Ni   &  -483.6935 (-0.0343) &  -483.6593 (-0.0001) &  -483.6598 (-0.0004)  &  -483.6592 \\
$\mathrm{^{78}}$Ni   &  -641.2669 (-0.0325) &  -641.2364 (-0.0020) &  -641.2346 (-0.0002)  &  -641.2344 \\
$\mathrm{^{100}}$Sn  &  -828.6610 (-0.1600) &  -828.5256 (-0.0246) &  -828.5001 ( 0.0009)  &  -828.5010 \\ 
$\mathrm{^{132}}$Sn  & -1103.6085 (-0.0224) & -1103.6138 (-0.0277) & -1103.5859 ( 0.0002)  &  -1103.5861 \\
$\mathrm{^{208}}$Pb  & -1638.1732 (-0.4591) & -1637.7423 (-0.0282) & -1637.7112 ( 0.0290)  &  -1637.7141 \\
$\mathrm{^{304}120}$ & -2134.7740 (-0.4803) & -2134.6843 (-0.3906) & -2134.2822 ( 0.0115)  &  -2134.2937 \\  \hline
\end{tabular}
\end{table*}
%%%%%%%%%%%%%%%%%%%%%%%%%%%%%%%%%%%%%%%%%%%%%%

%%%%%%%%%%%%%%%%%%%%%%%%%%%%%%%%%%%%%%%%%%%%%
\begin{table*}[htb]
\centering
\caption{The same as Table \ref{table-asympt-DD-MEX} but for the DD-PC1 functional.
\label{table-asympt-DD-PC1}
}
\begin{tabular}{| c| c| c| c| c|}
\hline
Nuclei & $B_{\infty}(N_F=16)$ [MeV] & $B_{\infty}(N_F=20)$ [MeV] &  $B_{\infty}(N_F=36)$ [MeV] & $B(N_F=38)$ [MeV] \\ 
 1 & 2 & 3 & 4 & 5 \\ 
$\mathrm{^{40}}$Ca   &  -345.1465 (0.0111) &  -345.1697 ( 0.0121) &  -345.1581  ( 0.0050) &  -345.1576 \\
$\mathrm{^{48}}$Ca   &  -418.0405 (0.0049) &  -418.0542 (-0.0089) &  -418.0431  ( 0.0022) &  -418.0453 \\
$\mathrm{^{56}}$Ni   &  -481.5399 (0.0322) &  -481.5659 ( 0.0062) &  -481.5732  (-0.0011) &  -481.5721 \\
$\mathrm{^{78}}$Ni   &  -645.8577 (0.1720) &  -646.0103 ( 0.0194) &  -646.0309  (-0.0012) &  -646.0297 \\
$\mathrm{^{100}}$Sn  &  -827.4990 (0.5906) &  -827.9659 ( 0.1237) &  -828.0948  (-0.0052) &  -828.0896 \\
$\mathrm{^{132}}$Sn  & -1108.0432 (0.4649) & -1108.2864 ( 0.2217) &  -1108.5157 (-0.0076) &  -1108.5081 \\
$\mathrm{^{208}}$Pb  & -1640.4863 (0.0940) & -1640.1875 ( 0.3920) &  -1640.6014 ( 0.0211) & -1640.5803 \\
$\mathrm{^{304}120}$ & -2131.8989 (1.4768) & -2133.2219 ( 0.1538) &  -2133.4097 (-0.0340) & -2133.3757 \\  \hline
\end{tabular}
\end{table*}
%%%%%%%%%%%%%%%%%%%%%%%%%%%%%%%%%%%%%%%%%%%%%%%%

%%%%%%%%%%%%%%%%%%%%%%%%%%%%%%%%%%
\subsubsection{Asymptotic values of binding energies: the analysis of
the procedure from non-relativistic DFTs.}
\label{fermion-sect-asymptot}
%%%%%%%%%%%%%%%%%%%%%%%%%%%%%%%%%%

   To circumvent the numerical problem of finding $B_{\infty}$ in very large 
basis it was suggested in non-relativistic  DFT calculations to use 
the following approximation \cite{HG.07} 
\begin{eqnarray}
B_{\infty}  \approx 2 B(N_F+2) - B(N_F)
\label{B-infty}
\end{eqnarray} 
to estimate the asymptotic value of the binding energy\footnote{Note that in
non-relativistic calculations only the number $N$ of harmonic oscillator 
shells is defined because of the absence of bosonic sector. It is equivalent 
to  $N_F$ in relativistic DFT  calculations. Thus the discussion of non-relativistic
results is presented here in terms of $N_F$.} $B_{\infty}$.
It follows from the following 
relation
\begin{eqnarray} 
\left[ B(N_F+2) - B(N_F)\right] \approx 2 \times \left [B(N_F+4) - B(N_F+2) \right] \nonumber \\
\label{differ}
\end{eqnarray}
which was tested in numerical HFB calculations of various nuclei with Gogny 
D1S force \cite{HG.07}. For example, the prescription of Eq.\ (\ref{B-infty})
was used in the fitting protocols of the D1M \cite{D1M} and D1M* \cite{D1M*}
Gogny EDFs  and the BCPM functional \cite{BCPM}.  However, the numerical
accuracy of this approximation has not been defined in either publication.

   Note that such an approach has not been used in the CDFT 
calculations  of binding energies so far.  In addition, no detailed and systematic 
numerical analysis of this approach to the  calculations of $B_{\infty}$ and its 
numerical errors has been published so far in non-relativistic DFTs. Thus,  it is 
important to understand whether this approach works in CDFT and 
what are related numerical errors.

   It turns out that Eqs.\ (\ref{differ}) and (\ref{B-infty}) can be easily 
reduced  to 
\begin{eqnarray} 
B_{dif}(N_F) \approx 2 B_{dif} (N_F+2)
\label{B-dif-rel}
\end{eqnarray}
and 
\begin{eqnarray}
B_{\infty} \approx B(N_F) + 2 B_{dif} (N_F),
\label{B-infty-new} 
\end{eqnarray}
respectively, and the discussion of numerical accuracy of the approximations under 
consideration can be reduced to the analysis of the evolution of  $B_{dif}(N_F)$ with 
increasing $N_F$.

%%%%%%%%%%%%%%%%%%%%%%%%%%%%%%%%%%%% 
\begin{figure*}[htb]
\centering
\includegraphics*[width=8.5cm]{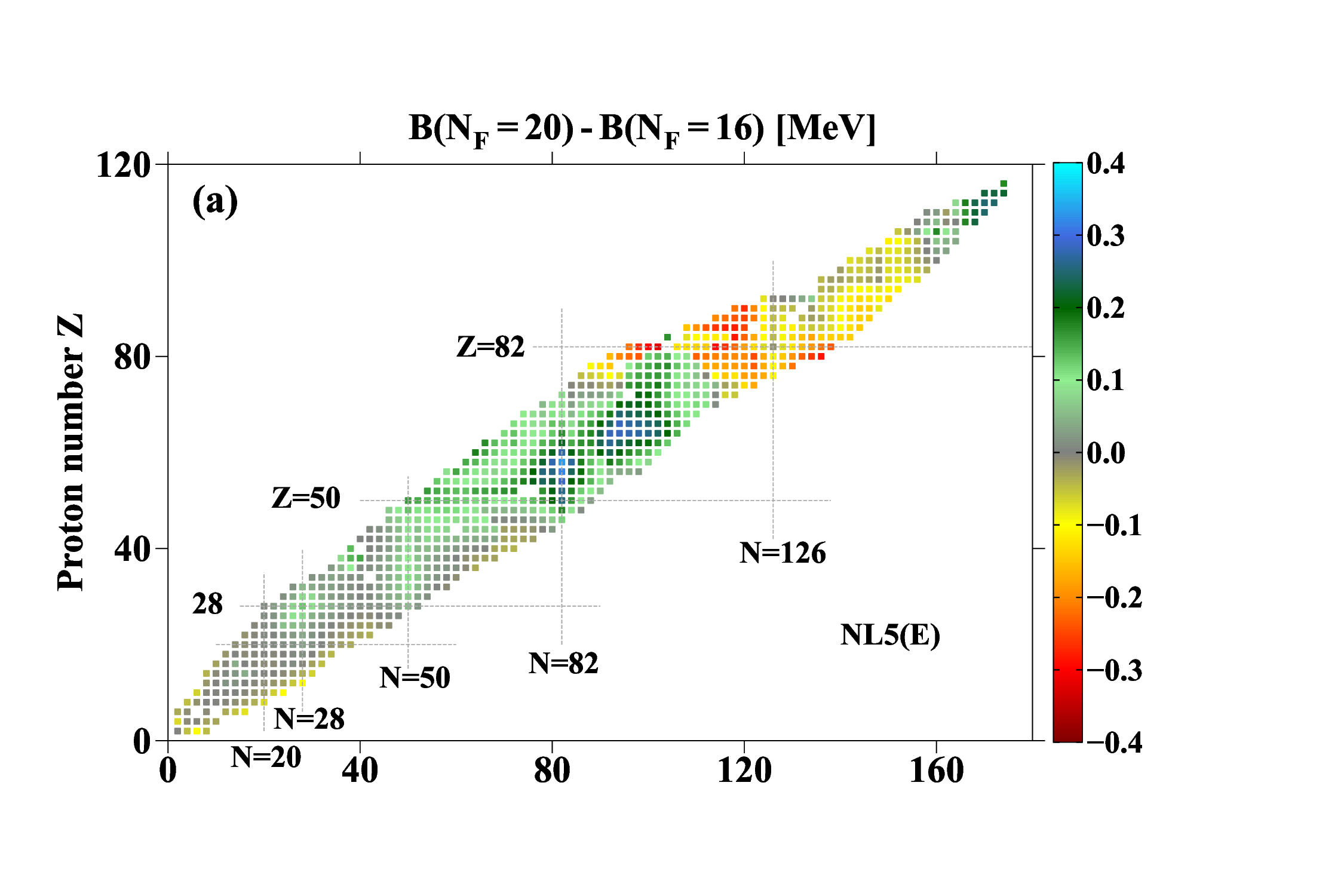}
\includegraphics*[width=8.5cm]{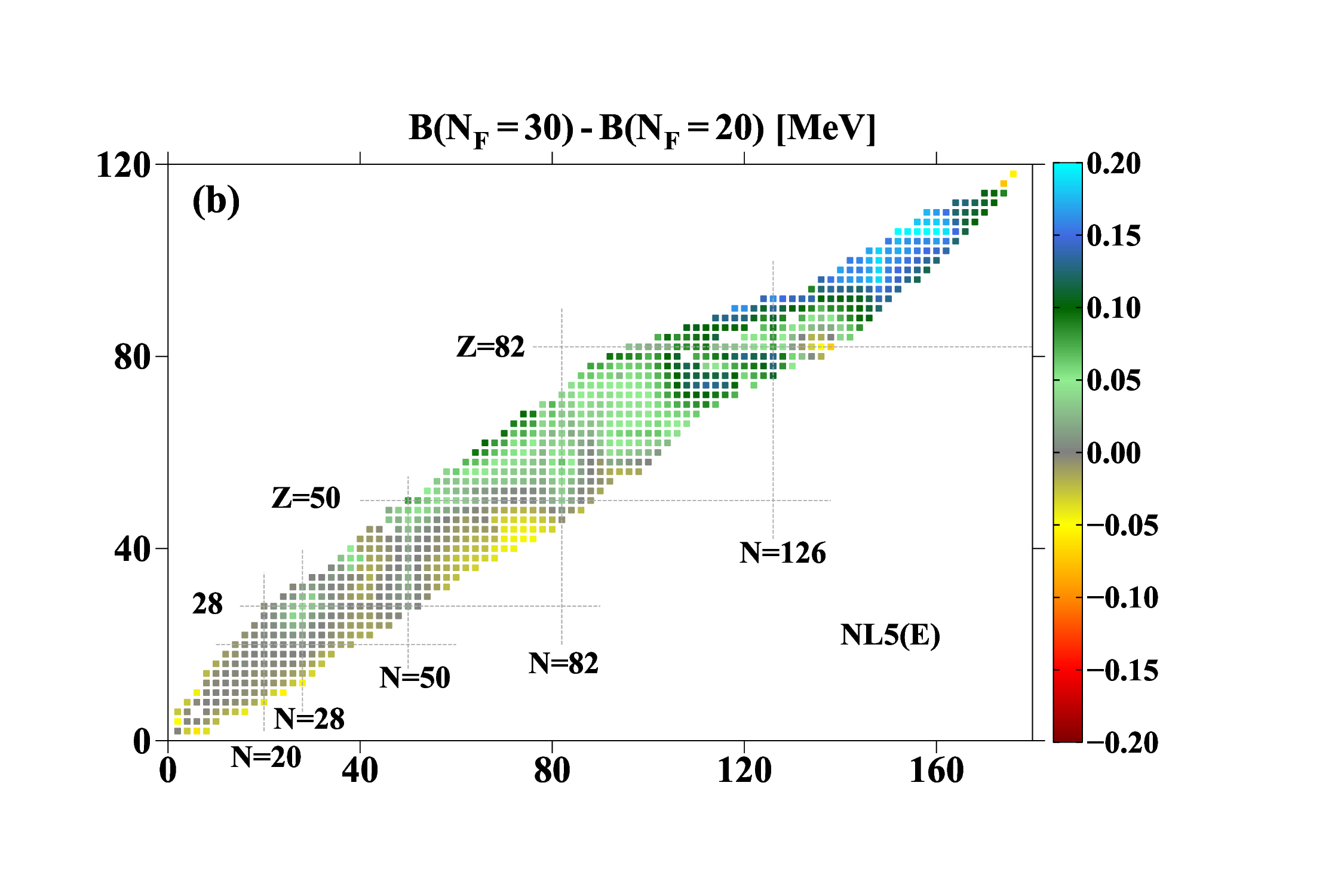}
\includegraphics*[width=8.5cm]{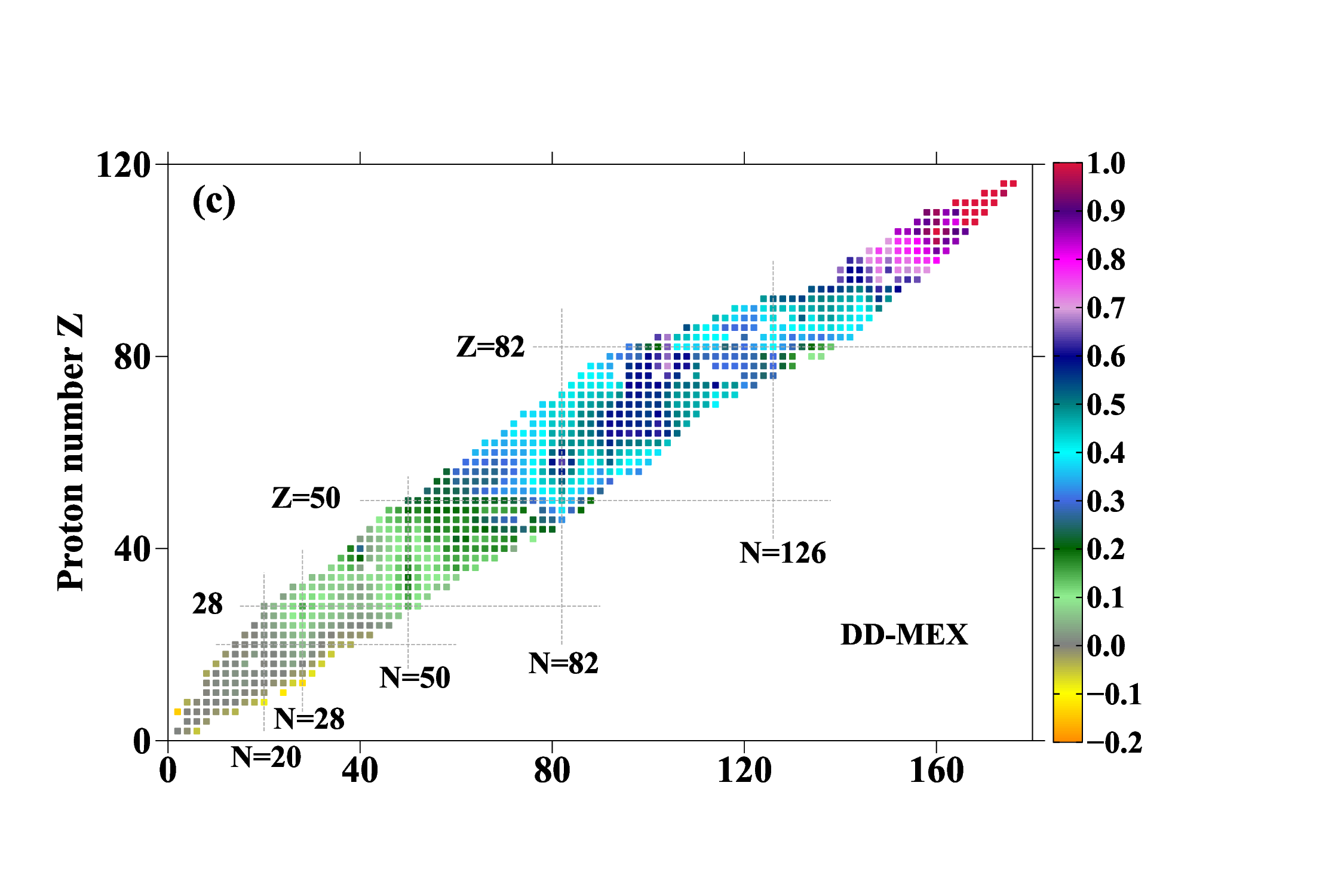}
\includegraphics*[width=8.5cm]{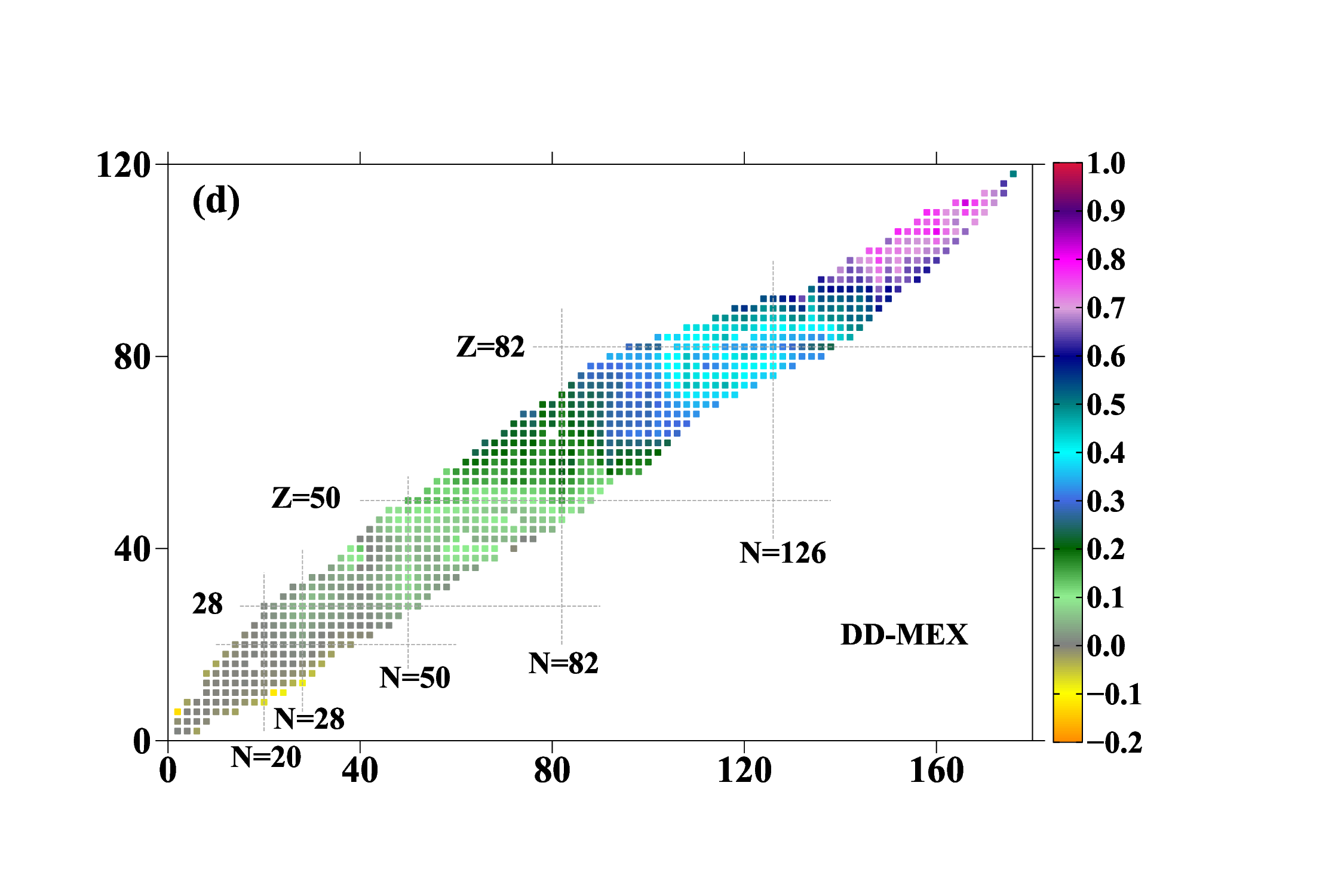}
\includegraphics*[width=8.5cm]{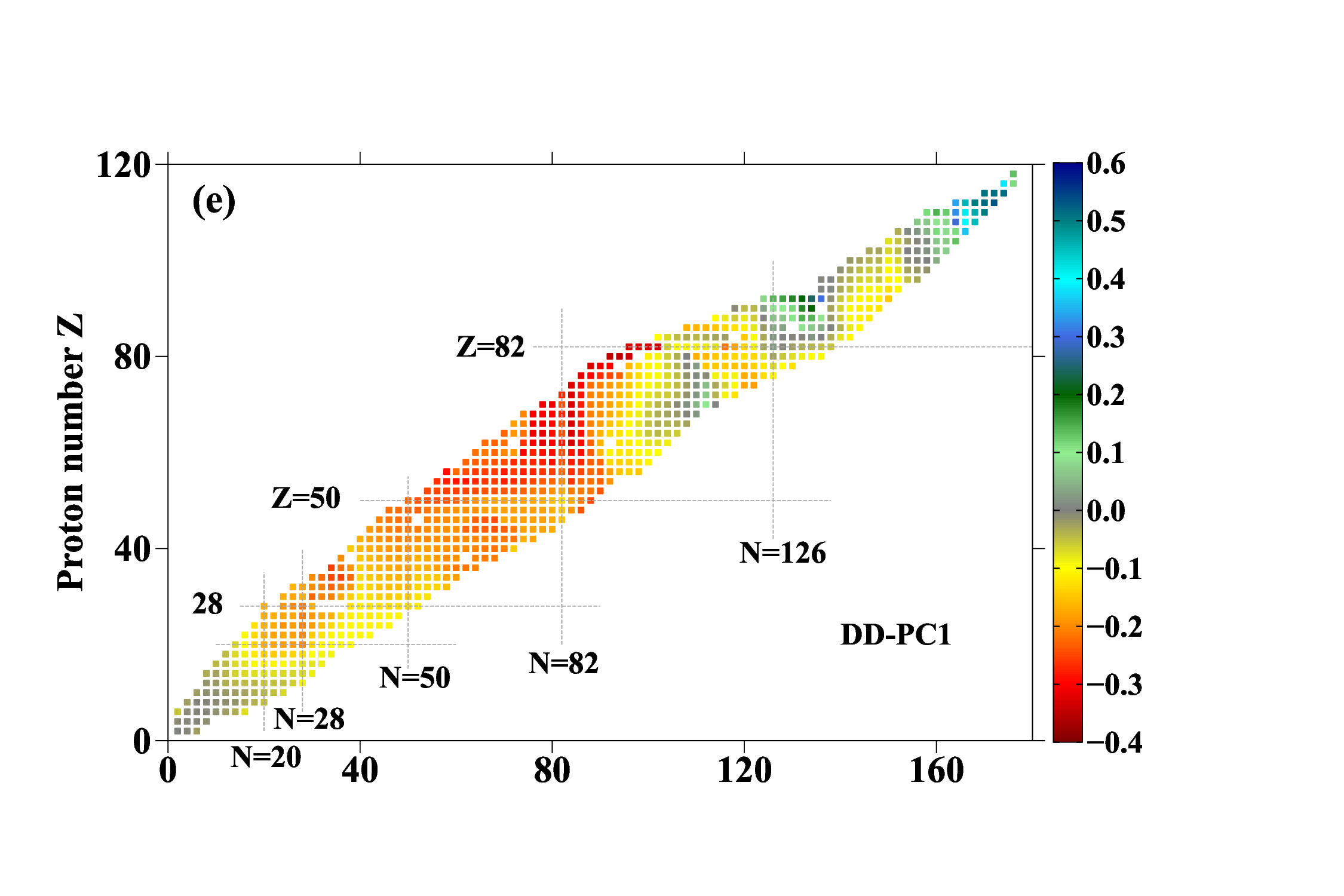}
\includegraphics*[width=8.5cm]{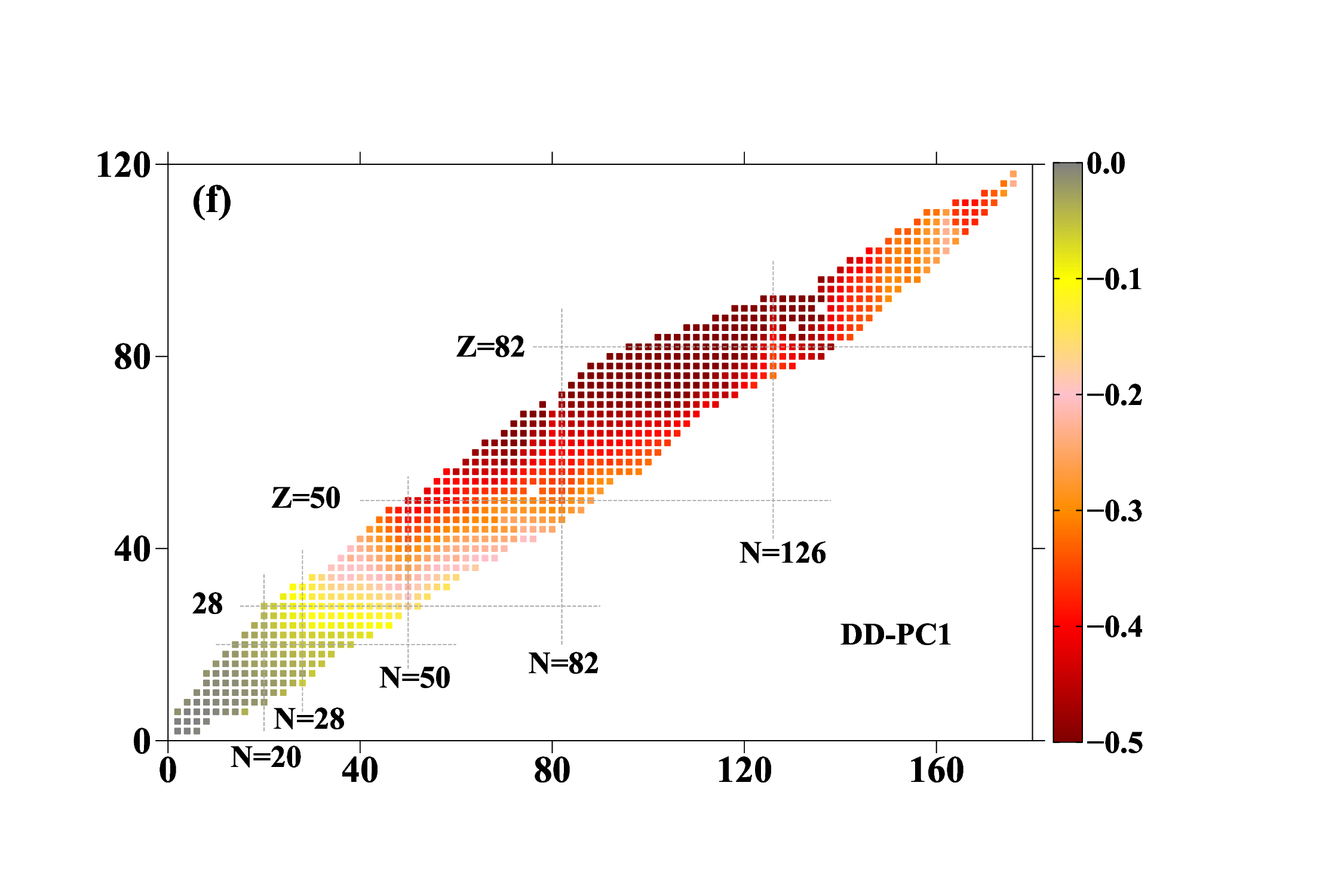}
\includegraphics*[width=8.5cm]{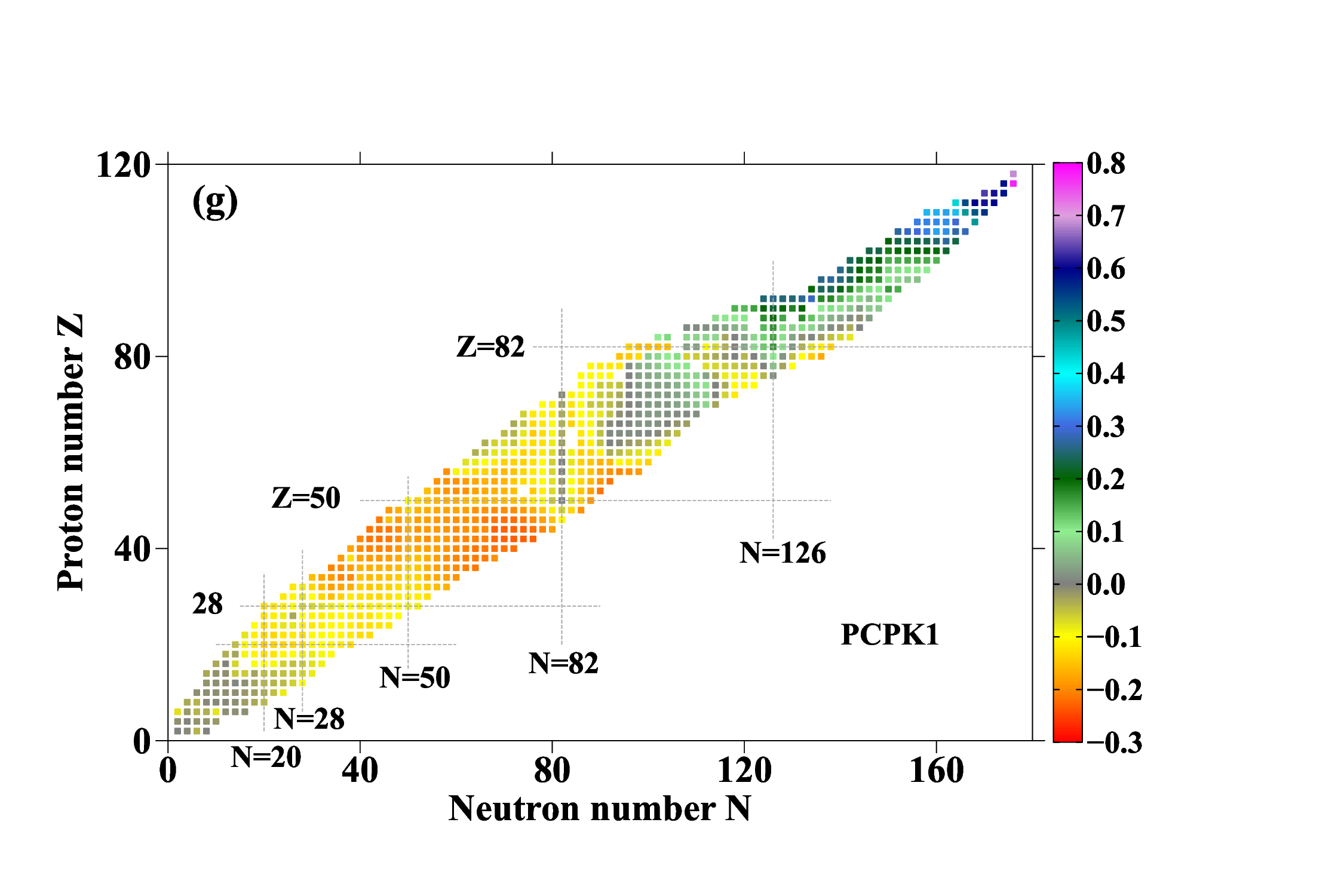}
\includegraphics*[width=8.5cm]{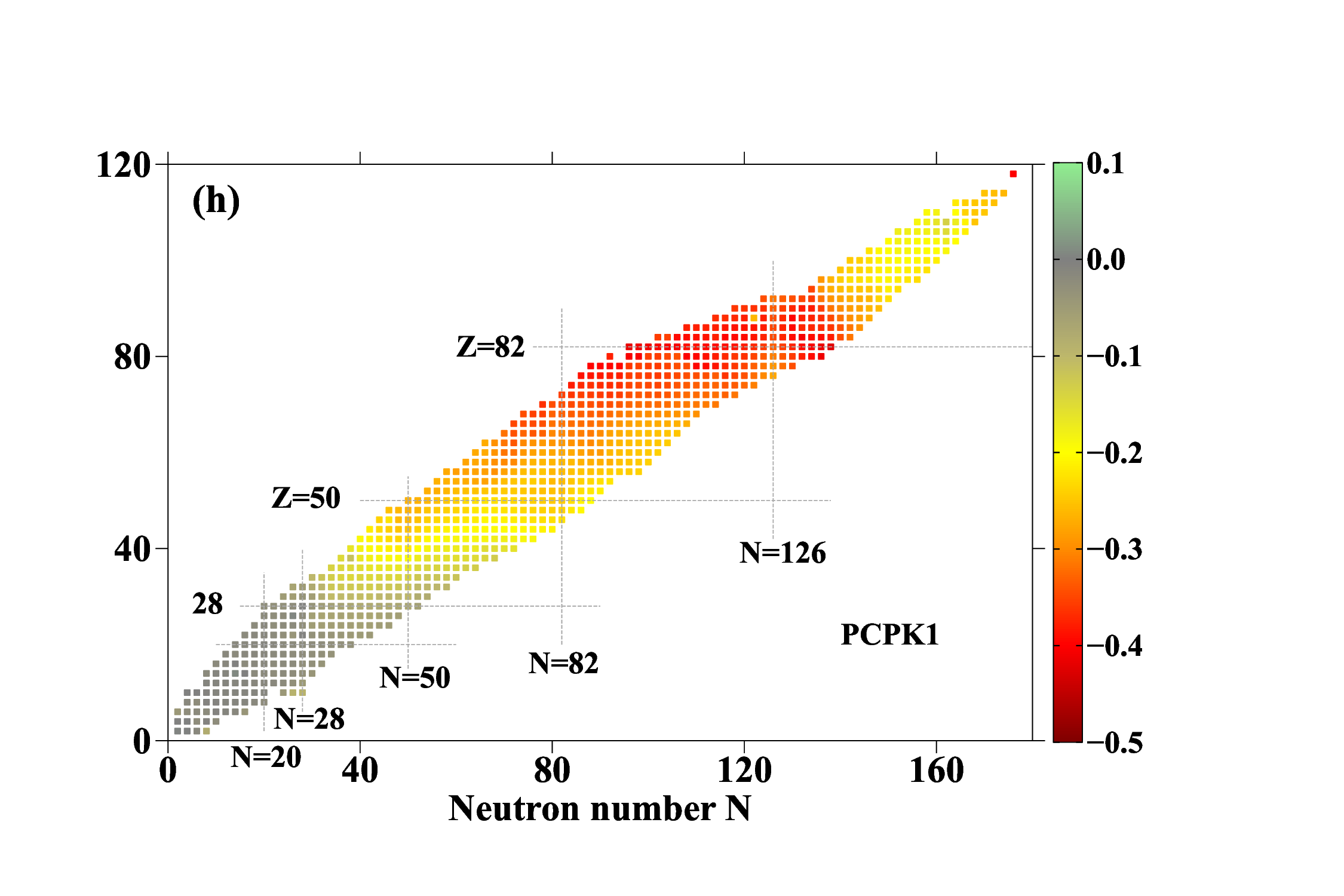}
\caption{The difference between binding energies $B(N_F)-B(N_F')$ 
calculated with indicated truncations of the fermionic basis for the 855 even-even 
nuclei for which experimental data exists in Ref.\ \cite{AME2016-third}. White 
squares are used for the nuclei in which the calculations with compared truncations 
of the basis bring the quadrupole deformations $\beta_2$ which differ by more than 
0.01. Note that the colormaps are different in different panels.
\label{global-results}
}
\end{figure*}
%%%%%%%%%%%%%%%%%%%%%%%%%%%%%%%%%%%%%%

     Figs.\ \ref{E-ferm-conv-sph} and \ref{E-ferm-conv-def} show the evolution 
of  $B_{dif}(N_F)$ with $N_F$ in selected set of spherical and deformed nuclei,
respectively. One can see in the left panels of these figures that  in general 
$B_{dif}(N_F)$ does not follow the rules defined by Eq.\ (\ref{B-dif-rel}). Moreover,  
there are some local fluctuations in $B_{dif}$ at some low to medium values of 
$N_F$ for some functionals [see Figs.\ \ref{E-ferm-conv-sph}(b),(d) and (f) 
and   Figs.\ \ref{E-ferm-conv-def}(b),(d) and (f)] and some drastic changes in the 
slope of $B_{dif}(N_F)$ in superheavy nuclei at $N_F \approx 16$  [see Figs.\ 
\ref{E-ferm-conv-sph}(f) and Figs.\ \ref{E-ferm-conv-def}(f)].

   These results suggest that in reality $B_{\infty}$ as defined by Eq.\ 
(\ref{B-infty}) [or alternatively, by Eq.\ (\ref{B-infty-new})] depends on $N_F$:
thus, further we will denote it as $B_{\infty}(N_F)$. In 
order to illustrate its dependence on $N_F$ the 
$B_{\infty}(N_F) - B_{\infty}(N_F=36)$ quantity is 
shown as a function of $N_F$ for selected set of spherical nuclei in Fig.\ 
\ref{ferm-sect-conv-sph}. Since $B_{\infty}(N_F=36)$ is very close to
asymptotic value of binding energy,  the $B_{\infty}(N_F) - B_{\infty}(N_F=36)$ 
quantity is a measure of numerical error in $B_{\infty}$ due to the use of finite $N_F$.
 In addition, Tables \ref{table-asympt-DD-MEX}
and \ref{table-asympt-DD-PC1} show the numerical values of this quantity at
$N_F=16$, 20 and 36 for the DD-MEX and DD-PC1 CEDFs and compare 
them with the binding energies calculated at $N_F=38$.

   Based on the results presented in these figures and tables one can make 
the following conclusions. First, the $B_{\infty}$ quantity as defined by
Eqs.\ (\ref{B-infty}) and (\ref{B-infty-new}) indeed depends on $N_F$ 
and this dependence is especially pronounced at low $N_F$ values. For example, the 
use of the $N_F=16$ introduces into $B_{\infty}$ the numerical error of the order of 
500 keV for the $^{208}$Pb and  $^{304}$120 nuclei in the calculations with the 
DDME functionals  [see Figs. \ref{ferm-sect-conv-sph}(g) and (h) and Table 
\ref{table-asympt-DD-MEX}].  The errors are smaller for lighter nuclei. In reality, the 
evolution of these numerical errors in $B_{\infty}$ with $N_F$ is similar to those in 
binding energies. The numerical errors in $B_{\infty}$ in general decrease with increasing
$N_F$ but there are some local fluctuations in some nuclei which disturb this general trend.
For example,  in the calculations with DD-PC1 for the $^{304}$120 nucleus
the  $B_{\infty}(N_F) - B_{\infty}(N_F=36)$ quantity is equal to
approximately 200 keV at $N_F=20$ but it increases to $\approx 700$
keV at $N_F=22$ and then more or less gradually decreases with 
increasing $N_F$ [see Fig.\ \ref{ferm-sect-conv-sph}(h)].

  Second, the convergence of $B_{\infty}$ with $N_F$ strongly depends
on the functional. It turns out that as a function of $N_F$, the $B_{\infty}$ 
quantity fully converges at lower $N_F$ for the DDME functionals than 
for the DD-PC1 one.  For example, in $^{132}$Sn  a full convergence 
(within a few keVs of the $N_F=36$ results) of  the DD-MEX and  DD-ME2 
functionals is reached at $N_F=26$, while to achieve that in the DD-PC1 
functional one should have $N_F=34$ [see Fig.\ \ref{ferm-sect-conv-sph}(f)].

  Third, Fig.\ \ref{ferm-sect-conv-sph} and Tables  \ref{table-asympt-DD-MEX} 
and \ref{table-asympt-DD-PC1} clearly illustrate that for a given functional the
size of fermionic basis required for a full (or nearly full) convergence increases 
with mass number of the nucleus.

%%%%%%%%%%%%%%%%%%%%%%%%%%%%%%%%%%
\subsubsection{Global comparison of binding energies at different 
                          truncation of fermionic basis.}
\label{fermion-sect-asymptot-rel}
%%%%%%%%%%%%%%%%%%%%%%%%%%%%%%%%%%
                     
   Fig.\ \ref{global-results} provides a global comparison of the differences 
of binding energies obtained in the calculations with $N_F=16$, $N_F=20$ 
and $N_F=30$.  The first two truncations of the fermionic basis were frequently 
used in the RHB and RMF+BCS calculations during the last decade, and the 
last one is the maximum achievable in the deformed RHB  calculations at 
high-performance computing facility used by our group.
    
       One can see in Fig.\ \ref{global-results} that local fluctuations  in the 
 $B(N_F=20)-B(N_F=16)$  values are quite substantial in some parts of the nuclear 
 chart. Moreover, these fluctuations can have different signs in different
 parts of the nuclear landscape. For example, for the NL5(E) CEDF the rare-earth 
 and superheavy nuclei  are less bound and the nuclei in lead region and actinides 
 are more bound in the calculations with $N_F=20$ than those in the calculations 
 with $N_F=16$ [see Fig.\ \ref{global-results}(a)]. However, the absolute values of 
 $B(N_F=20)-B(N_F=16)$ very rarely exceed 0.3 MeV. In contrast, the $Z>20$  
 nuclei are less bound in the $N_F=20$ calculations with 
DD-MEX functional and the $B(N_F=20)-B(N_F=16)$  difference increases with 
increasing mass number reaching the vicinity of 1.0 MeV  in actinides and 
superheavy nuclei [see Fig.\ \ref{global-results}(c)]. However, the $B(N_F=20)-B(N_F=16)$  
difference is not a smooth function of mass or particle numbers since substantial local
deviations from a general trend are present [see Fig.\ \ref{global-results}(c)]. 
For the DD-PC1 and PC-PK1 functionals, the nuclei located between  neutron numbers 
$N\approx 20$ and $N\approx 90$ are more bound in the  calculations with $N_F=20$  
[see Fig.\ \ref{global-results}(e) and (g)]. In contrast heavier  nuclei are typically less 
bound and this trend is more pronounced in the PC-PK1 functional.  In addition, 
local fluctuations in the $B(N_F=20)-B(N_F=16)$ values are seen for 
the $N\geq 90$ nuclei and they are somewhat more pronounced in the 
calculations with DD-PC1.

  The $B(N_F=30)-B(N_F=20)$  values behave significantly smother as 
a function of particle number than the $B(N_F=20)-B(N_F=16)$ ones 
(compare right with left panels in Fig.\ \ref{global-results}). This clearly
suggest that the local fluctuations seen  in the $B(N_F=20)-B(N_F=16)$ 
values are due to the impact on underlying shell effects on the convergence 
of binding energies as a function of $N_F$ which is significantly more
pronounced at low value of $N_F=16$.

  In all studied functionals the $B(N_F=30)-B(N_F=20)\approx 0$ in light
nuclei (see Fig.\ \ref{global-results}). However, with increasing mass number 
the nuclei gradually become less and less bound in the NL5(E) and DD-MEX 
functionals and the $B(N_F=30)-B(N_F=20)$ values gradually approach 
$\approx 0.2$ and $\approx 1.0$ MeV in superheavy nuclei, respectively
[see Fig.\ \ref{global-results}(b) and (d)]. The situation in the calculations
with PC functionals is different: the nuclei located between proton numbers 
$Z\approx 40$ and $Z\approx 90$ are more bound  (by up to 0.5 MeV) 
in the calculations with $N_F=30$ than those in the calculations with
$N_F=20$ but that difference decreases on moving away from this region [see 
Figs.\ \ref{global-results}(f) and (h)].      
      
%%%%%%%%%%%%%%%%%%%%%%%%%%%%%%%%%%
\section{Asymptotic values of binding energies: alternative
procedure.}
\label{fermion-sect-asymptot-rel}
%%%%%%%%%%%%%%%%%%%%%%%%%%%%%%%%%%
 
%%%%%%%%%%%%%%%%%%%%%%%%%%%%%%%%%%%% 
\begin{figure*}[htb]
\centering
\includegraphics*[width=12.5cm]{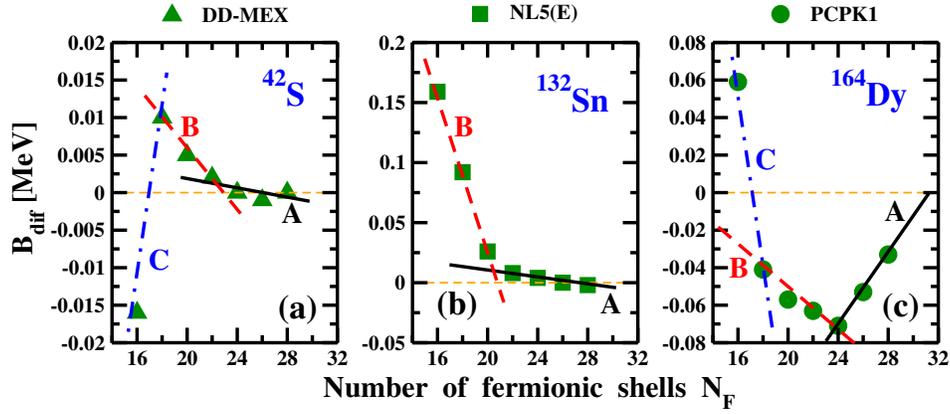}
\caption{Typical patterns of the evolution of $B_{dif}$ as a function 
of the number of fermionic shells $N_F$ illustrated by a few selected
cases. They are based on the 
analysis of the results of the calculations for a  significant number of 
nuclei, the small part of which is presented in Figs.\ \ref{E-ferm-conv-sph} 
and \ref{E-ferm-conv-def}. Thin orange dashed lines show $B_{dif}=0$. Thick 
blue dot-dashed, red dashed and black solid lines show the interpolations of 
the evolution of $B_{dif}$ on the subsets of calculated data points. See text 
for further details. The employed functionals are indicated.
\label{New-scheme}
}
\end{figure*}
%%%%%%%%%%%%%%%%%%%%%%%%%%%%%%%%%%%%%
 
   The detailed analysis of the prescription for finding asymptotic value of 
binding energy in the modest size basis which is used in  non-relativistic DFTs 
presented in Sec.\ \ref{fermion-sect-asymptot} clearly indicates that this method 
does not provide high numerical accuracy and predictive power in the 
case of CDFT. Thus, an alternative procedure is required.

   We suggest to base this procedure on the analysis of the evolution of $B_{dif}$ 
as a function of the number of fermionic shells $N_F$.  The typical patterns of 
such evolution are shown in Fig.\ \ref{New-scheme}.  The full convergence of 
the calculated binding energies as a function of $N_F$ is reached when $B_{dif}=0$. 
Blue dot dashed lines labeled as C in Figs. \ \ref{New-scheme}(a) and (c) show 
linear interpolations for $B_{dif}$  obtained based on the points at $N_F=16$ and
$N_F=18$. One can see  that linear extrapolations based on these interpolations 
will lead to divergent results. The use of the data points at $N_F=18, 20$ and 22 
in the case of the $^{42}$S and $^{132}$Sn nuclei and $N_F=18-24$ in the case 
of $^{164}$Dy leads to the red dashed interpolation lines labelled as B in Fig.\ 
\ref{New-scheme}.  One can see that linear extrapolations based on these interpolations 
will lead to convergent results  for binding energies in the cases of the $^{42}$S and 
$^{132}$Sn nuclei but to divergent results in the case of the $^{164}$Dy nucleus.
Finally, the linear interpolations obtained based on the $N_F=22-28$ data points in 
the cases of the $^{42}$S and $^{132}$Sn nuclei and with $N_F=24-28$ in the case 
of the $^{164}$Dy nucleus are shown by thick solid black lines labelled as A in Fig.\ 
\ref{New-scheme}.  One can see that numerical results fully converge in the first
two nuclei and show approach of full convergence (which according to linear 
extrapolation is expected around $N_F\approx 31$) in $^{164}$Dy.

  The absence of a unique smooth convergence pattern of $B_{dif}$ as a function
of $N_F$ and the drastic changes of their slopes at some values of $N_F$ which
are seen in the change of the slopes of the interpolations C, B and A in Fig.\ 
\ref{New-scheme} clearly indicate that the information on binding energies 
for $N_F>20$ is needed  in order to have a numerical accuracy of calculated 
binding energies within a few tens of keV.  Thus the following calculational 
scheme for the definition of asymptotic binding energies is suggested and 
used in the present paper:

\begin{itemize} 
\item
   The global calculations for binding energies of all nuclei included into fitting
protocol are performed in deformed RHB code with $N_F=20, 22,  24, 26, 28$ and 30. 
This\footnote{ Although such calculations are numerically expensive (see  
Table \ref{table-compt-time-deform}), 
they are feasible at modern high-performance computers. The computational
time presented in this table can be further reduced by at least 30\% 
if the calculations at $N_F$ are carried from the fields defined at $N_F-2$. Note that
the data presented in Table  \ref{table-compt-time-deform} has been obtained
in numerical calculations starting from the fields given by the Woods-Saxon potential
to guarantee the same initial conditions. An extra time required for
such calculations of the 855 nuclei ranges from approximately 15000 CPU-hours
for the DD-PC1 functional to approximately 21000 CPU-hours for the DD-MEX 
one (the numbers are based on the data of Table \ref{table-compt-time-deform}).
This is less than 10\% of the calculational time of full ABOA and one iteration
in the RGA.
} is done after several initial steps of the optimization of the functional with 
$N_F=20$ when  the global rms deviations between experimental and calculated 
binding energies reach $2-3$ MeV.   This guarantees that the corrections for binding
energies between $N_F=20$ and $N_F=\infty$ bases are almost the same for
this somewhat non-optimal solution and the optimal one\footnote{This expectation is based 
on the fact that the evolution of $B_{dif}$ with $N_F$ in a given nucleus is very similar 
for the functionals representing the same class of CEDFs.  For example, 
the analysis of right panels in Figs.\ \ref{E-ferm-conv-sph} and \ref{E-ferm-conv-def} 
clearly shows that this is a case for the DD-ME2 and DD-MEX functionals which 
represent the DD-ME* class of the functionals and for the DD-PC1 and PC-PK1 
CEDFs which are representative examples  of the PC functionals.  This
clearly indicates that for a given class of the functionals the variations of the
parameters of the functional within a reasonable range do not affect substantially
the convergence of binding energies as a function of $N_F$.}. 
 As a consequence,
such calculations for corrections in binding energies have to be done only 
once.

\item
   The results of such calculations allow to define $B_{dif}[Z,N](N_F)$ for each 
nucleus included into the fitting protocol and consequently the infinite basis 
correction  $\Delta B_{\infty} (Z,N) = B[Z,N](N_F=20) - B[Z,N] (N_F=\infty)$ 
using the following procedure:

\begin{itemize}
\item
(A) The infinite basis correction is equal to $\Delta B_{\infty} (Z,N) =  B[Z,N](N_F=20) - B(Z,N) 
(N_F=30)$ for the cases when (i) $B_{dif}[Z,N](N_F) \leq \delta E_{err}$ for
$N_F=28, 26$ and 24 and (ii) the linear interpolation of $B_{dif}[Z,N](N_F)$
based on these $N_F$ [see line A in Figs.\ \ref{New-scheme}(a)
and (b)] clearly indicates the convergence of calculated binding energies.
Here the $\delta E_{err}$ [defined by the user - typically a few keV] is related to
the error in the definition of the value of asymptotic binding energy.

\item 
(B) In the cases when $B_{dif}$ clearly shows the trend of the convergence 
of binding energies with increasing $N_F$ [as for the interpolation line A in 
Fig.\ \ref{New-scheme}(c)]  the convergence point at $\tilde{N_F}$ is defined by 
linear extrapolation of the interpolation line defined at $N_F=24, 26$ and 28 to 
$B_{dif}=0$ [i.e. $B_{dif}(\tilde{N_F})=0$]. This allows to define extrapolated 
energy $B(\tilde{N_F})$, which is associated with infinite basis binding energy
$B_{\infty}(Z,N)=B(\tilde{N_F})$,  and infinite basis  correction  
$\Delta B_{\infty} (Z,N) =  B[Z,N](N_F=20) - B[Z,N](\tilde{N_F})$.

\item 
If the conditions given in (A) and (B) are not satisfied [as for interpolation line 
B in Fig.\ \ref{table-compt-time-deform}(c)], this means that full convergence
of $B_{dif}$ cannot be defined at $N_F=30$.  This is typically happening
in the PC models for very heavy nuclei (see discussion below).  To define 
$\Delta B_{\infty} (Z,N)$ for such cases one should carry the numerical 
calculations at the $N_F$ values which are beyond $N_F=30$.

\end{itemize}

\end{itemize}

  Such procedure allows to define the map in the $(Z,N)$ plane of the infinite 
basis corrections which can be used at all iterations of the fitting procedure. In 
addition, it allows to evaluate the numerical errors in the definition of such
corrections. 

\begin{figure*}[htb]
\centering
\includegraphics*[width=7.8cm]{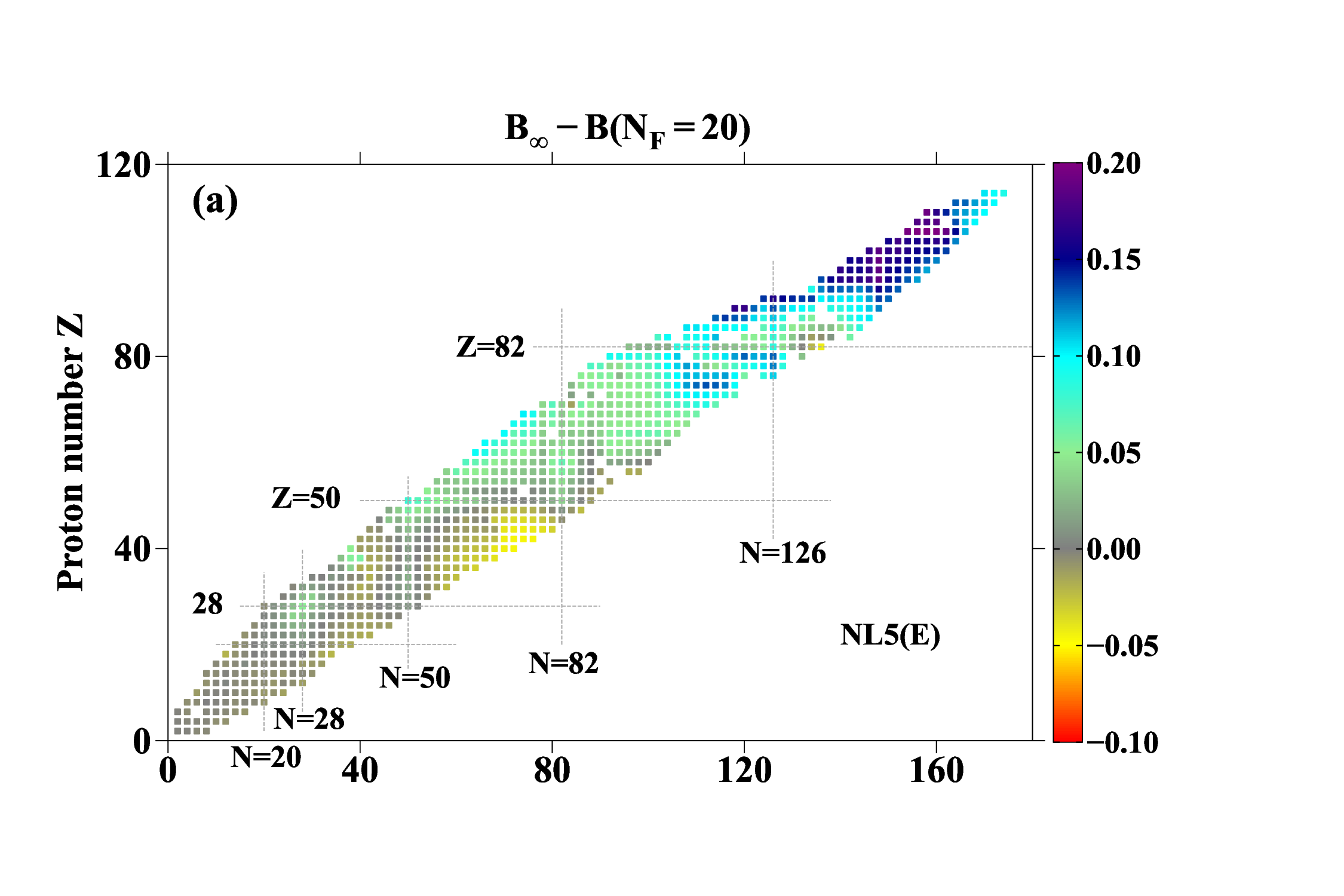}
\includegraphics*[width=7.8cm]{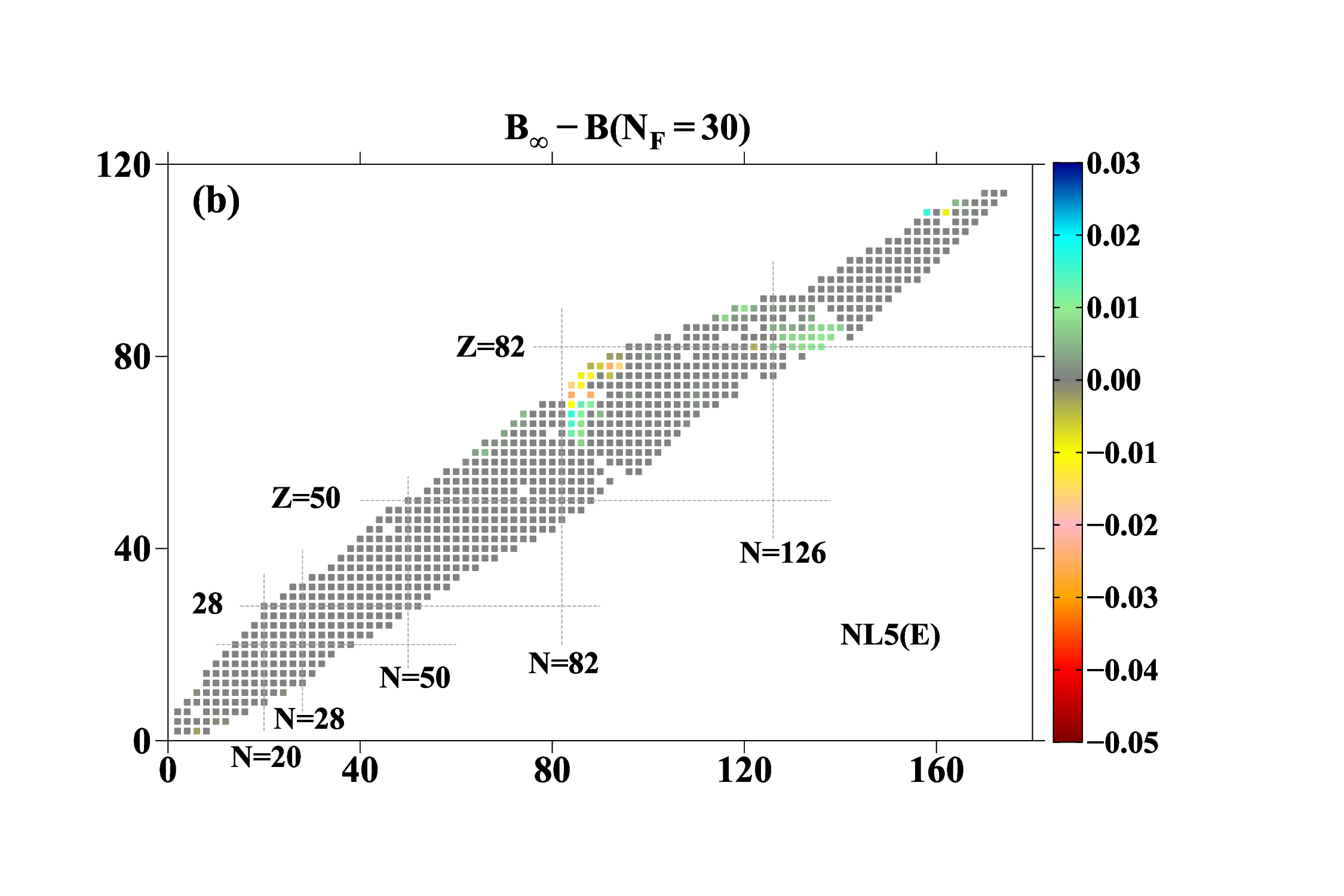}
\includegraphics*[width=7.8cm]{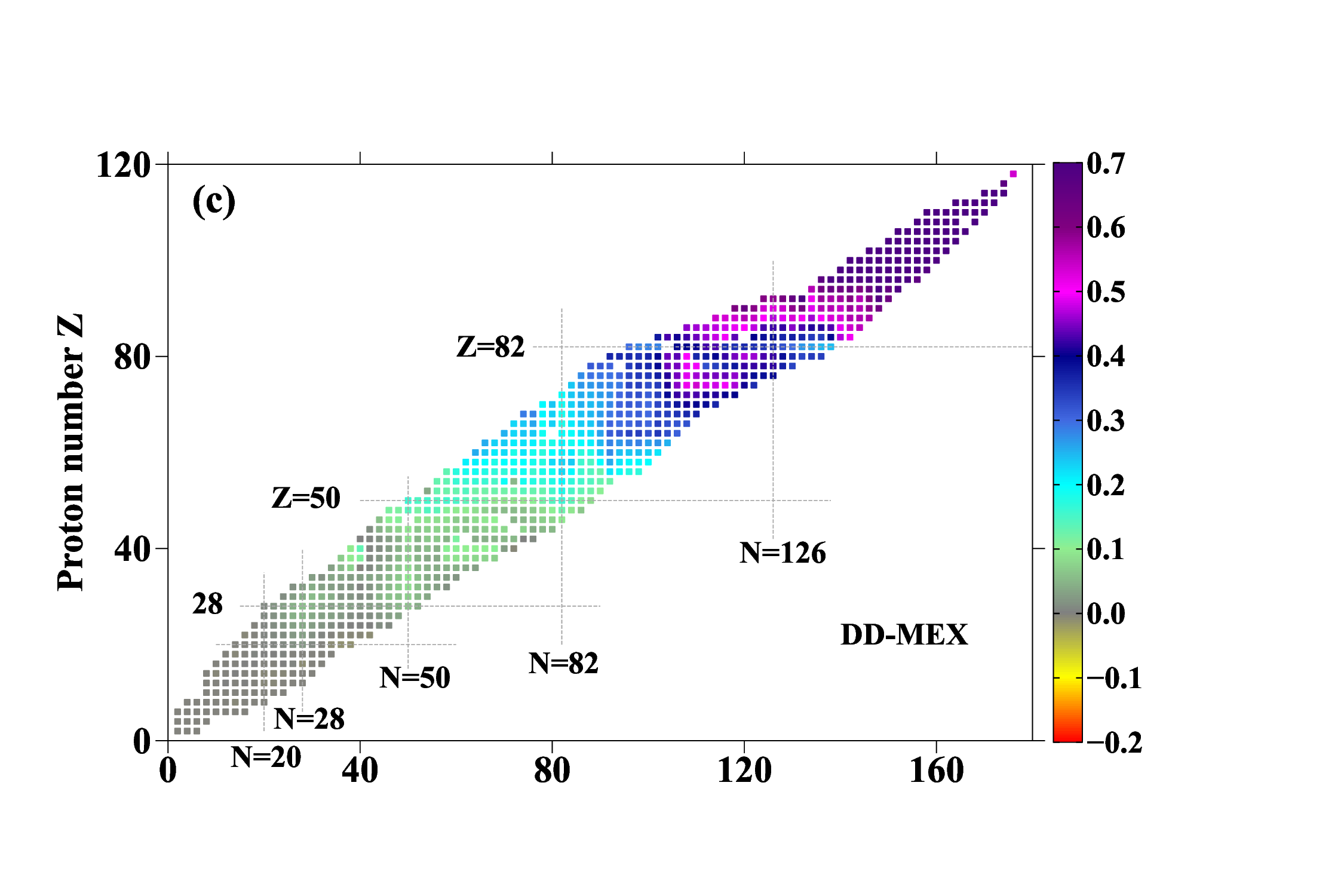}
\includegraphics*[width=7.8cm]{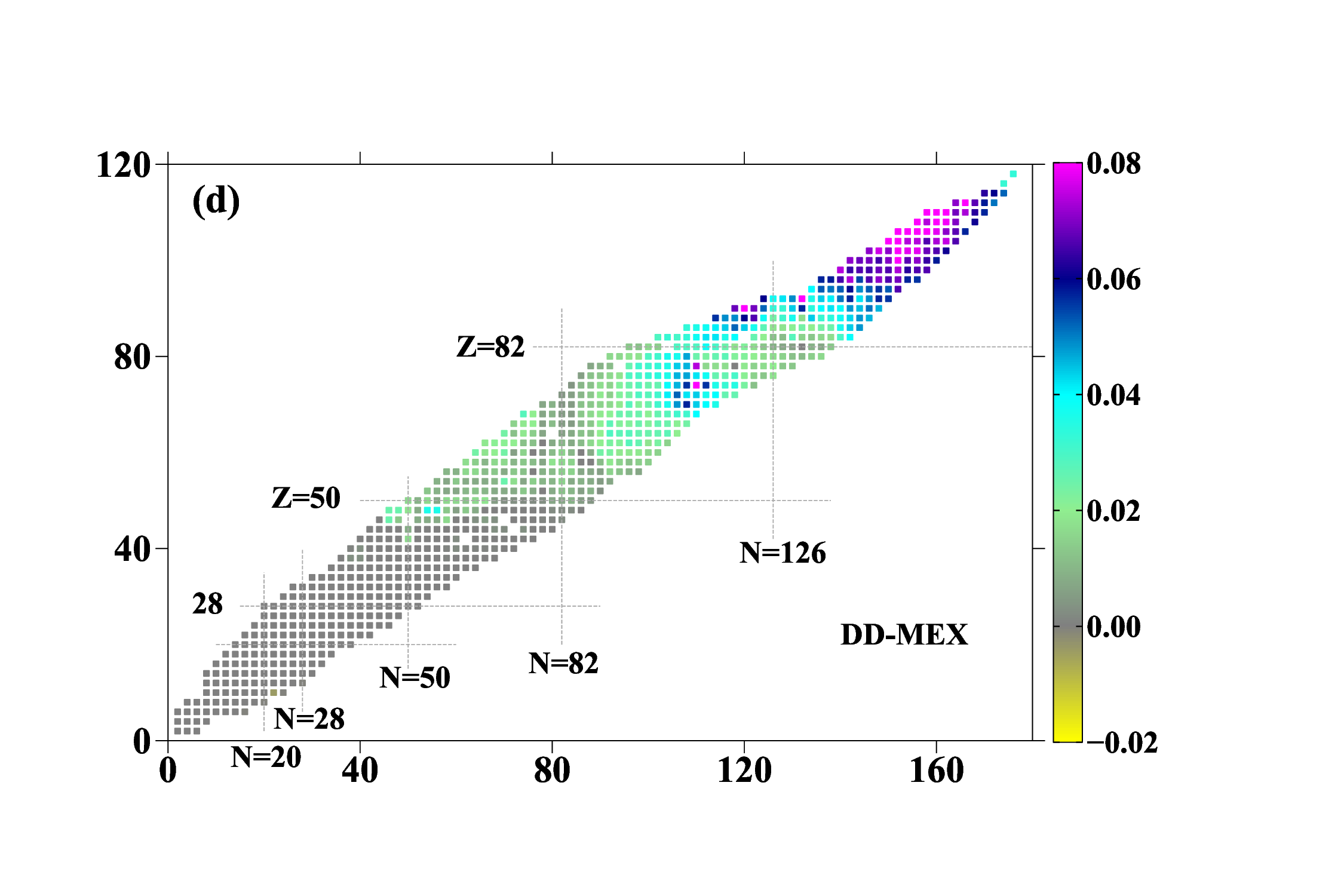}
\includegraphics*[width=7.8cm]{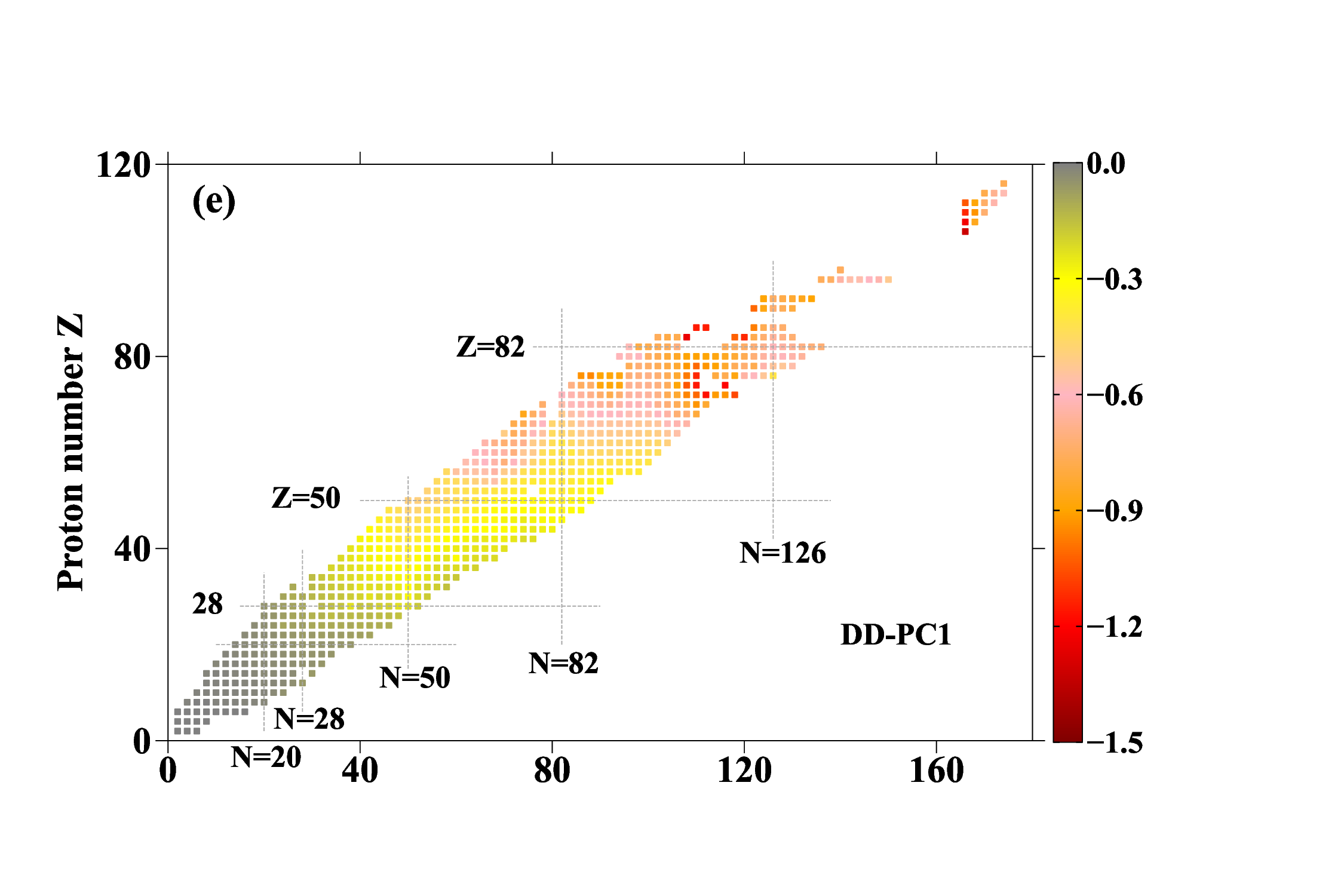}
\includegraphics*[width=7.8cm]{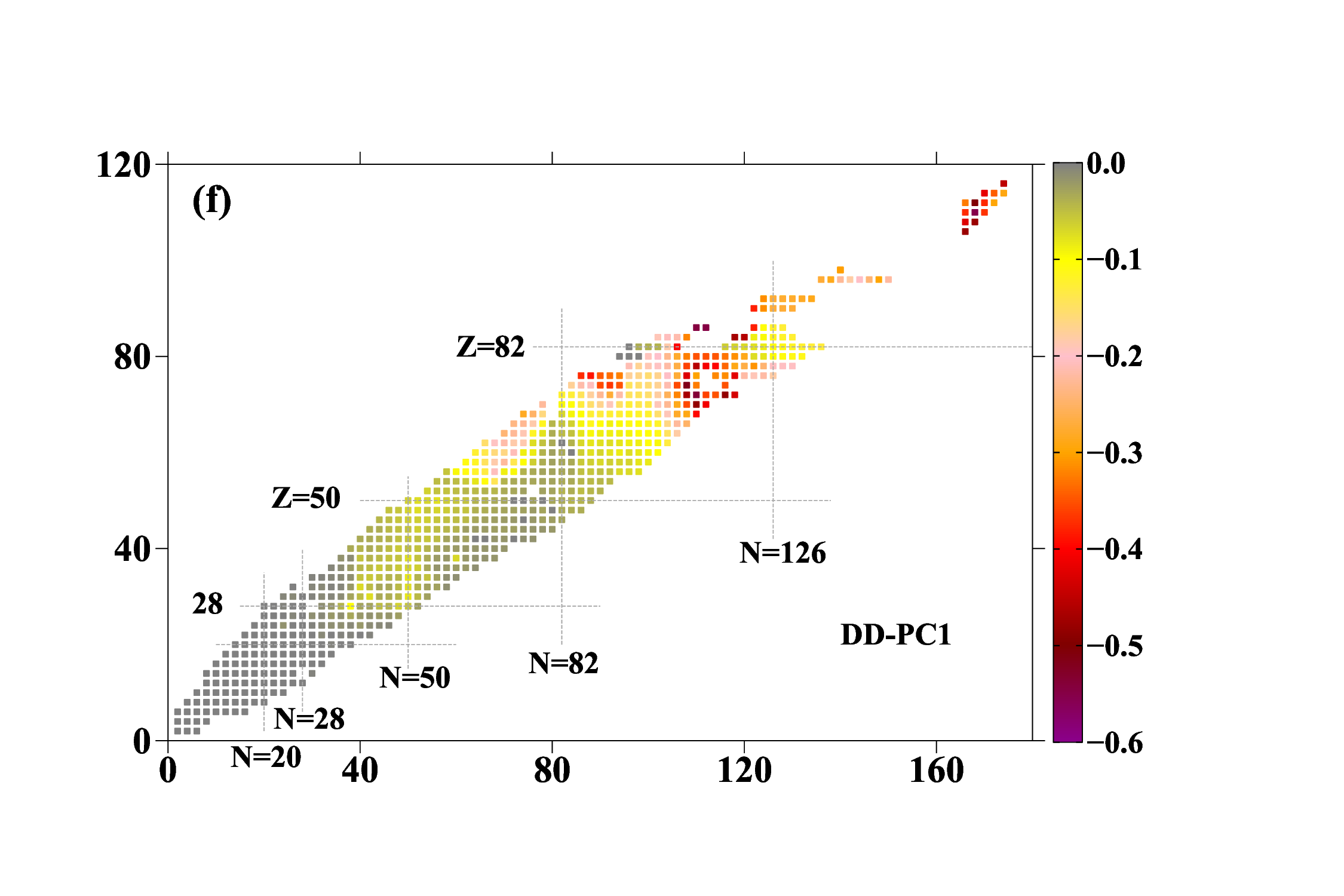}
\includegraphics*[width=7.8cm]{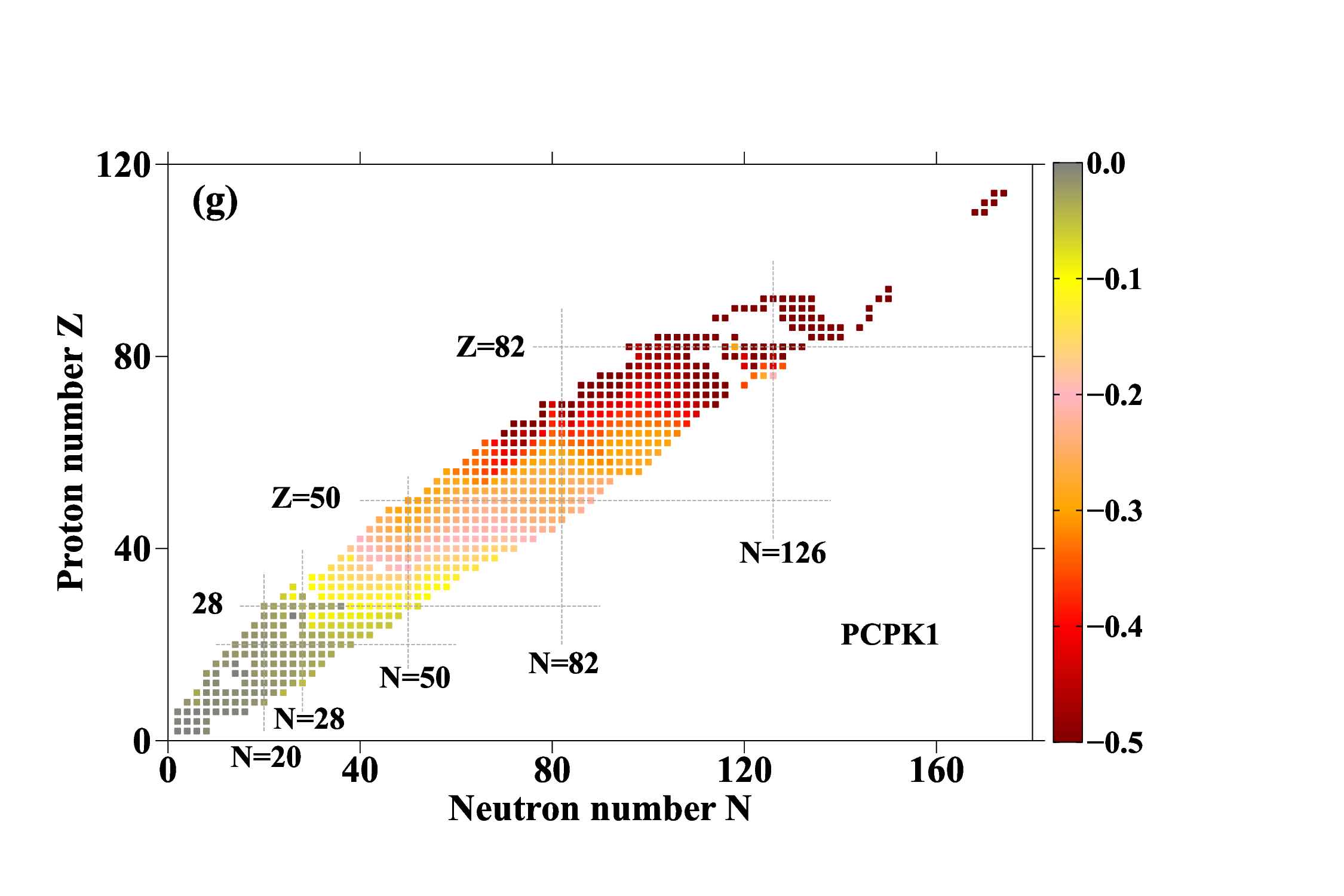}
\includegraphics*[width=7.8cm]{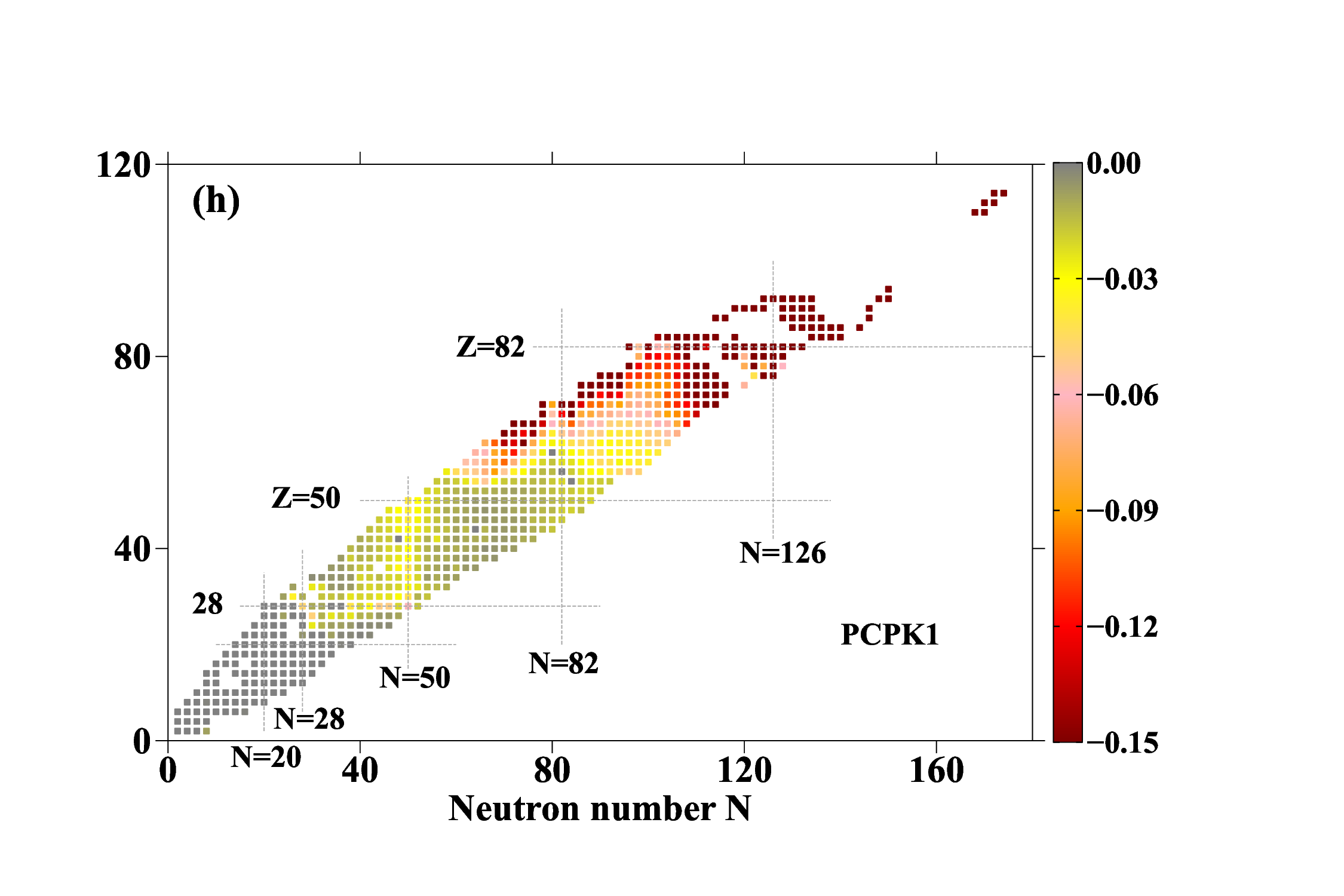}
\caption{The comparison of asymptotic binding energies with those obtained
in the calculations with $N_F=20$ and $N_F=30$. White squares are used for 
the nuclei in which either the calculations with indicated truncations of the basis 
bring the quadrupole deformations $\beta_2$ which differ by more than 0.01
or the definition of $B_{\infty}$ is not numerically possible. 
Note that the colormaps are different in different panels.
\label{B-infinity}
}
\end{figure*}

   The infinite basis binding energies $B_{\infty}(Z,N)$ obtained with discussed above 
procedure are compared with those obtained in the calculations with $N_F=20$ and 
$N_F=30$ in Fig.\ \ref{B-infinity}. The analysis of this figure reveals some important
features.

   First, the infinite basis binding energies can be defined for the NL5(E) and DD-MEX functionals 
for all nuclei of interest.  In contrast, in the case of the DD-PC1 and PC-PK1 functionals,
they can be defined only for sub-lead region and for some nuclei which have spherical 
shape in lead and superheavy region. Note that the definition of $B_{\infty}$ for latter 
nuclei is done in spherical RHB code which allows the calculations with $N_F=38$. To 
define $B_{\infty}$ for the PC functionals for transitional and deformed nuclei in the lead 
region, actinides and superheavy region one should carry out numerical calculations in 
deformed RHB code with $N_F>30$ which are numerically impossible at present. 
However, based on the examples of the DD-MEX functional [see Figs.\ \ref{B-infinity}(c) 
and (d)] and PC functionals for $Z<80$ region [see Figs.\ \ref{B-infinity}(e), (f), (g) and 
(h)], it is reasonable to expect that infinite basis  corrections $\Delta B_{\infty} (Z,N)$ for 
the deformed and spherical nuclei in the regions under discussion will be comparable 
and will not differ by more than 100-200 keV also for the PC functionals.

  Second,  the results presented in the right column of Fig.\ \ref{B-infinity}, which
compares the $B(N_F=30)$ and $B_{\infty}$ values, show that the NL5(E) CEDF
is characterized by the best convergence among considered functionals. Indeed,
only in a few nuclei the absolute value of $B_{\infty} - B(N_F=30)$ exceeds 10 
keV [see Fig.\ \ref{B-infinity}(b)].  The next best convergence is obtained for the 
DD-MEX functional for which the $B_{\infty} - B(N_F=30)$ stays below 100 keV 
even in superheavy nuclei [see Fig.\ \ref{B-infinity}(d)].  In contrast, the worst 
convergence is obtained for the functionals which contain point coupling: some 
heavy nuclei in the infinite basis are more bound than the ones in  the $N_F=30$ 
basis by around 500 keV and 150 keV in the DD-PC1 and PC-PK1 functionals, 
respectively. In addition, no $B_{\infty}$ values can be defined for actinides and
superheavy nuclei in deformed RHB calculations with $N_F\leq 30$. These
features are rather general: the analysis of other less systematic calculations 
reveals that the best convergence as a function of $N_F$ is obtained in the 
simplest class of the functionals, namely, NLME one, followed by the DDME 
CEDFs and the worst convergence exist in the PC functionals. The analysis
of the structure of the functionals suggests that the addition of derivative terms
to the functional makes convergence of the functional as a function of
$N_F$ worse.

  Third, the analysis of the results presented in the left column of Fig.\ 
\ref{B-infinity} clearly illustrates that on average the $B_{\infty}-B(N_F=20)$ 
values evolve reasonably smoothly with increasing mass number.  Note 
that the fits of many CEDFs carried out within the last two decades are 
performed  in the $N_F=20$ basis. For such functionals the part of the 
effects related to the truncation of the basis seen in the smooth trend of 
the evolution of the $B_{\infty}-B(N_F=20)$ values with mass number 
is built into the model parameters during the fitting procedure\footnote{This 
implies that for a best description of binding energies  their calculations 
with the functionals created in such a way should be performed  in the 
$N_F$ basis used in the fitting protocol. Note that such philosophy also 
applies for non-relativistic functionals fitted in restricted basis without 
infinite basis corrections (such as UNEDF class of the Skyrme functionals) 
(see Ref.\ \cite{Kort-private}).}. However, some local fluctuations and 
isospin trends seen in the $B_{\infty}-B(N_F=20)$ values are not taken 
care in such an approach. To resolve them the mapping of the binding 
energies to infinite basis is needed. This can be done by mapping 
infinite basis  corrections $\Delta B_{\infty} (Z,N)$  and using them in 
the fitting protocol which is based on the $N_F=20$ basis. These 
corrections are expected to be only marginally affected by the local 
variations of the parameters of the functional and thus their map in 
the $(Z,N)$ plane can be calculated only once.
 
\begin{figure*}[htb]
\centering
\includegraphics*[width=8.5cm]{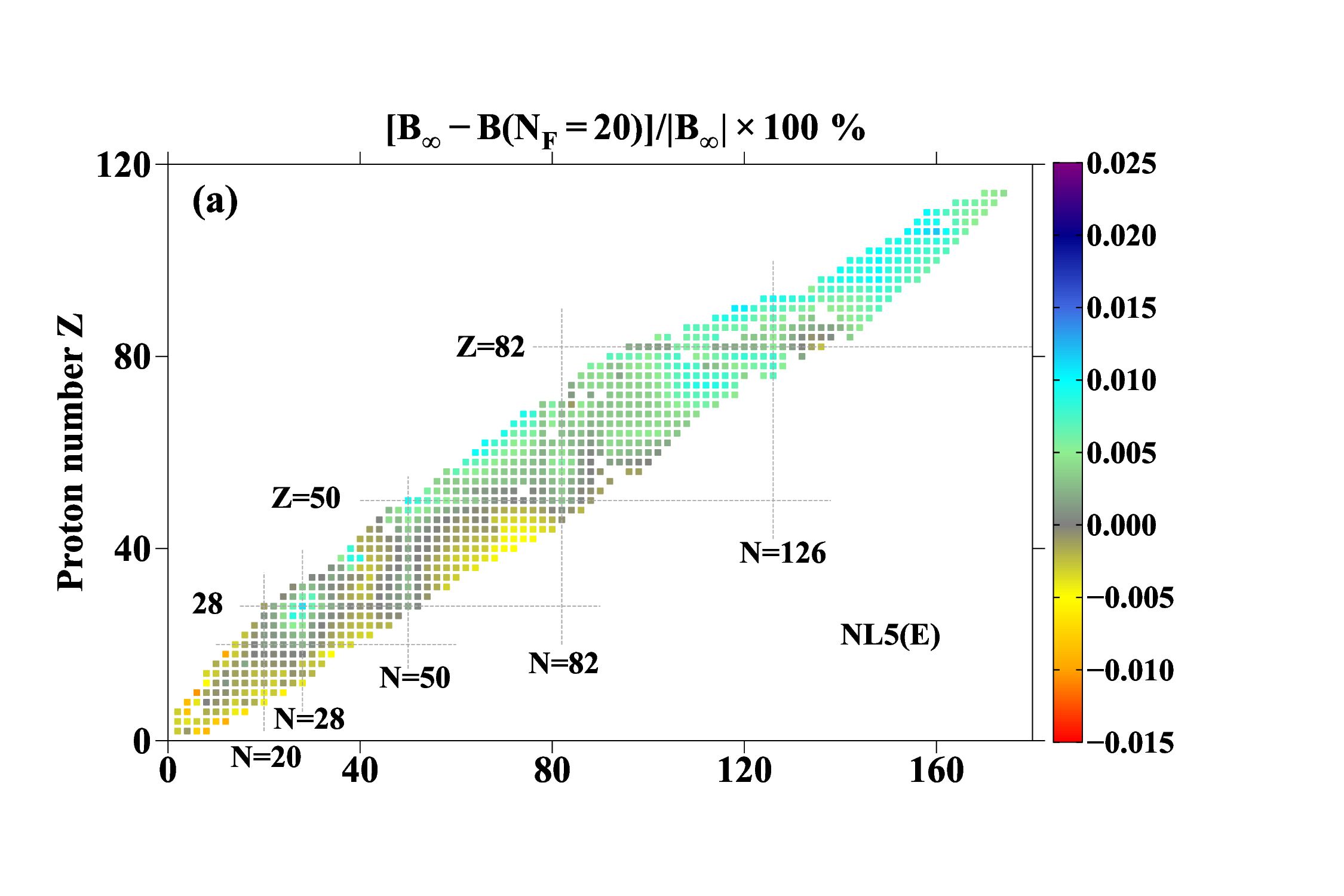}
\includegraphics*[width=8.5cm]{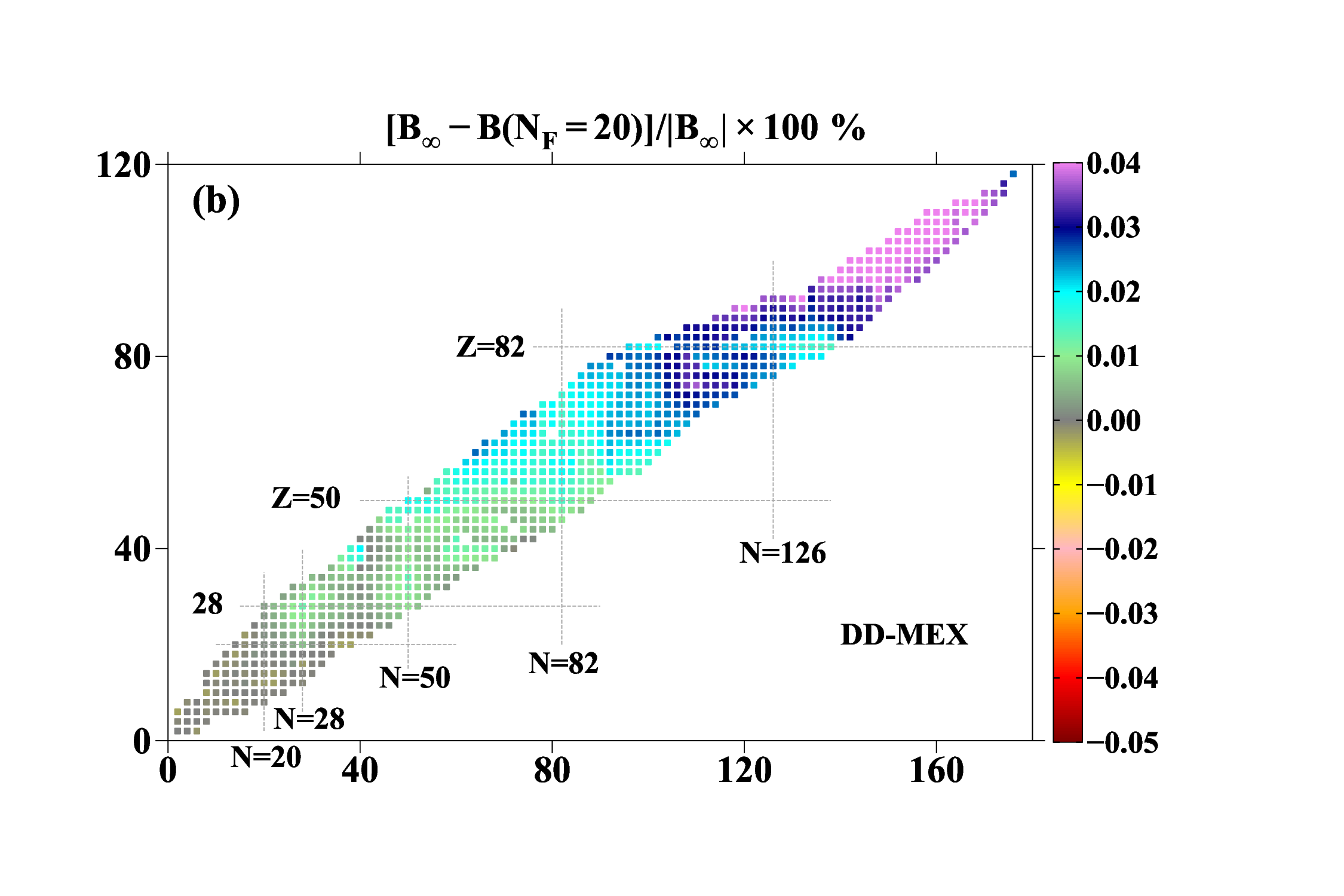}
\includegraphics*[width=8.5cm]{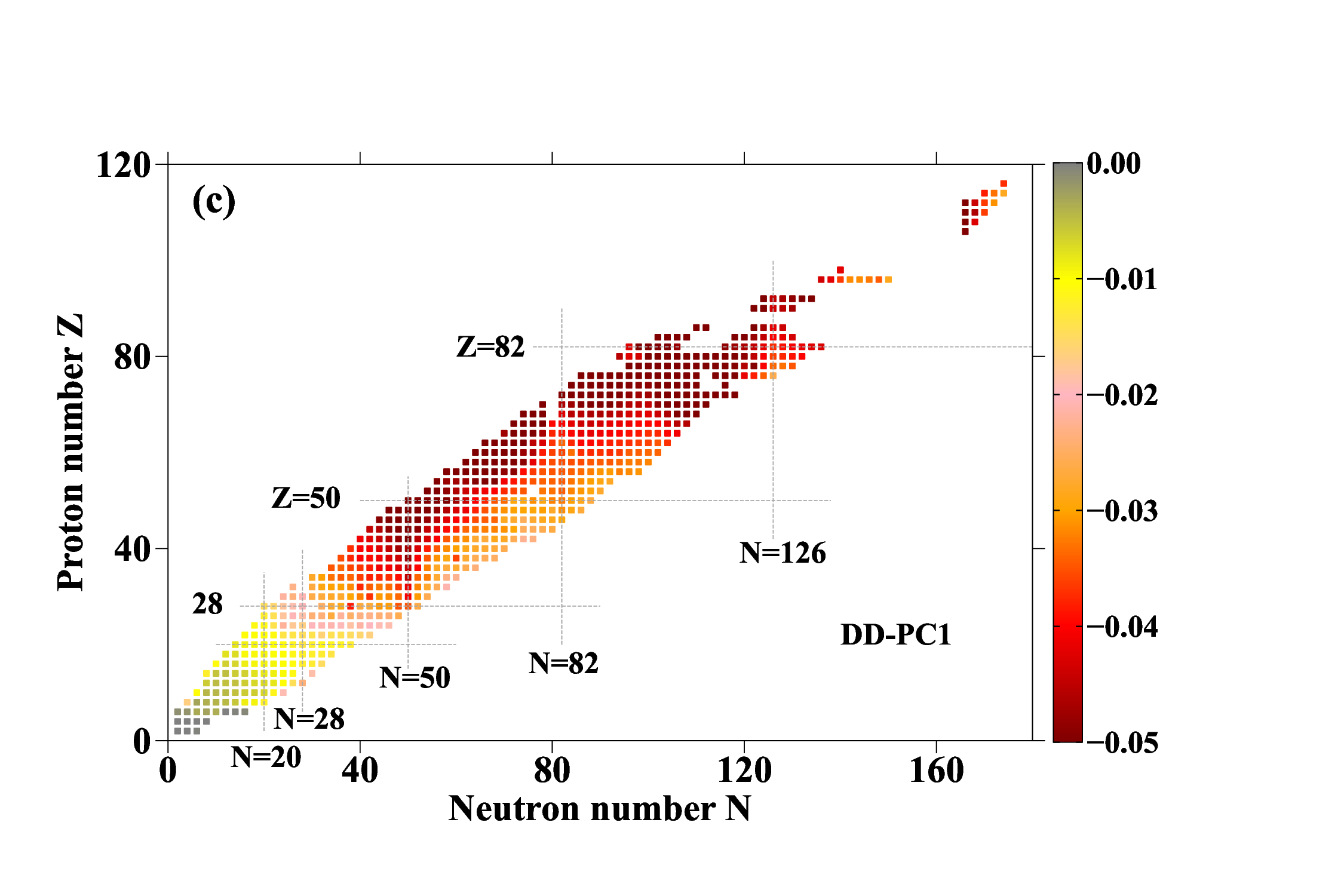}
\includegraphics*[width=8.5cm]{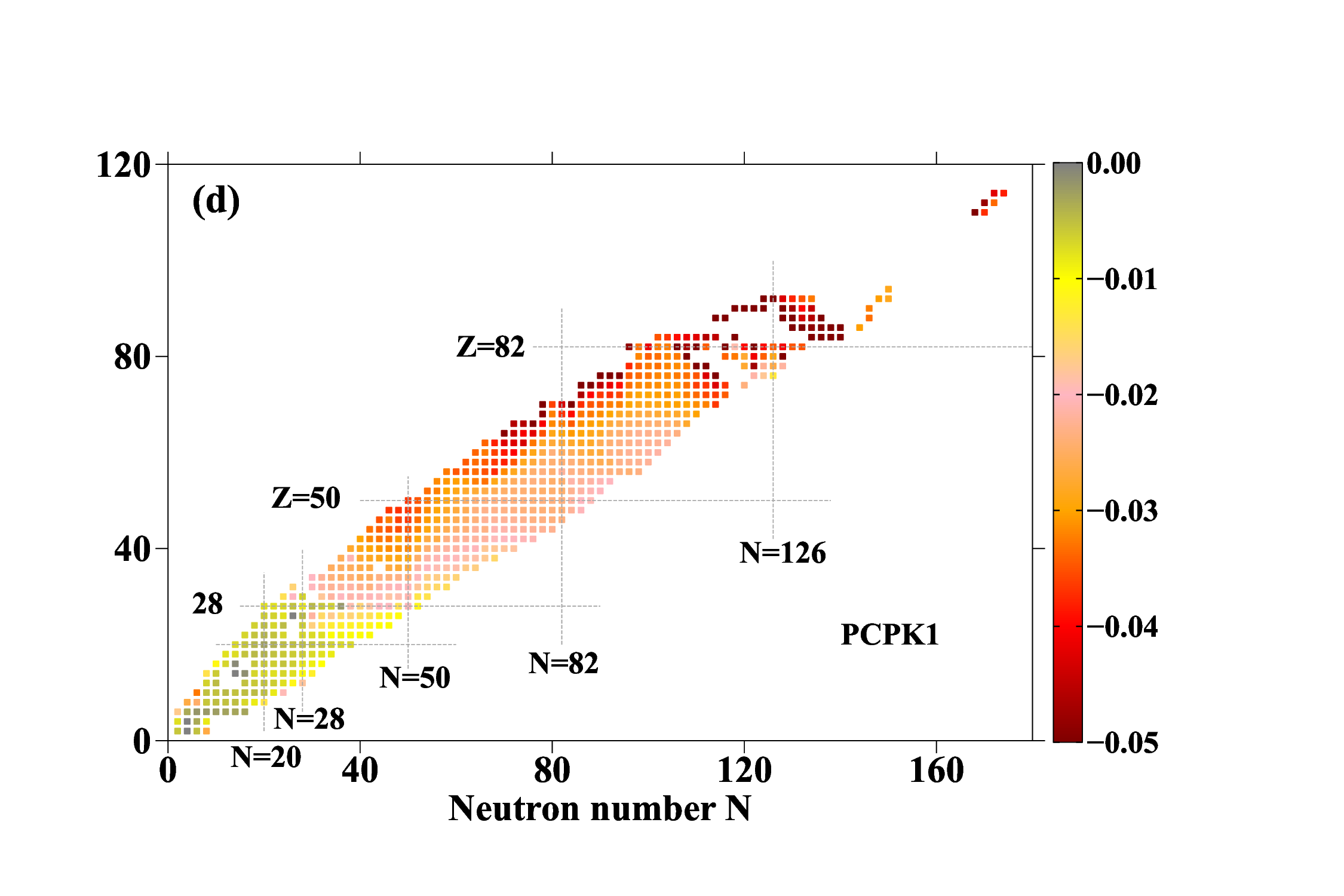}
\caption{Relative errors (in \%) in the calculations of binding energies $B(N_F=20)$ 
in the  truncation of basis with $N_F=20$ as compared with those ($B_{\infty}$)
obtained in the infinite basis. Note that the colormaps have different ranges: only
in the case of the DD-PC1 and PC-PK1 functionals the same ranges are used.
White squares are used for the nuclei in which either the calculations with the 
$N_F=20$ and $N_F=30$ bases bring the quadrupole deformations 
$\beta_2$ which differ by more than 0.01 or the definition of $B_{\infty}$ is 
not numerically possible. 
\label{rel-errors}
}
\end{figure*}

   Relative errors in the calculations of the binding energies in the
$N_F=20$ basis as compared with infinite basis binding energies 
are presented in 
Fig.\ \ref{rel-errors}. For absolute majority of the nuclei they are
better than 0.05\%. The impact of the transition from the basis with
$N_F=20$ to the infinite basis is very small and below experimental
uncertainties for other physical 
observables such as deformations, single-particle energies,
moments of inertia etc.  This is a reason why the $N_F=20$ basis 
is sufficient for the calculations of such physical observables.
The only exceptions are few transitional or 
shape-coexisting nuclei (shown by white squares in Fig.\ \ref{rel-errors}) 
in which  such a transition triggers either moderate change of 
deformation because of softness of potential energy surface 
or the transition from one to another minimum in potential energy 
surface. For such nuclei, infinite basis  correction $\Delta B_{\infty} (Z,N)$
can be defined as an average of such corrections in neighboring nuclei.

%%%%%%%%%%%%%%%%%%%%%%%%%%%%%%%%%%%% 
%%%%%%%%%%%%%%%%%%%%%%%%%%%%%%%%%%%%

%%%%%%%%%%%%%%%%%%%%%%%%%%%%%%%%%%%% 
%%%%%%%%%%%%%%%%%%%%%%%%%%%%%%%%%%%%%%

%%%%%%%%%%%%%%%%%%%%%%%%%%%%%%%%
\section{Conclusions}
\label{Concl}
%%%%%%%%%%%%%%%%%%%%%%%%%%%%%%%%  
 
   The main goal of the present study is further development of
covariant energy density  functionals towards more accurate 
description of the binding energies across the nuclear chart. It is 
focused on detailed analysis of the anchor based optimization 
approach (ABOA), its comparison with alternative global fitting 
protocols and on the global analysis of the errors related to the
truncation of bosonic and fermionic bases in the calculation of 
binding energies and on the design of new fitting protocols which 
will allow to eliminate such errors.  The main results of this study 
can be summarized as follows.

\begin{itemize}

\item
  The detailed comparison of the anchor based optimization approach 
(ABOA)  of global optimization of energy density functionals
with fully global (FGA) and reduced global (RGA) approaches  has been 
presented. The practical realization of each of these approaches is always 
a compromise between numerical accuracy in the calculation of binding
energies and the number of the nuclei included into the fitting protocol.
This limitation is due to the restrictions in the availability of computational 
power. In such a situation, an ABOA emerges as a reasonable alternative 
to FGA and RGA.  It provides a solution which is close to that obtained in 
RGA but at small portion of computational time required for RGA.
The correction function of Eq.\ (\ref{corr-func}) provides a significant push 
to a minimum within one to three iterations of ABOA and the use of a softer 
correction function of Eq.\ (\ref{eq-abg=mod}) guaranties the smoothness 
of the convergence process at further iterations.

\item
   Our analysis indicates that ten spherical anchor nuclei with 
experimental  information on charge radii are sufficient for constraining  the 
properties of  charge radii globally in ABOA. Because of their single and doubly 
magic nature they are characterized by reduced theoretical uncertainties as 
compared with a global set of the data theoretical description of which will bring 
larger theoretical uncertainties. However, spin-orbit contributions to charge radii 
and the $\frac{N}{Z}r_n^2$ term of Eq.\ (\ref{r_charge_general}) have to be 
taken  into account in the fit of the next generation of CEDFs. Note that unresolved 
{\it proton radius puzzle} puts a limit on the accuracy of the description of charge
radii.

\item
  For the first time,  the numerical errors in binding
energies  related to the truncation 
of bosonic basis have been investigated in spherical and deformed nuclei
with respect of asymptotic values corresponding to the infinite basis.
It was shown that  these errors increase with increasing the mass of the 
nucleus. They also increase on transition from spherical to deformed nuclei.
The  dependence of these errors on the functional is weak.  The truncation
of bosonic basis at $N_B=20$ provides a numerical error which is better
than 50 keV in all nuclei with exception of superheavy ones with $Z\approx 120$
and $N\approx 184$. The increase of bosonic basis to $N_B=28$, which
is easily achievable at modern computers, will reduce this error to below
10 keV for almost all nuclei of interest.

\item
  The numerical errors in binding energies related to the truncation of the 
fermionic basis have been systematically and globally investigated for 
the first time. A number of new results has been obtained. First, the way 
on how the calculated binding energies approach as a function of $N_F$  
their asymptotic values depends on the type of the functional. 
 For example, for $N_F\geq 18$, the absolute 
values of binding energy increase with increasing $N_F$ for the functionals which 
contain point coupling (PC-PK1 and DD-PC1) while it decreases for CEDFs 
which contain meson exchange (NL5(E), DD-ME2 and DD-MEX). Second, the 
convergence as a function of $N_F$ depends on the class of CEDF: in a given 
nucleus, the fastest (at lower $N_F$) approach of asymptotic values of binding
energies is seen in the NLME functionals, followed by the DDME ones and
the PC functionals are characterized by the slowest convergence. Third, for a given
functional, the increase of either mass or the deformation of nucleus requires
the increase of the size of fermionic basis for obtaining asymptotic value of
binding energy.

\item
  The present paper clearly indicates the need for accounting of infinite
basis corrections both for calculating the binding energies and their 
comparison with experimental data and for fitting of next generation
of CEDFs. The current generation of  CEDFs ignores these 
corrections.  Although partially these corrections are built into the 
parameters of the functionals,  this leads to some uncontrollable 
deviations in binding energies as compared with those obtained
in the infinite basis. The existing prescription for finding asymptotic values 
of  binding energy in the modest size basis which is used in  
non-relativistic DFTs does not provide high numerical accuracy and 
predictive power in the case of CDFT.  An alternative procedure for
finding the asymptotic values of binding energies has been suggested
for the first time in the present paper (see Sec.\ \ref{fermion-sect-asymptot-rel}). 
It is free from above mentioned deficiencies of non-relativistic prescription 
and allows better control of numerical errors in the definition of asymptotic 
binding energies.
 
 \end{itemize}

   The present paper clearly indicates that the binding energies of the
nuclei represent the most numerically demanding type of physical observable
among the ones considered at the mean  field level. This is due to both
their slow convergence as a function of $N_F$, the request for extremely 
high precision in their calculation in different subfields of nuclear physics
and nuclear astrophysics as well as high precision of their experimental
measurements. In contrast, accurate description of other physical 
observables (such as deformations, moments of inertia, etc) requires 
substantially smaller size of fermionic basis.  The present paper also indicates
the need for a new generation of CEDFs which takes into account
infinite basis corrections in the calculation of binding energies.  The work 
in defining such functionals is in progress and the results of such study
will be reported later.
 
  Note that the functionals studied in the present paper are restricted to the ones 
defined at the mean field level. However, the results obtained here can be generalized 
to the approaches which include correlations beyond mean field. For example, for that 
the RHB  approach in the anchor based optimization method  has to be replaced by an 
appropriate beyond-mean-field method (such as a five-dimensional collective Hamiltonian 
\cite{NLVPMR.09,LNVMLR.09,SALM.19}).  This will allow  to bypass the existing challenge
of extreme computational cost of fitting  EDFs at  the beyond-mean-field level and 
generate such functionals.  It is expected that 
similar to non-relativistic  (see, for example, Ref.\ \cite{D1M})  and relativistic (see Refs.\ 
\cite{ZNLYM.14,YWZL.21}) DFT approaches this will lead to a further improvement of the 
description of binding energies.  An alternative approach to the improvement of the
description of binding energy is Bayesian Neural Network (BNN) approach in which 
the fluctuating part of binding energy is described with the help of statistical methods
(see, for example, Refs.\ \cite{UPP.16,NCNV.18}). The BNN approaches improve the accuracy of the 
prediction of the binding energy at relatively modest numerical cost, provide  
statistical  errors for predicted binding energies but do not generate a microscopic insight 
on the origin of the fluctuating part of binding energy which they improve.
 
    Significant numerical cost associated with the solution of the Dirac equation
in the CDFT framework calls for a search of alternative approaches to its solution. 
The use of Dirac oscillator basis (see Refs.\ \cite{Moshinsky_1989,YP.20}) could 
be  one of these alternatives. This, however, requires the development of the RHB
code in the Dirac oscillator basis for axially deformed nuclei. It would be also 
interesting to investigate whether the reduced basis method for building of efficient 
and accurate emulators  \cite{BGKD.22,AOP.22} can be useful in the reduction of 
the computational cost. However, so far this method has been applied only to the 
single-particle energies and wavefunctions in spherical nuclei.  Thus,  the extension 
of this method to axially deformed nuclei and the investigations of both its accuracy in 
the calculation of binding energies and computational gains as compared with the 
method applied in the present paper are needed to establish its feasibility.

%%%%%%%%%%%%%%%%%%%
\section{ACKNOWLEDGMENTS}
%%%%%%%%%%%%%%%%%%%

  This material is based upon work supported by the U.S. Department of Energy,  
Office of Science, Office of Nuclear Physics under Award No. DE-SC0013037.

\bibliography{references-44-next-gen-CEDFs.bib}

\end{document}